\documentclass[11pt,letter]{revtex4-1}
\pdfoutput=1
\usepackage[latin1]{inputenc}
\usepackage{array}
\usepackage{url}
\usepackage{graphicx}
\usepackage{setspace}
\usepackage{amsmath}
\usepackage{times}
\usepackage{fullpage}
\usepackage{color}
\usepackage{sidecap}
\usepackage{wrapfig}
\usepackage{tabularray}

\usepackage[normalem]{ulem}
\usepackage{xcolor}
\newcommand\redsout{\bgroup\markoverwith{\textcolor{red}{\rule[0.5ex]{2pt}{0.4pt}}}\ULon}

\SetCell{preto=\hspace{1pt}}
\newcommand\mc[1]{\textcolor{black}{#1}}
\begin{document}

\title[TRANSP code]{TRANSP integrated modeling code for interpretive and predictive analysis of tokamak plasmas}

\author{A.Y Pankin$^1$, J. Breslau$^1$, M. Gorelenkova$^1$, R. Andre$^1$, B. Grierson$^2$, J. Sachdev$^1$, M. Goliyad$^1$, G. Perumpilly$^1$}
\affiliation{$^1$ Princeton Plasma Physics Laboratory,  Princeton, NJ 08543-0451, USA}
\affiliation{$^2$ General Atomics, San Diego, CA 92121, USA}
\email{pankin@pppl.gov}

\begin{abstract}
This paper provides a comprehensive review of the TRANSP code, a sophisticated tool for interpretive and predictive analysis of tokamak plasmas, detailing its major capabilities and features. It describes the equations for particle, power, and momentum balance analysis, as well as the poloidal field diffusion equations. The paper outlines the spatial and time grids used in TRANSP and details the equilibrium assumptions and solvers. Various models for heating and current drive \mc{and radiation}, including updates to the NUBEAM model, are discussed. The handling of large-scale events such as sawtooth crashes and pellet injections is examined, along with the predictive capabilities for advancing plasma profiles. The integration of TRANSP with the ITER Integrated Modeling and Analysis Suite (IMAS) is highlighted, demonstrating enhanced data access and analysis capabilities. Additionally, the paper discusses best practices and continuous integration techniques to enhance TRANSP's robustness. The suite of TRANSP tools, designed for efficient data analysis and simulation, further supports the optimization of tokamak operations and coupling with other tokamak codes. Continuous development and support ensure that TRANSP remains a major code for the analysis of experimental data for controlled thermonuclear fusion, maintaining its critical role in supporting the optimization of tokamak operations and advancing fusion research.

\vspace{1pc}  
\end{abstract}

\maketitle

\section{Introduction\label{intro}}
The development of controlled thermonuclear fusion is important because it offers the potential for clean, abundant, and sustainable energy that can mitigate climate change, enhance energy security, and provide long-term solutions to global energy needs. Tokamaks are one of the most widely used and studied devices in magnetic confinement fusion research due to their ability to confine plasma efficiently and their potential for achieving practical fusion energy. Experimental studies of tokamaks play a crucial role in advancing fusion physics and refining fusion reactor designs. These experiments validate theoretical concepts, optimize plasma parameters, identify and address plasma instabilities, and develop mitigation techniques for plasma instabilities. Important to this progress are reliable interpretive codes, which analyze experimental data to provide insight into the physical processes that took place during the experiment. 

One of the most actively used interpretive codes worldwide for analysis of tokamak discharges is the TRANSP code~\cite{transp,hawryluk81,ontega98,grierson18}. In continuous development since the late 1970s, TRANSP has undergone significant improvement over the years, now supporting 55 tokamak configurations and performing around 10,000 simulations per year to support current and future fusion energy experiments. TRANSP handles the time-dependent evolution of plasma profiles including the density, temperature, and rotation profiles, while considering a selection of heating and current drive sources, along with complex boundary conditions, to thoroughly and accurately determine the transport of energy during a tokamak discharge. Over time, TRANSP has undergone continuous refinement, evolving into a sophisticated code with a rich set of features. Recent modernization efforts include integrating interfaces with the ITER Integrated Modeling and Analysis Suite (IMAS) to directly access data stored in the IMAS format. Although these advancements are frequently discussed at various meetings, they have not been systematically documented in a paper until now. This paper addresses a critical gap in comprehensively describing TRANSP capabilities and features.

\begin{wrapfigure}{l}{0.55\textwidth}
    \includegraphics[width=0.54\textwidth]{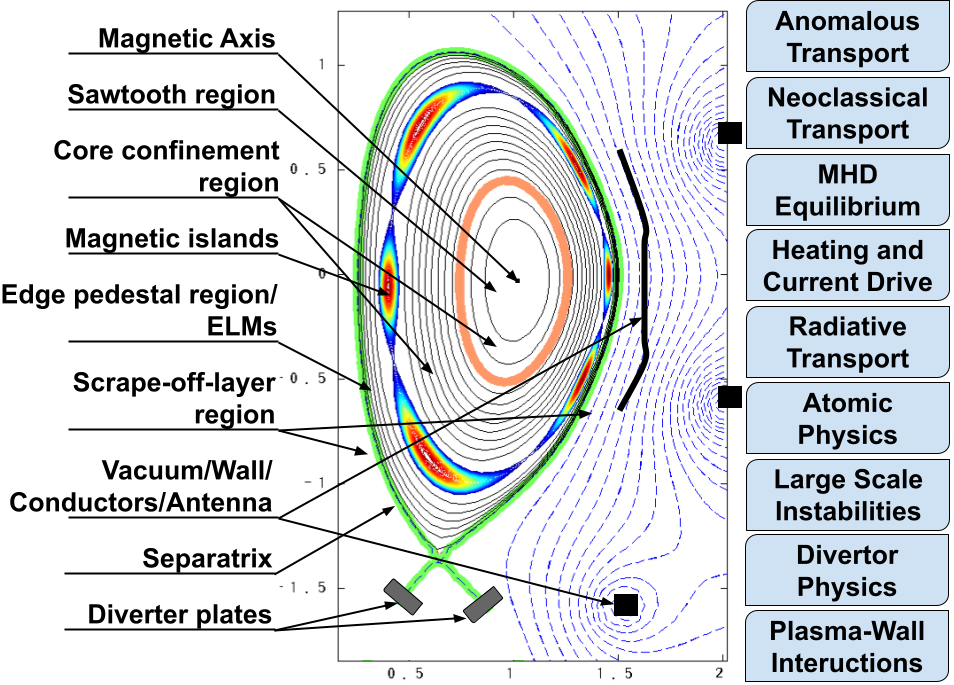} 
    \singlespacing
    \caption{Different regions, regions of large-scale macroscopic events in tokamaks, and main physics modules for the description of the tokamak physics in an integrated modeling code. } \label{fig:tokamak_cross}
\end{wrapfigure}

The tokamak is a complex fusion device including distinct regions such as the core, edge, and wall as shown in Fig.~\ref{fig:tokamak_cross}. Each of these regions plays a crucial role in the overall behavior and performance of the tokamak. To accurately describe and understand the intricate dynamics within this system, it is important to integrate multiple physics models that account for the diverse spatial and temporal scales present. These models must effectively capture phenomena occurring on scales ranging from the microscopic interactions of individual particles to the macroscopic behavior of the plasma and magnetic field as a whole. By interfacing various physics models in an integrated modeling code, it is possible to gain insight into the effects associated with different physical interactions and to optimize tokamak operation. As an integrated modeling code, TRANSP includes all components shown on the right of Fig.~\ref{fig:tokamak_cross} with the exception of plasma-wall interactions and divertor physics. 

Originally developed as an interpretive code, TRANSP is widely used in this capacity. The interpretive analysis of tokamak experiments holds significant importance for several key reasons. First, it ensures internal consistency, guaranteeing that computed results align with each other and providing a reliable foundation for understanding plasma behavior. Second, it can be used for the validation of theoretical or derived empirical models, thereby enhancing the accuracy and reliability of the descriptions of plasma behavior within the tokamak. While interpretive analysis can be challenging due to the difficulty in isolating separate aspects of the problem, it provides a comprehensive understanding of the complex interactions within the tokamak plasma, which is crucial for optimizing its performance. The interpretive analysis plays a crucial role in advancing our understanding and improving tokamak technology.

\begin{figure}[hb]
    \centering
    \includegraphics[width=0.8\textwidth]{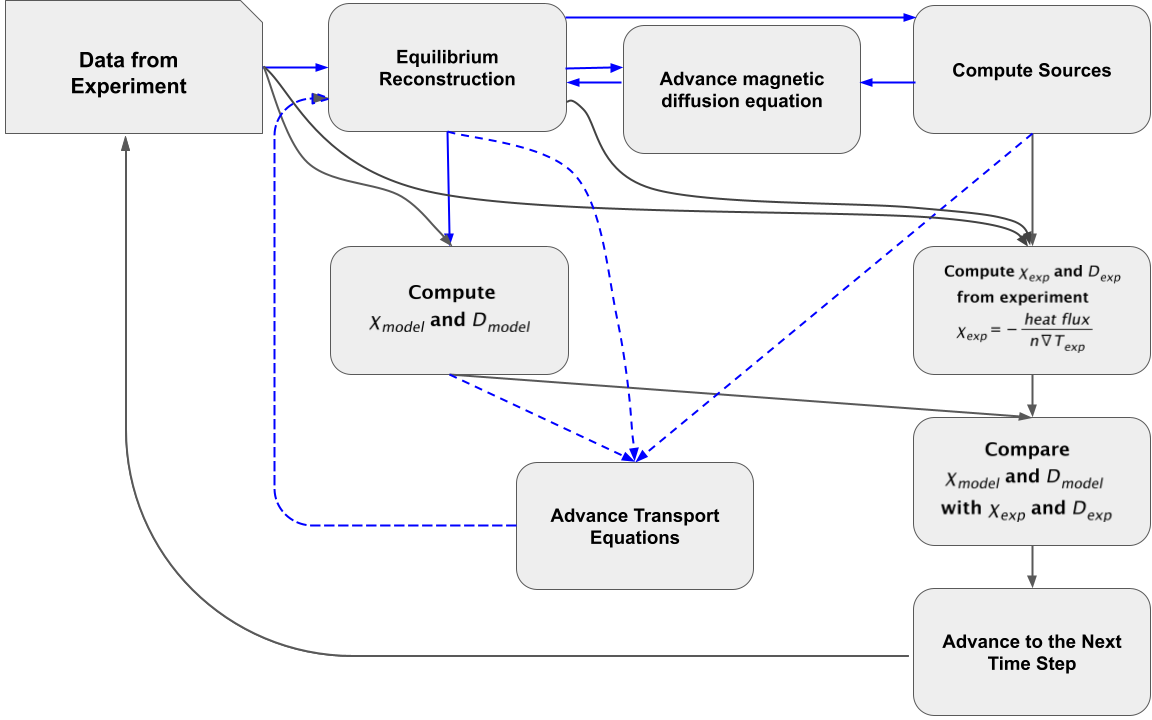}
	\caption{\mc{A simplified diagram of the TRANSP interpretive and predictive dataflows. Connections between modules that are shared between the interpretive and predictive dataflows are depicted with bold solid arrows. The connections between the modules in the interpretive mode only are shown as solid thin arrows. The connections between modules in the predictive mode only are shown as dashed arrows. The models for sawteeth, ripples, pellets and some other are not shown in the diagram. Not all transport models are available in the interpretive mode.}}
    \label{fig:dataflow}
\end{figure}

In addition to its utility as an interpretive code, another key feature of TRANSP is its predictive capability. In development since the 1990s, predictive TRANSP employs theory-based transport models for anomalous and neoclassical transport to conduct time-dependent simulations to predict tokamak behavior in diverse scenarios, allowing users to assess the impact of operational parameters on plasma performance, crucial for optimizing tokamak operation. As shown in Fig.~\ref{fig:dataflow}, the predictive and interpretive frameworks share most of the same components. This commonality helped to motivate the addition of predictive capabilities to TRANSP. The diagram in Fig.~\ref{fig:dataflow} that shows the predictive and interpretive dataflow in TRANSP is significantly simplified. It omits some models and physics effects that are included in TRANSP, such as the magnetic diffusion equation, sawtooth crashes, ripples, pellets and other effects. Nevertheless, the diagram highlights the major connections between different modules in the code in both the interpretive and predictive modes. 

Due to its versatility and comprehensive suite of capabilities, TRANSP has become an indispensable tool for the global tokamak research community. Its extensive usage spans the analysis of experimental data, validation and development of transport models, exploration of advanced tokamak scenarios, and the design of future fusion devices. TRANSP's continual refinement and adaptability contribute significantly to advancing our understanding of tokamak plasmas and, consequently, the pursuit of practical fusion energy.

The primary aim of this paper is to review the capabilities and features of the TRANSP code. The structure of the remainder of this paper is as follows: Section~\ref{sec:grids} describes the spatial and time grids used in TRANSP, highlighting the advantages of these grid systems for accurately modeling plasma behavior. Section~\ref{sec:equilibrium} discusses the assumption of nested axisymmetric flux surfaces satisfying force balance and covers the related quantities, including the free- and fixed-boundary equilibrium solvers in TRANSP. Section~\ref{sec:balance} outlines the analysis of power, particle, and momentum balance in TRANSP, using the 1.5D balance equations as the fundamental time-dependent equations governing the evolution of Maxwellian plasma. In Section~\ref{sec:mdiffusion}, the derivation and implementation of the poloidal field diffusion equation in TRANSP are presented. Section~\ref{sec:sources} describes the various models used for heating and current drive, including neutral beam injection and RF heating, and details changes to the NUBEAM model~\cite{goldston81,pankin04} developed over the past 20 years. \mc{Section~\ref{sec:radiation} details TRANSP's calculations of radiative contributions to the electron energy balance.} Section~\ref{sec:events} addresses the handling of large-scale events like sawtooth crashes and pellet injections in TRANSP simulations. Section~\ref{sec:predictive} explores the predictive options in TRANSP for advancing plasma profiles, including electron and ion temperature, electron density, and toroidal rotation. Section~\ref{sec:data} highlights the tools and methods for analyzing TRANSP simulation results. Section~\ref{sec:imas} discusses recent efforts to integrate TRANSP with the ITER Integrated Modeling and Analysis Suite (IMAS) for accessing IMAS data. Section~\ref{sec:practices} provides guidelines and best practices for using TRANSP effectively, ensuring accurate and reliable simulation results. Finally, Section~\ref{sec:summary} summarizes TRANSP's evolution into a sophisticated code for interpretive and predictive tokamak plasma analysis.

\section{Description of grids\label{sec:grids}}
\subsection{Spatial grid\label{sec:sgrid}}
TRANSP can use an arbitrary time-dependent 2D magnetic geometry based on the toroidal flux $\Phi(\rho) = \pi \rho^2 B_0$, where $\rho$ is the absolute flux surface label and $\Phi_{sep} = \pi \rho_{sep}^2 B_0$ is the flux enclosed by the plasma. Using this definition, the relative flux surface label $\xi = \rho / \rho_{sep} = \sqrt{\Phi / \Phi_{sep}}$ is the fundamental spatial coordinate in TRANSP, covering the range from $\xi=0$ to $\xi=1$.

The advantages of this grid selection are:
\begin{enumerate}
    \item \textbf{Time-invariant range:} While $\Phi_{sep}$ and $\rho_{sep}$ may vary in time, $\xi_{sep} = 1$ remains constant. These fixed $\xi$ zones always cover the core plasma to the prescribed boundary just inside the separatrix.
    \item \textbf{Independence from $B_0$:} $\xi$ does not depend on the choice of $B_0$, which is particularly useful in compression experiments.
    \item \textbf{Straightforward coordinate transformations:} For a quantity $a(\rho(t), t)$, the transformations between absolute $\rho$ and relative $\xi$ flux are simple. For example, defining $\dot{l} = \frac{1}{2\Phi_{sep}} \frac{d\Phi_{sep}}{dt}$, we get $\frac{\partial}{\partial t} \big|_{\rho} = \frac{\partial}{\partial t} \big|_{\xi} - \xi \dot{l} \frac{\partial}{\partial \xi}$. Therefore, the transport equations are solved on the $\xi$-grid but the results can be viewed on either the $\xi$ or $\rho$ grids.
\end{enumerate}

\begin{figure}[hb]
    \includegraphics[width=0.85\textwidth]{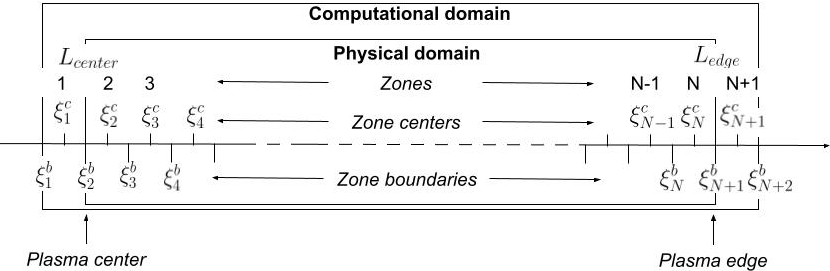}
    \justify
    \caption{Spatial grid in the TRANSP code.} \label{fig:grid}
\end{figure}

The 1D grid for flux-averaged quantities is shown in Fig.~\ref{fig:grid}. The code operates in one dimension along the minor radius. The plasma is divided into $N$ evenly spaced zones of size ${\Delta}{\xi} = 1/N$ from the magnetic axis at $\xi = 0$ to the edge at $\xi = 1$. These plasma zones are numbered sequentially from 2 to $N + 1$ with increasing $\xi$. Additionally, there are two dummy zones: one located inside the axis and one outside the edge, resulting in a total of $N + 2$ zones. Each zone is characterized by two boundary points $\xi_j^b$ and $\xi_{j+1}^b$ and one center point $\xi_j^c$.

The zone and boundary points next to the axis and edge are defined as follows:
\begin{eqnarray*}
\xi_2^c &=& {\Delta}{\xi}/2 \\
\xi_{N+1}^c & = & 1 - {\Delta}{\xi}/2 \\
\xi_2^b & = & 0 \\
\xi_3^b & = & {\Delta}{\xi} \\
\xi_{N+1}^b & = & 1 - {\Delta}{\xi} \\
\xi_{N+2}^b & = & 1 \\
\end{eqnarray*}

Some TRANSP variables are defined on zone centers, such as densities, temperatures, and energies. Other variables, such as magnetic field components, are defined on zone boundaries. The quantities defined at zone centers are interpolated to the zone boundaries:

$$
F_j^b = \frac{(\xi_j^c - \xi_j^b) F_{j-1}^c + (\xi_j^b - \xi_{j-1}^c) F_j^c}{\xi_j^c - \xi_{j-1}^c}
$$

Gradients of quantities defined at zone centers are computed from values at zone centers to give gradients defined at zone boundaries.

Table~\ref{tab:grid} defines some major TRANSP variables associated with the spatial grid. The poloidal and toroidal fluxes are closely connected to the equilibrium formulations described in Sec.~\ref{sec:equilibrium}. The mapping of the $\xi$ grid to these variables as well as to the various radii is performed only when the input equilibrium is present.

\begin{table}[ht]
\centering
\caption{Major TRANSP input/output variables related to the spatial grid.}
\label{tab:grid}
\begin{tabular}{|c|c|c|c|}
\hline
\textbf{TRANSP variable} & \textbf{Quantity} & \textbf{Description} & \textbf{Unit} \\ 
\hline
\texttt{X}       & $\xi^c$ & Relative flux surface label at zone center& - \\
\texttt{XB}      & $\xi^b$ & Relative flux surface label at zone boundary& - \\
\texttt{TRFLX}  & $\rho$ & Toroidal flux used in the code & Wb \\
\texttt{TRFCK}  & $\rho$ & Toroidal flux computed & Wb \\
\texttt{TRFLXD} & $\rho$ & Toroidal flux (input) & Wb \\
\texttt{NZONES} & $N$ & Number of zones & - \\
\texttt{ELDOT}  & $\dot{l}$ & Grid motion & $1/s$ \\
\texttt{RMNMP} & $r$ & Midplane minor radius & cm \\
\texttt{RMJMP} & $R$ & Midplane major radius & cm \\
\texttt{RZON}   & $r^c$ & Zone radius & cm \\
\texttt{RBOUN}  & $r^b$ & Radius of zone boundary & cm \\
\texttt{PLFLXA}  & $\psi$ & Enclosed poloidal flux used in the code & Wb/rad \\
\texttt{PLFLXD } & $\psi$ & Enclosed poloidal flux (input) & Wb/rad \\
\texttt{PSI0\_TR} & $\psi_0$ & Poloidal flux on axis used in the code & Wb/rad \\
\texttt{PSI0\_DATA} & $\psi_0$ & Poloidal flux on axis (input) & Wb/rad \\
\hline
\end{tabular}
\end{table}

\subsection{Time grid\label{sec:tgrids}}
The time step for the transport analysis is constrained within the ranges set by the TRANSP input. The actual time step is adjusted before each time cycle. The variables \texttt{DTMINT}, \texttt{DTMAXT}, and \texttt{DTINIT} set the minimum, maximum and and initial time steps for transport analysis. The individual modules in TRANSP can have different time steps that can be controlled with similar sets of input parameters. The plasma quantities are interpolated to a corresponding time within a particular module. Table~\ref{tab:tgrid} lists some of time control variables and their default values. The transport time step is reduced if the transport analysis has failed on the current step. In this case, the plasma parameters are restored to the old values and the step is repeated with a reduced time step. The time step for the ECH calculations \texttt{DTECH} does not have a default value and it needs to be defined through the TRANSP input.

\begin{table}[ht]
\centering
\caption{Time step controls.}
\label{tab:tgrid}
\begin{tabular}{|c|c|c|c|}
\hline
\textbf{TRANSP variable} & \textbf{Description} & \textbf{Default value} & \textbf{Unit} \\ 
\hline
\texttt{DTMINT} & Minimum time step for transport analysis & $10^{-7}$ & s \\
\texttt{DTMAXT} & Maximum time step for transport analysis & $2\times10^{-3}$ & s \\
\texttt{DTINIT} & Initial time step for transport analysis & $10^{-3}$ & s \\
\texttt{DTMING} & Minimum geometry (equilibrium) time step & $10^{-5}$ & s \\
\texttt{DTMAXG} & Maximum geometry (equilibrium) time step & $10^{-2}$ & s \\
\texttt{DTMINB} & Minimum time step for magnetic diffusion equation & $10^{-7}$ & s \\
\texttt{DTMAXB} & Maximum time step for magnetic diffusion equation & $2\times10^{-3}$ & s \\
\texttt{DTBEAM} & Time step for neutral beam & $5\times10^{-3}$ & s \\
\texttt{DTLH} & Time step for the  LH calculations & $5\times10^{-3}$ & s \\
\texttt{DTICRF} & Time step for the ICRF calculations & $5\times10^{-3}$ & s \\
\texttt{DTECH} & Time step for the ECH calculations & $0.0$ & s \\
\hline
\end{tabular}
\end{table}

\section{Equilibrium\label{sec:equilibrium}}
\subsection{Flux surface quantities\label{sec:fluxsurf}}
\begin{figure}[h]
  \includegraphics[width=2.0in]{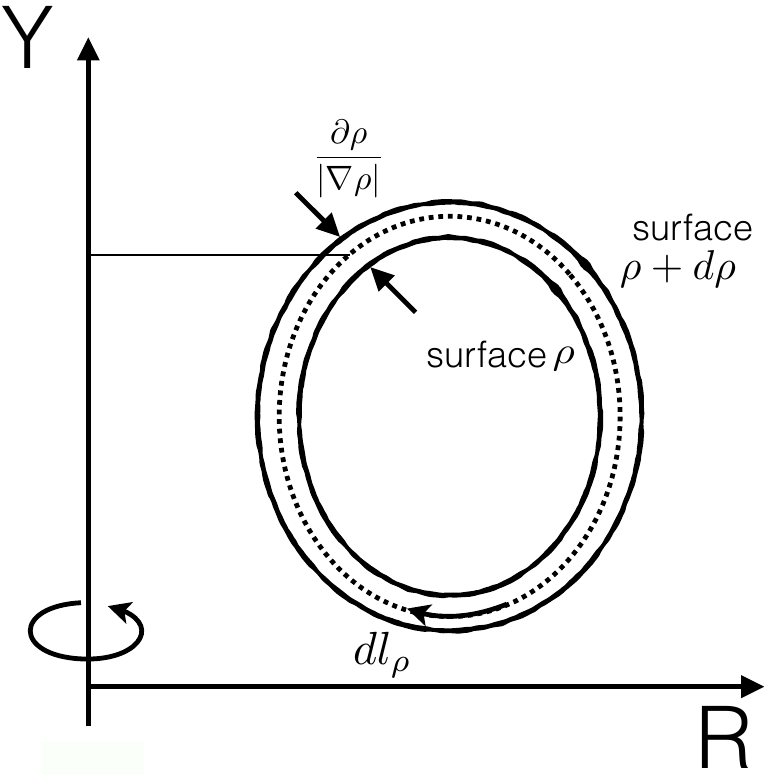}
  \caption{\label{fig:transp_surface}General flux surface.}
\end{figure}

TRANSP assumes that the plasma is composed of a sequence of nested axisymmetric flux surfaces satisfying force balance $\mathbf{J} \times \mathbf{B} = \nabla P$ without islands, separatrices, ergodic regions, or other singularities. In this model, $\nabla P$ is perpendicular to the flux surface, while $\mathbf{J}$ and $\mathbf{B}$ lie within the surface. Consequently, several quantities become constant on a flux surface, including pressure $p$, density $n$, and temperature $T$, toroidal flux $\Phi$, poloidal flux $\psi$, rotational transform $\iota = 1/q$, the product of major radius and toroidal field $RB_T$, and loop voltage $V_l = 2\pi(RE_T)$. Additionally, geometric quantities derived from the equilibrium model, such as shifted circle or \texttt{gEQDSK}, are generalized for flux surface $j$ and include formulas for concentric toroid equivalents: zone volume $\Delta V_j$, zone cross-sectional area $\Delta A_j$, surface area $S_j$, poloidal circumference $L_j$, gradient operator $\langle | \nabla \xi | \rangle \frac{\partial}{\partial \xi}$, zone averages of $1/R$ and $1/R^2$, and the zone average of $\langle \frac{| \nabla \xi |^2}{R^2} \rangle$. These general flux surface zones are conceptually represented in Fig.~\ref{fig:transp_surface}, showing a surface of constant toroidal flux label $\rho$ or $\xi$, where the $\rho$ surfaces coincide with the $\xi$ surface and $\partial \rho = \rho_{sep} \partial \xi$.

The differential volume element is defined as
\begin{equation}
  \frac{1}{\rho_{sep}}\frac{\partial V}{\partial \xi} = \frac{\partial V}{\partial \rho} = 2\pi \oint dl_\rho \frac{R}{|\nabla \rho|}
\end{equation}
making the differential volume average of quantity $f$
\begin{eqnarray}
  \langle f \rangle_V & = &\bigg(\frac{\partial V}{\partial \rho}\bigg)^{-1} 2\pi \oint dl_\rho f \frac{R}{|\nabla \rho|} \\
  & = & \bigg(\frac{\partial V}{\partial \xi}\bigg)^{-1} 2\pi \oint dl_\rho f \frac{R}{|\nabla \xi|}
\end{eqnarray}
where the real-space mapping that is used comes from a Fourier moment expansion of the flux surfaces from an equilibrium solver given here:
\begin{eqnarray}
  R(\rho,\theta) = R_0(\rho) + \sum_{j=1}^{J \le 64} R_j^c(\rho) \cos(j\theta) + R_j^s(\rho) \sin(j\theta) \\
  Y(\rho,\theta) = Y_0(\rho) + \sum_{j=1}^{J \le 64} Y_j^c(\rho) \cos(j\theta) + Y_j^s(\rho) \sin(j\theta)
\end{eqnarray}
It is these Fourier coefficients that are used when performing the numerical integration of quantities $\langle f \rangle_V$ needed for the transport and poloidal field diffusion calculations (such as neutral beam quantities that are not flux functions).
TRANSP works with a grid of $N$ discrete radial zones set in the input namelist, and therefore the quantities $\langle f \rangle_V$ need to be integrated across finite zone width as well as around a surface.
For zone $j$ spanning toroidal flux $\xi_j \rightarrow \xi_{j+1}$ the zone volume is
\begin{equation}
  \Delta V_j =
  \int_{\xi_j}^{\xi_{j+1}} d\xi \bigg(\frac{\partial V}{\partial \xi}\bigg) =
  \int_{\xi_j}^{\xi_{j+1}} d\xi \oint dl_\rho \frac{2\pi R}{| \nabla \xi |}
\end{equation}
which is used in integrating the particle, energy and momentum
densities in the balance equations.  The zone cross-sectional area is
\begin{equation}
  \Delta A_J = \int_{\xi_j}^{\xi_{j+1}} d\xi \oint dl_\rho \frac{1}{| \nabla \xi |}
\end{equation}
which is used in integrating the toroidal current densities for component contributions to the total current.
Two further useful surface quantities are the surface area of a flux toroid
\begin{equation}
  S_j = 2 \pi \oint R dl_\rho
\end{equation}
which is used in the continuity equation in the divergence of the cross-field flux $\langle \nabla \cdot \Gamma \rangle_j = (S_{j+1}\langle \Gamma \rangle_{j+1} - S_j \langle \Gamma \rangle_j)/\Delta V_j$.
The poloidal path length
\begin{equation}
  L_j = \oint dl_\rho
\end{equation}
is used in the average poloidal magnetic field where $|B_p| = (|\nabla \xi|/R) \partial \psi/\partial \xi$ and its volume average is
\begin{eqnarray}
  \langle B_p \rangle_V &=& 2\pi \bigg(\frac{\partial V}{\partial \xi}\bigg)^{-1} \oint dl_\rho B_p \frac{R}{|\nabla \xi|} \\
  &=& 2\pi \bigg(\frac{\partial V}{\partial \xi}\bigg)^{-1} \frac{\partial \psi}{\partial \xi} L_j
\end{eqnarray}
because $\partial \psi/\partial \xi$ is a flux function.
The volume average of the flux surface label gradient $|\nabla \xi|$ is
\begin{equation}
  \langle |\nabla \xi| \rangle = \frac{1}{\Delta V_j} 2 \pi \int_{\xi_j}^{\xi_{j+1}} d\xi \oint dl_\rho R
\end{equation}
which is used in defining the gradient of flux functions $f(\xi)$ where $\nabla f = (\partial f/\partial \xi) \nabla \xi$ so $\langle |\nabla f|\rangle = \partial f/\partial \xi \langle | \nabla \xi |\rangle$.
In the transport equations, this is used to define the zone averaged gradient to combine with diffusivities.
The volume average of $1/R$ is
\begin{equation}
  \langle 1/R \rangle = \frac{1}{\Delta V_j} 2 \pi \int_{\xi_j}^{\xi_{j+1}} d\xi \oint dl_\rho \frac{1}{|\nabla \xi|} = \frac{2\pi \Delta A_j}{\Delta V_j}
\end{equation}
which is used in the definition of the toroidal electric field because the toroidal loop voltage $V_l \sim R E_T$ is a flux function: $\langle E_T \rangle = R E_T \langle 1/R \rangle$.
The volume average of $1/R^2$ is
\begin{equation}
  \langle 1/R^2 \rangle = \frac{1}{\Delta V_j} 2 \pi \int_{\xi_j}^{\xi_{j+1}} d\xi \oint dl_\rho \frac{1}{R|\nabla \xi|}
\end{equation}
which is used in the average toroidal field energy density $\langle B_T^2 / 2 \mu_0 \rangle = \frac{(RB_T)^2}{2\mu_0}\langle 1/R^2 \rangle$ where $(RB_T)^2$ is a flux function.
The final volume averaged quantity presented is the $1/R^2$ weighted squared flux surface label gradient
\begin{equation}
  \langle |\nabla \xi|^2 / R^2 \rangle =
  \frac{1}{\Delta V_j} \int_{\xi_j}^{\xi_{j+1}} d\xi \oint dl_\rho \frac{2 \pi |\nabla \xi|}{R}
\end{equation}
which is used in the average poloidal field energy density $\langle B_p^2/2\mu_0 \rangle = (1/2\mu_0)(\partial \psi/\partial \xi)^2 \langle | \nabla \xi |^2/R^2\rangle$.

In summary, evaluation of the geometric factors $\Delta V_j, \ldots \langle 1/R \rangle \ldots$ with sufficient accuracy for time differentiation enables solution of the 1.5D generalized transport and poloidal field equation.
The transport and field equations are developed on a flux surface $\rho$ grid, but in TRANSP they are solved on the normalized $\xi=\rho/\rho_{sep}=(\Phi/\Phi_{sep})^{1/2}$ grid, which is time invariant (spanning $\xi=[0,1]$).
The coordinate transformation is straightforward with $\partial \rho = \rho_{sep} \partial \xi$.

\subsection{Equilibrium solvers\label{sec:eqbmsolvers}}
In interpretive mode, TRANSP supports three distinct options for advancing its representation of the plasma MHD equilibrium in time.  {\mc{ These are selected by the choice of the integer namelist variable \texttt{LEVGEO}.}}  The first and most fundamental {\mc{ option ($\texttt{LEVGEO} = 8$)}} is to read in equilibrium profiles computed by an external code, performing interpolation in time as necessary, and not recomputing the equilibrium internally. The second option {\mc{($\texttt{LEVGEO} = 11$)}} is to invoke the fixed-boundary inverse equilibrium solver TEQ to solve for the equilibrium at each geometry timestep given the pressure and $q$ profiles and the value of $R\cdot B_{toroidal}$ at the boundary.  The edge $q$ is adjusted iteratively to match the total plasma current.

The third and most versatile option {\mc{($\texttt{LEVGEO} = 12$)}} is for TRANSP to use the ISolver free boundary solver to advance the equilibrium. ISolver maintains a detailed model of the geometry and material characteristics of the poloidal field coil set for each device for which it is run, and can either perform a least-squares fit for the PF coil currents to match a prescribed plasma boundary; or solve a circuit equation incorporating coil current data, feedback circuits, and induced vessel currents, with the shape of the plasma boundary computed self-consistently. ISolver can also advance the $q$ profile self-consistently instead of using TRANSP's poloidal field diffusion equation, enabling modeling of the inductive coupling of the coils and vessel to the plasma.

In addition to toroidal and poloidal variables defined in Table~\ref{tab:grid}, TRANSP output includes the variables described in Table~\ref{tab:equil}. These quantities are independent of the equilibrium solver as long as equilibrium options are enabled in the TRANSP inputs.

\begin{table}
\begin{tabular}{|c|c|c|}
\hline
\textbf{TRANSP Variable} & \textbf{Description} & \textbf{Units} \\ \hline
\verb+Q+ & Safety factor $q$ & - \\ \hline
\verb+DVOL+ & Flux surface zone volume $\Delta V_j$ & $cm^3$ \\ \hline
\verb+DAREA+ & Flux surface zone cross-sectional area $\Delta A_j$ & $cm^2$ \\ \hline
\verb+SURF+ & Flux surface area $S_j$ & $cm^2$ \\ \hline
\verb+LPOL+ & Poloidal path length $L_j$ & $cm$ \\ \hline
\verb+GRI+ & Inverse radius $\langle 1/R \rangle$ & $cm^{-1}$ \\ \hline
\verb+GR2+ & Squared radius $\langle R^2 \rangle$ & $cm^2$ \\ \hline
\verb+GR2I+ & Inverse squared radius $\langle 1/R^2 \rangle$ & $cm^{-2}$ \\ \hline
\verb+GB1+ & Magnetic field $\langle B \rangle$ & $T$ \\ \hline
\verb+GB2+ & Squared magnetic field $\langle B^2 \rangle$ & $T^2$ \\ \hline
\verb+GB2I+ & Inverse squared magnetic field $\langle 1/B^2 \rangle$ & $T^{-2}$ \\ \hline
\verb+GBR2+ & $R^2$ weighted magnetic field $\langle BR^2 \rangle$ & $T \cdot cm^2$ \\ \hline
\verb+GXI+ & $\xi$ gradient $\langle |\nabla \xi| \rangle$ & $cm^{-1}$ \\ \hline
\verb+GXI2+ & Squared $\xi$ gradient $\langle |\nabla \xi|^2 \rangle$ & $cm^{-2}$ \\ \hline
\verb+GR2X2+ & $R^2$ squared $\xi$ gradient $\langle R^2 |\nabla \xi|^2 \rangle$ & - \\ \hline
\verb+GX2R2I+ & Inverse $R^2$ squared $\xi$ gradient $\langle |\nabla \xi|^2/R^2 \rangle$ & $cm^{-4}$ \\ \hline
\verb+GX2B2I+ & Inverse $B^2$ squared $\xi$ gradient $\langle |\nabla \xi|^2/B^2 \rangle$ & $cm^{-2}/T^2$ \\ \hline
\end{tabular}
\caption{Major TRANSP output variables related to equilibrium.}
\label{tab:equil}
\end{table}

\subsubsection{External Equilibrium}

The information that must be provided to TRANSP to represent an equilibrium solution includes the 2D flux surface geometry as a function of minor radius (normalized toroidal flux), poloidal angle, and time; 1D profiles for $q$, $R\cdot B_{toroidal}$, and pressure $p$ as functions of minor radius and time; and 0D time histories of total toroidal and poloidal flux.  A TRANSP utility can extract this data from EFIT equilibria in an MDS+ database tree and write them to legacy format UFiles that can be read by the TRANSP preprocessor.  TRANSP now also has the ability to retrieve this information directly from an IMAS database.

Because TRANSP was co-developed with PPPL's TFTR tokamak, which had a circular plasma cross-section, efficiency dictated the choice of a moments representation of the equilibrium geometry, with cylindrical $R$ and $Z$ for a plasma surface provided as cosine and sine expansions in poloidal mode number.  This moments representation is still supported by the code. Since 2002, however, TRANSP has also supported the alternative to read in and use $R(x,\theta,t)$ and $Z(x,\theta,t)$, which avoids truncation errors in the representation of the more highly-shaped cross-sections of modern tokamaks. TRANSP can also read the poloidal flux function $\psi(R,Z,t)$, {\mc{ as reconstructed by EFIT,}} which provides the field solution outside the last closed flux surface.  These last three functions are always present in the representation read from IMAS{\mc{, but can also be supplied in UFiles}}.

When $\psi(R,Z,t)$ is provided, TRANSP performs additional analysis on the equilibrium.  It attempts to locate all X-points, the magnetic axis, and the bounding point of the plasma, which may be either an X-point or a location on the limiter.  For X-point-bounded plasmas, it has the option to trace the separatrix for interface with scrape-off-layer codes.  It can also compute the normalized inductances at the analysis boundary and the separatrix.

The presence of a free-boundary $\psi(R,Z,t)$ also provides the option to directly compute the locations of the flux surfaces (overriding the provided geometry data) and the enclosed poloidal flux within each surface. When this poloidal flux profile is combined with the provided $q$ and $R\cdot B_{toroidal}$ profiles, the equilibrium is overdetermined, which can be resolved in a number of ways determined by the user. By default, TRANSP will use its self-evolved $q$ profile combined with $\psi(R,Z,t)$ to recalculate the toroidal flux profile, which in turn affects the geometry mapping.  Another alternative is to use the input $q$ profile for this calculation.  To give priority to $R\cdot B_{toroidal}$, users can instead elect to recompute $q$ based on the poloidal and toroidal fluxes; or to use the provided $R\cdot B_{toroidal}$ and $q$ and recompute the equilibrium to match $\psi(R,Z,t)$, ignoring the provided pressure profile.

{\mc{ For the $\texttt{LEVGEO} = 8$ option in general, it should be noted that the pressure profile evolved by TRANSP will not exactly match that originally used by EFIT to provide the input.  If poloidal field diffusion is turned on, the $q$ profile will differ as well.  In both cases, $g = R B_{T}$ will be internally computed by surface averaging the Grad-Shafranov solution and will not exactly match the input EFIT profile.  TRANSP equilibria in this mode are therefore evidently not entirely self-consistent, and are not suitable for analysis with MHD stability codes.}}

\subsubsection{TEQ fixed boundary equilibrium solver}

The equilibrium solver TEQ originally formed the basis of Lawrence Livermore's Corsica transport code~\cite{CORSICA}.
Given a plasma boundary with a known vacuum $F = R B_{T}$ along with pressure and $q$ profiles, the fixed-boundary form of the solver begins with an initial guess for the Grad-Shafranov solution

\begin{equation}
\Delta^\star\psi = -\mu_0 R^2 \frac{\partial p}{\partial\psi} -F \frac{\partial F}{\partial\psi}
\end{equation}

discretized on a $(\xi,\theta)$ grid,
and then iteratively adjusts this solution, changing $q_{edge}$ to match the total plasma current with $F$ as a boundary condition until the residual average Grad-Shafranov error is small.
Some user-accessible controls for TEQ are listed in Table~\ref{tab:teqcontrols}.
Output quantities for plotting include the achieved average G-S error \verb+GSERROR+ and the TEQ residual \verb+TEQRESID+.

\begin{table}
\begin{tabular}{|c|c|c|c|}
\hline
\textbf{TRANSP Variable} & \textbf{Description} & \textbf{Default value} \\
\hline
\verb+NTEQ_STRETCH+ & radial point distribution & 0 (no change) \\
\verb+NTEQ_NRHO+ & number of radial points & 71 \\
\verb+NTEQ_NTHETA+ & number of poloidal points & 127 \\
\verb+NTEQ_CONFIG+ & select from multiple configurations & 0 \\
\verb+NTEQ_MODE+ & select free parameters & 5 ($Q$, edge $F$; match $I_p$) \\
\verb+TEQ_SMOOTH+ & half-width in $\xi$ for smoothing & -1.5 \\
\verb+TEQ_AXSMOOTH+ & smoothing near axis & 0.05 \\
\verb+SOFT_TEQ+ & max tolerated avg Grad-Shafranov error & 0.3 \\
\hline
\end{tabular}
\caption{TEQ controls}
\label{tab:teqcontrols}
\end{table}

TEQ requires an existing inverse equilibrium solution for a given device as an initial guess for its calculation{\mc{, $i.e.$, one providing $R$ and $Z$ as functions of poloidal angle and minor radius}}. Nearly 40 devices are currently supported, and new ones are regularly added by the TRANSP developers in response to user requests.

\subsubsection{ISOLVER free-boundary equilibrium solver}

The ISolver TRANSP equilibrium module is derived from Jon Menard's original standalone IDL version of ISolver~\cite{ISOLVER1}.  As described above, it performs free-boundary equilibrium calculations incorporating information on the evolving currents in PF coils and passive structures surrounding the plasma and optionally any feedback circuits controlling them, which allows it to evolve the $q$ profile self-consistently for predictive calculations.  As of early 2024, TRANSP had ISolver device data available for 18 different tokamaks and spherical tori, including DIII-D, NSTX-U, SPARC, MAST, Asdex-U, EAST, KSTAR, and ITER.  New devices can be added by providing information on coil geometry and composition, circuits, and feedback controllers in a specified format and pre-processing it through the make\_eqtok utility; this task is normally performed by a TRANSP developer at users' request{\mc{, but training can be provided by the developers to users who wish to experiment with device configurations themselves.  Alternatively, if the user can supply an IMAS database containing the pertinent information, a recently added option exists to pre-process this data through make\_eqtok at runtime without the need for developer involvement.}}

Internally, ISolver uses Picard iteration to converge on a new equilibrium solution at each geometry time step.  It begins with the existing toroidal current density distribution on an $(R, z)$ grid and the $q$ profile either input or, at the user's discretion, evolved according to the flux diffusion equation on a flux surface,

\begin{equation}
\left.\frac{\partial\psi}{\partial t}\right|_\phi = -\frac{V_{loop}}{2\pi} =
-\eta_{\parallel} J^{OH} = -\eta_{\parallel} \frac{\left<\vec{J}\cdot\vec{B}\right> -
\left<\vec{J}^{drive}\cdot\vec{B}\right>}{\left<\vec{B}\cdot\nabla\phi\right>}
\end{equation}

where $\eta_{\parallel}$ is the plasma resistivity and $\left<\vec{J}^{drive}\cdot\vec{B}\right>$  represents driven currents.

Boundary conditions for this equation can be chosen either to match the plasma current or the surface voltage in the input data. Because of the relatively coarse resolution of the typical $(R, z)$ mesh, ISolver is poor at handling equilibria containing current sheets.

With $q$ updated, the toroidal field flux function can be evaluated according to

\begin{equation}
FV'\left<\frac{1}{R^2}\right> = 4\pi^2 q \Delta\psi
\end{equation}

where $V' = dV/d\bar\psi $ is the differential volume and $\Delta\psi$ is the enclosed poloidal flux. Next the plasma current is updated according to

\begin{equation}
J_\phi^{p,F}(R,z)=\frac{R}{\Delta\psi}\frac{dp}{d\bar\psi} +
\frac{F}{\mu_0 R \Delta\psi}\frac{dF}{d\bar\psi},
\end{equation}

\begin{equation}
J_\phi^{new}(R,z)=\omega J_\phi^{p,F}(R,z) + (1-\omega)J_\phi^{old}(R,z),
\end{equation}

where $\omega$ is the relaxation parameter for the iteration.  This step is followed by solving for the plasma poloidal flux, given by the elliptic equation

\begin{equation}
\Delta^\star\psi^{plasma} = -\mu_0 R J_\phi^{new}(R,z)
\end{equation}

The total poloidal flux is then the sum of this plasma flux and that due to the coils.  The coil currents may be set from the input data, but this usually produces poor results.  The default behavior is to determine the currents based on a least-squares fit of the resulting plasma boundary shape to a reference boundary provided. The free-boundary alternative is to solve the circuit equation to advance these currents. ISolver will search for the magnetic axis, X points, and flux surfaces based on the new plasma+coil flux solution, and will repeat the plasma current update step with successively smaller relaxation parameters until good surfaces are found.

Once an updated current distribution has been found to produce good flux surfaces, the solution is checked for convergence.  If this has been been achieved, ISolver returns the updated equilibrium to TRANSP; if not, it loops back to the $q$ profile update step with the new current distribution and the iteration continues.

\section{Power, particle and momentum balance analysis\label{sec:balance}}
\mc{\subsection{1.5D Balance Equations}}

This section introduces the fundamental time-dependent equations governing the evolution of a Maxwellian plasma, which are formulated on the $\xi$ grid. These equations represent the core manipulations performed within TRANSP. Here, no differentiation is made between known and unknown quantities; the code outputs all variables, some of which are inputs while others are outputs based on available data and user-defined parameters. For instance, one can either utilize a transport model for $Q_i$, specify $\chi_i$, and have TRANSP compute $T_i$, or input $T_i$ and derive $\chi_i$.

\begin{figure}
    \includegraphics[width=0.9\textwidth]{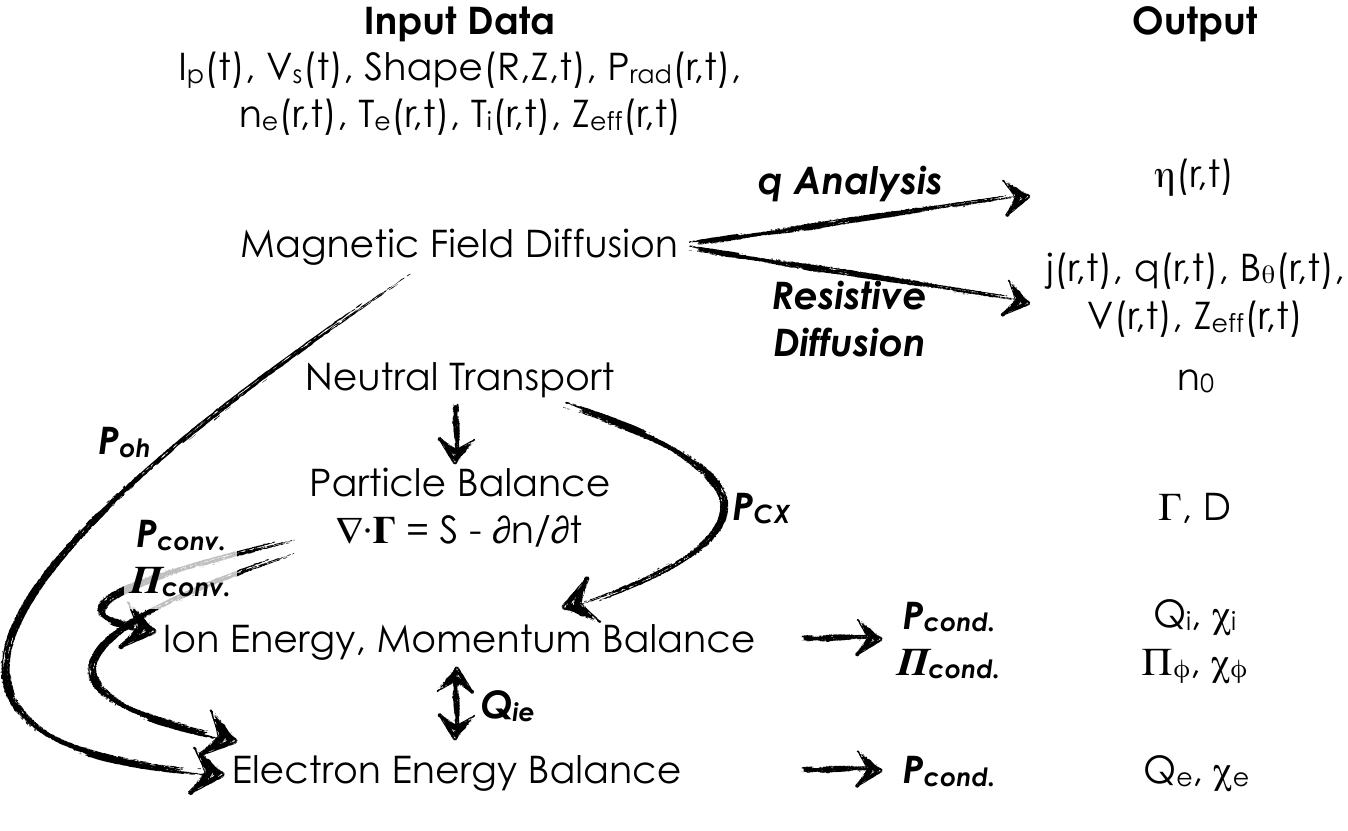}
    \justify
    \caption{Data analysis chart in TRANSP interpretive runs.} \label{fig:balance}
\end{figure}

To determine the plasma transport coefficients, a comprehensive series of steps, outlined in Fig.~\ref{fig:balance}, is executed. Regardless of the specific scenario, the general transport equation for $a(\rho(t),t)$ is employed:
\[
\frac{\partial a}{\partial t} + \nabla \cdot \mathbf{b} = c
\]
where $a$ represents a quantity such as particles, energy, or momentum, $\mathbf{b}$ denotes a transport flux, and $c$ encompasses the sum of sources and sinks. This equation undergoes a flux-surface averaging procedure to accommodate arbitrary time-evolving geometries. Subsequently, it is rearranged and numerically integrated to compute the flux, eliminating divergence and yielding effective transport coefficients. This process facilitates the derivation of quantities like diffusion coefficients, as exemplified by $b_r = -D\nabla_r a = (1/A)\int dV(c -\partial a/\partial t)$, where $A$ denotes area and $V$ denotes volume.

In the absence of sources and sinks, the ideal gas law holds true:
\[
\left(\frac{3}{2}\right) nT\left(\frac{\partial V}{\partial \rho}\right)^{5/3} = \text{const.}
\]
where $V$ signifies volume and the time derivative is zero. This equation can be expressed in two parts by multiplying both sides by $\left(\frac{\partial V}{\partial \rho}\right)^{-5/3}$:
\begin{equation}
  \frac{\partial}{\partial t}
  \left\{
  \frac{3}{2} n T
  \left( \frac{\partial V}{\partial \rho} \right)
  \left( \frac{\partial V}{\partial \rho} \right)^{2/3}
  \right\}\bigg|_{\rho} = 0
\end{equation}

\begin{equation}
  \left(\frac{\partial V}{\partial \rho}\right)^{-1}
  \frac{\partial}{\partial t}
  \left\{
  \frac{3}{2} n T
  \left( \frac{\partial V}{\partial \rho} \right)
  \right\}\bigg|_{\rho} =
  -\left(\frac{\partial V}{\partial \rho}\right)^{-1}
  n T
  \frac{\partial}{\partial t}
  \left(\frac{\partial V}{\partial \rho}\right)\bigg|_{\rho}
\end{equation}
where the LHS is the time rate of change of energy density due to heating and losses, and the RHS is the compression power density due to the changing volume.
When converting to the $\xi$ grid, this yields
\begin{eqnarray}
  &&\left(\frac{\partial V}{\partial \xi} \right)^{-1}
  \left\{
    \frac{\partial}{\partial t}\bigg|_{\xi}
    \left(\frac{3}{2}n T\frac{\partial V}{\partial \xi}\right)
    - \dot{l}\frac{\partial}{\partial \xi}
    \left(\xi \frac{3}{2} n T \frac{\partial V}{\partial \xi}\right)
  \right\} \nonumber \\
  &&= - \left(\frac{\partial V}{\partial \xi}\right)^{-1} n T
  \left\{ \frac{\partial}{\partial t}\bigg|_{\xi}\left(\frac{\partial V}{\partial \xi}\right)
  - \dot{l}\frac{\partial}{\partial \xi} \left( \xi \frac{\partial V}{\partial \xi}\right)\right\}
\end{eqnarray}
The above equation is the basic time evolution equation that incorporates the change in energy and geometry in TRANSP.
The time rate of change of the geometric volume is one of the considerations when the time advance step size is determined.

\subsection{\label{sec:pbalance}Particle Balance}
The continuity equation for particle density $n$ reads
\begin{equation}
  \frac{\partial n}{\partial t} + \langle \nabla \cdot \Gamma \rangle = R(\rho)
\end{equation}
where $n$ is the density taken as a flux function, $\Gamma$ is the particle flux and $R$ is the net source.
In TRANSP the electron and ion source functions are computed from neutral beam, pellet and thermal neutral transport models.
The time rate of change of the density comes from input electron density, $Z_{eff}$ measurements and quasi-neutrality.
Therefore the surface averaged particle flux is computed by integrating the known quantities and using the divergence theorem $\int \nabla \cdot \mathbf{X} dV = \int \mathbf{X} \cdot d\mathbf{A}$.
\begin{equation}
  \langle \Gamma_r \rangle_S =
  \frac{1}{S}
  \int_0^\xi dV
  \bigg( R -  \bigg[ \frac{\mathrm{d} n}{\mathrm{d} t} \bigg] 
  \bigg)
\end{equation}
where $ [ \frac{\mathrm{d} n}{\mathrm{d} t} ] $ represents the time rate of change of particle density while accounting for motion of the flux surfaces.  It is given by,
\begin{equation}
  \bigg[ \frac{\mathrm{d} n}{\mathrm{d} t} \bigg] = \left(\frac{\partial V}{\partial \xi}\right)^{-1}
  \left( \frac{\partial}{\partial t} \left[ n \left(\frac{\partial V}{\partial \xi}\right) \right]
  -  \dot{l} \frac{\partial}{\partial \xi} \left[ n \xi \left(\frac{\partial V}{\partial \xi}\right) \right]
  \right)
\end{equation}
where $\langle \rangle_S$ denotes surface average, $S$ is the area of the flux surface and $V$ is the volume.
In TRANSP this is done over discrete surface areas $S_j$ for surface $j$ with zone volume $\Delta V_j$.
The calculated particle flux is used to determine the convected energy flux in the energy balance equations, and is also used to determine the effective particle diffusivity from $\langle \Gamma_r \rangle = - D_{eff} \nabla n$ or effective radial velocity $\langle \Gamma_r \rangle = n \langle v_r \rangle_S$. \mc{Both the effective diffusivity and the effective radial velocity contribute to the total particle flux. Additionally, there exist specific effective velocity and diffusivity terms associated with the neoclassical Ware pinch. The Ware velocity is given by
\begin{equation}\label{eq:vware} v_{W} = \frac{E_z}{B_p} \frac{(1 - r/R)^{3/2}}{(1 + r/R)^{1/2} \left(1 + 1.46 (r/R)^{1/2}\right)} \end{equation}
where $E_z$ is the toroidal electric field, and $B_p$ is the poloidal magnetic field. Unlike the total effective particle diffusivity, the Ware effective particle diffusivity explicitly accounts for the Ware velocity contribution:
\begin{equation} D_{W} = -\frac{\langle\Gamma_r\rangle - n_e v_W}{\nabla n_e} \end{equation}
where $\langle\Gamma_r\rangle$ represents the electron particle flux.}

Information on TRANSP outputs related to particle, energy, and momentum balances is consolidated in Table~\ref{tab:balance}. \mc{While not included in Table~\ref{tab:balance} for particle balance quantities, TRANSP also outputs effective diffusivities and radial velocities for individual species if they are present in the simulation. For instance, if the simulation includes deuterium, TRANSP provides the outputs \texttt{DIFFD} and \texttt{VELD} for the effective diffusivity and radial velocity, respectively, computed in the same manner as for electrons. Similarly, TRANSP generates outputs for other species, including hydrogen (\texttt{DIFFH} and \texttt{VELH}), helium-3 (\texttt{DIFFHE3} and \texttt{VELHE3}), helium-4 (\texttt{DIFFHE4} and \texttt{VELHE4}), lithium (\texttt{DIFFLI} and \texttt{VELLITH}), and tritium (\texttt{DIFFT} and \texttt{VELT}).} Depending on the specific inputs, the output may include additional terms describing other particle sources, such as those related to pellets. There are convenient multigraph sets available for use with the \texttt{rplot} tool (see Sec.~\ref{sec:rplot}) for the electron particle balance (\texttt{EPBAL}). Equivalent quantities for ions and impurities are provided in the multigraphs \texttt{IPBAL} and \texttt{IMBAL}.

The ``rolling volume integral'' or accumulation integral of the divergence of the flux $\int_0^\xi (\nabla \cdot \Gamma)(\xi') \delta V(\xi') d\xi'$ is the total particle flow in $\#/s$ which is commonly output by gyrokinetic simulations.

\begin{table}[h!]
    \centering
    \begin{tblr}{colspec={|Q[3.5cm]|Q[1.3cm]|Q[1.45cm]|Q[1.8cm]|Q[7.6cm]|}, rows={abovesep=0pt,belowsep=0pt}, colsep=2pt}
        \hline
        \textbf{Term} & \textbf{Symbol} & \textbf{Variable} & \textbf{Units} & \textbf{Description} \\ \hline
        \SetCell[c=5]{c} \textbf{Particle Balance Variables} \\ \hline
        Time derivative of electron density & $ \frac{\mathrm{d} n}{\mathrm{d} t}  $ & \texttt{DNEDT} & $\#/s/cm^3$ & Rate of change of electron density over time. \\ \hline
        Wall neutrals & $R_W$ & \texttt{SCEW} & $\#/s/cm^3$ & Contribution of wall neutrals. \\ \hline
        Volume neutrals & $R_V$ & \texttt{SCEV} & $\#/s/cm^3$ & Contribution of volume neutrals. \\ \hline
        Impurity ionization & $R_Z$ & \texttt{SCEZ} & $\#/s/cm^3$ & Contribution of impurity ionization. \\ \hline
        Neutral beam injection & $R_{NBI}$ & \texttt{SBE} & $\#/s/cm^3$ & Contribution of neutral beam injection. \\ \hline
        Divergence of flux & $\nabla \cdot \Gamma$ & \texttt{DIVFE} & $\#/s/cm^3$ & Divergence of particle flux. \\ \hline
        \mc{Effective ion particle diffusivity} & $D$ & \texttt{DIFFI} & $cm^2/s$ & \mc{Derived from the ion particle flux.} \\ \hline
        \mc{Effective electron particle diffusivity} & $D$ & \texttt{DIFFE} & $cm^2/s$ & \mc{Derived from the electron particle flux.} \\ \hline
        \mc{Electron radial velocity} & $v_e$ & \texttt{VELE} & $cm/s$ & \mc{Derived from the particle flux.} \\ \hline
        \mc{Effective Ware particle diffusivity} & $D_W$ & $cm^2/s$ & \texttt{DIFWE} & \mc{Derived from particle flux and Ware convective velocity (Eq.~\ref{eq:vware})} \\ \hline
        \mc{Ware radial velocity} & $v_W$ & $cm/s$ & \texttt{VELWE} & \mc{Computed from neoclassical expression for Ware pinch} \\ \hline
        \SetCell[c=5]{c} \textbf{Power Balance Variables} \\ \hline
        Electron gain & $\partial W/\partial t$ & \texttt{GAINE} & $W/cm^3$ & Rate of change of electron energy over time. \\ \hline
        Total heating & $Q_{aux}$ & \texttt{EHEAT} & $W/cm^3$ & Total auxiliary heating power. \\ \hline
        Electron/ion coupling & $Q_{ie}$ & \texttt{QIE} & $W/cm^3$ & Power transfer from electrons to ions. \\ \hline
        Radiated power & $Q_{rad}$ & \texttt{PRAD} & $W/cm^3$ & Power lost due to radiation. \\ \hline
        Neutral ionization & $Q_{0,iz}$ & \texttt{PION} & $W/cm^3$ & Power consumed in neutral ionization processes. \\ \hline
        Convection & $q_{conv}$ & \texttt{PCNVE} & $W/cm^3$ & Power loss due to convective transport. \\ \hline
        Conduction & $q_{cond}$ & \texttt{PCNDE} & $W/cm^3$ & Power loss due to conductive transport. \\ \hline
        Thermal diffusivity & $\chi_e$ & \texttt{CONDE} & $cm^2/s$ & Derived from conduction term $q_{cond}$. \\ \hline
        \SetCell[c=5]{c} \textbf{Momentum Balance Variables} \\ \hline
        Momentum gain & $\partial p_\varphi/\partial t$ & \texttt{MDOT} & $N$-$m/cm^3$ & Rate of change of angular momentum over time. \\ \hline
        Total torque & $T$ & \texttt{TQIN} & $N$-$m/cm^3$ & Total applied torque. \\ \hline
        Net charge- exchange loss & $T_{CX}$ & \texttt{M0NET} & $N$-$m/cm^3$ & Net loss of momentum due to charge-exchange processes. \\ \hline
        Convection & $\Pi_{conv}$ & \texttt{MCONV} & $N$-$m/cm^3$ & Momentum loss due to convective transport. \\ \hline
        Conduction & $\Pi_{cond}$ & \texttt{MVISC} & $N$-$m/cm^3$ & Momentum loss due to conductive transport. \\ \hline
        Momentum diffusivity & $\chi_\phi$ & \texttt{CHPHI} & $cm^2/s$ &Derived from conductive term $\Pi_{cond}$. \\ \hline
    \end{tblr}
    \caption{TRANSP outputs that describe the particle, energy, and momentum balances. The sign sources come from the following equations: for particle balance $\nabla \cdot \Gamma = -\partial n/\partial t + \sum_i R_i$, for energy balance $\nabla \cdot Q = -\partial W/\partial t + \sum_i R_i$, and for the momentum balance $\nabla \cdot \Pi = -\partial p_\varphi/\partial t + \sum_i R_i$.}
    \label{tab:balance}
\end{table}

\subsection{Energy Balance}
Recalling the equation that describes an evolving Maxwellian with changing volume, sources, sinks, and the transport flux $\nabla \cdot Q$ can be added to present the energy transport equation for $(3/2)n T \partial V/\partial \rho$.
\begin{align}
 \bigg(\frac{\partial V}{\partial \xi} \bigg)^{-1} &
  \bigg\{
    \frac{\partial}{\partial t}\bigg|_{\xi}
    \bigg(\frac{3}{2} n T\frac{\partial V}{\partial \xi}\bigg)
     - \dot{l}\frac{\partial}{\partial \xi}
    \bigg(\xi \frac{3}{2} n T \frac{\partial V}{\partial \xi}\bigg)
  \bigg\} \notag \\
  &= - \bigg(\frac{\partial V}{\partial \xi}\bigg)^{-1} n T
  \bigg\{ \frac{\partial}{\partial t}\bigg|_{\xi}\bigg(\frac{\partial V}{\partial \xi}\bigg)
  - \dot{l}\frac{\partial}{\partial \xi} \bigg( \xi \frac{\partial V}{\partial \xi}\bigg)\bigg\} 
  - \bigg(\frac{\partial V}{\partial \xi}\bigg)^{-1} \frac{\partial}{\partial \xi}
  \bigg\{S\langle q_{cond}\rangle_S + S \langle q_{conv} \rangle_S \bigg\} \notag \\
  & \quad + \langle Q_{ie} \rangle_V + \langle Q_{0,iz} \rangle_V  - \langle Q_{CX} \rangle_V + \langle Q_{aux.} \rangle_V - \langle Q_{rad.} \rangle_V
\end{align}
where the heat conduction and convection as well as the sources and sinks are introduced.  Here the heat conduction is $\langle q_{cond} \rangle_S = -n \chi \langle \nabla T \rangle = -n \chi \partial T/\partial \xi \langle | \nabla \xi | \rangle$ where $\chi$ is the diffusivity and $n\chi$ is the conductivity.
The heat convection is $\langle q_{conv} \rangle_S = \alpha (5/2) T \langle \Gamma_r \rangle_S$.
Common sources for electrons are ohmic and auxiliary (NBI, RF) heating with sinks from ionization and radiation.
For ions, common sources are NBI heating and charge-exchange losses.
Both species share the exchange heating $Q_{ie}$ which is a source for one species, and a sink for the other through collisional coupling. 

The TRANSP outputs related to the electron power balance are included in Table~\ref{tab:balance}. All terms that enter into the electron energy balance equation can be viewed in the multiplot \texttt{EEBAL} (see Sec.~\ref{sec:rplot}), which typically includes several key terms, some of which are the sum of multiple quantities such as the total electron heating from sources like ohmic, beams, RF, and compression, as seen in \texttt{EEHAT}.

Often, the total energy transport is desired in calculations such as those for gyrokinetic simulations. When the total ``observed'' electron energy transport is needed, the variable \texttt{EETR\_OBS} can be viewed and integrated over the volume to obtain the total power flow in Watts. A similar set of quantities is presented for ions in the multigraph \texttt{IEBAL}.

\subsection{Momentum Balance}

In this section, a basic expression for the momentum balance equation is provided following the derivation in Ref.~\cite{goldston85}. For simplicity, assume that all bulk ion species have the same $v_\varphi$ and $T_i$. Electrons are ignored in this momentum balance analysis. In a fixed, straight cylindrical system, the momentum conservation equation for the bulk plasma is given by:
\begin{eqnarray}
  \sum_i n_i m_i \frac{\partial v_\varphi}{\partial t} + v_\varphi \sum_i m_i \frac{\partial n_i}{\partial t}
  &&= F_{col} + F_{\mathbf{j}\times\mathbf{B}} + F_{b,th} + F_{iz} \nonumber \\
  &&- \sum_i n_i m_i v_\varphi \bigg(\frac{1}{\tau_{\varphi,CX}} + \frac{1}{\tau_{\varphi,\delta}}\bigg) \nonumber \\
  && + \nabla \cdot \sum_i n_i m_i \chi_\varphi \nabla v_\varphi - \nabla \cdot \sum_i m_i \mathbf{\Gamma}_i v_\varphi
\end{eqnarray}
where the forces are from collisions, $\mathbf{j}\times\mathbf{B}$, beam thermalization, ionization and recombination forces.
Time constants for charge-exchange $\tau_{\varphi,CX}$ and non-axisymmetric field perturbations $\tau_{\varphi,\delta}$ are also included.
The cross-field momentum diffusion is $\chi_\varphi$, which is the output when the rotation is specified from experiment, as well as the momentum flux associated with particle convection using the particle flux $\mathbf{\Gamma}$.

For the general axisymmetric time-dependent geometry, it is essential to consider a toroidal system and the conservation of angular momentum, which involves torques, moments of inertia, and angular velocities. The fundamental quantities in connection to the rotation energy evolution are as follows: the torque, $T$, is given by the integral $T = \int R F \, dV$; the moment of inertia, $I$, is defined as $I = \int \sum_i n_i m_i R^2 \, dV$; the angular momentum, $p_\varphi$, is the product of the moment of inertia and the angular velocity, $\Omega$, expressed as $p_\varphi = I \Omega$; and the rotational energy, \( W_{rot} \), is given by $W_{rot} = {\Omega^2 I}/{2}$. Here, $R$ is the major radius, $F$ is the force, \( n_i \) and \( m_i \) are the density and mass of species \( i \), respectively. The evolution of angular momentum is described by the equation 
$$
I \frac{\partial \Omega}{\partial t} = T - \Omega \frac{\partial I}{\partial t}
$$
which accounts for the change in angular momentum due to torque and the time-dependent variation in the moment of inertia, and the evolution of rotational energy is described by the equation 
$$
\frac{\partial W_{rot}}{\partial t} = \Omega T - \frac{\Omega^2}{2} \frac{\partial I}{\partial t}
$$
which incorporates the effects of torque and the time-dependent changes in the moment of inertia.

In addition to changing from velocity to angular momentum, the analysis involves flux surface averages and time-dependent geometry as introduced in this section. Assuming that $v_\varphi \propto R$ on a flux surface, the appropriate quantity to manipulate is the angular velocity $\Omega = v_\varphi / R$. In this form, the angular velocity gradient $\nabla \Omega$ drives momentum diffusion $\chi_\varphi$. Upon transforming to toroidal quantities and taking flux-surface averages, the angular momentum balance equation for $nmR^2\Omega$ becomes:
\begin{eqnarray}
  \sum_i && m_i n_i \left( \langle R^2 \rangle \frac{\partial \Omega}{\partial t}
  + \frac{\Omega \langle R^2 \rangle}{n_i} \frac{\partial n_i}{\partial t}
  +  \Omega \frac{\partial \langle R^2 \rangle}{\partial t} + 
  \langle R^2 \rangle \Omega \bigg(\frac{\partial V}{\partial \rho} \bigg)^{-1} \frac{\partial}{\partial t}\frac{\partial V}{\partial \rho}\right) = \nonumber \\
  &&  T_{col} + T_{\mathbf{j} \times \mathbf{B}} + T_{bth} + T_{iz} 
  + \bigg(\frac{\partial V}{\partial \rho}\bigg)^{-1}\frac{\partial}{\partial \rho} \frac{\partial V}{\partial \rho} \sum_i n_i m_i \chi_\varphi \langle R^2 (\nabla \rho)^2 \rangle \frac{\partial \Omega}{\partial \rho} 
   \nonumber \\
  && - \sum_i n_i m_i \langle R^2 \rangle \Omega \bigg(\frac{1}{\tau_{\varphi, CX}} + \frac{1}{\tau_{\varphi, \delta}}\bigg)
  - \bigg(\frac{\partial V}{\partial \rho}\bigg)^{-1}\frac{\partial}{\partial \rho} \frac{\partial V}{\partial \rho} \sum_i n_i m_i \Omega \langle R^2 (\nabla \rho)^2 \rangle \frac{\mathbf{v}_\rho}{\nabla \rho}
\end{eqnarray}
On the LHS are terms that contain the average angular momentum density on a flux surface where $\partial V/\partial \rho$ is a differential flux-surface volume and $\rho$ is a flux-surface label that moves with the toroidal flux as in Sec.\ref{sec:sgrid}.
On the RHS are first the direct collisional, fast-ion radial current, and thermalization torque densities, which can be determined from a neutral beam package.
The ionization term comes from both ionizing new neutral particles into the Maxwellian, and losing the rotating ions due to CX with beam neutrals.
Next are momentum loss terms due to edge neutrals and ripple.
Finally are the transport terms that represent momentum diffusion (perpendicular viscosity) in the form
\begin{equation}
  \Pi_\varphi = \sum_i n_i m_i \chi_\varphi R^2 \nabla \rho \frac{\partial \Omega}{\partial \rho}
\end{equation}
and convected flux with a term derived from the particle balance~\ref{sec:pbalance}, assuming that the convected particles carry the average angular momentum with them.
Other weightings could be treated as off-diagonal elements in the transport matrix.
A similar equation for the rotational energy density can be derived and placed into the energy conservation equation as an additional time-evolving term.

TRANSP outputs for the momentum balance are listed in Table~\ref{tab:balance}. All terms that enter into the ion angular momentum balance equation can be viewed in the multiplot \texttt{MOBAL} in the \texttt{rplot} tool (see Sec.~\ref{sec:rplot}), which typically includes several key terms. Some of these are the sum of multiple quantities, such as the total input torque, which includes contributions from collisional, $\mathbf{j}\times\mathbf{B}$, thermalization, and ripple effects, as seen in the multigraph \texttt{MOBALI}. The neutral torque sinks, including ionization and charge-exchange, can be viewed in the multigraphs \texttt{TQ0BA} and \texttt{TQ0BA\_*}.

Often, the total angular momentum transport is needed in calculations such as gyrokinetic simulations. When the total ``observed'' angular momentum transport is required, the variable \texttt{AMTR\_OBS} can be viewed and integrated over the volume to obtain the total flow of angular momentum in $Nt-m$. The contributions from rotation to the energy balance are presented in the multigraphs \texttt{ROBAL}, with sources detailed in \texttt{ROBALI}.

\section{Magnetic field diffusion\label{sec:mdiffusion}}
In this section, the poloidal field diffusion equation is presented and converted to TRANSP relative flux coordinates. The original derivation is from Ref.~\cite{hinton76}. The goal is to derive the evolution equation of the poloidal field on the $\rho$ grid, due to inductive and externally driven currents. Ultimately, we arrive at an equation relating the time and spatial derivatives of the poloidal flux amenable to a time-stepping algorithm using quantities known to TRANSP, as shown in Sec.~\ref{sec:equilibrium}. This magnetic field diffusion equation solver is historically called \verb+MAGDIF+, and the two options for resistivity as input or output are shown in  Fig.~\ref{fig:balance}.

From a high level, if the equilibrium is specified externally, then the plasma resistivity $\eta$ and ohmic contribution to the plasma current can be derived from the time-dependent evolution of the current profile and surface loop voltage. However, if the user chooses to evolve the current profile resistively (i.e., $\eta = \eta_{neo}$), then the current profile and ohmic current are determined by the total plasma current, external currents, and resistivity derived from the plasma profiles. 

To sketch the first situation of input equilibrium, TRANSP knows the magnetic field $\mathbf{B}(r,t)$. From Faraday's law, $\nabla \times \mathbf{E} = -\partial \mathbf{B}/\partial t$ is integrated in time to get $\nabla \times \mathbf{E}$, which is then integrated to get the loop voltage to within a constant (taken from experiment). In an ohmic plasma, $\langle \mathbf{E} \cdot \mathbf{B} \rangle = \eta \langle \mathbf{j} \cdot \mathbf{B} \rangle$, so we get the ohmic resistivity after removing the externally driven currents from $\eta = \langle \mathbf{E} \cdot \mathbf{B} \rangle / (\langle \mathbf{j} \cdot \mathbf{B} \rangle - \langle \mathbf{j} \cdot \mathbf{B} \rangle^{ext})$.

During resistive current diffusion, the poloidal flux is time-evolved based on an initial condition (typically given by the experimental $q$-profile or an assumption about the initial loop voltage profile). For both cases, a general equation for the time evolution of the poloidal flux is required and will be derived next. What follows is the derivation combining Faraday's law, Ampere's law, and Ohm's law to compute $-\partial \mathbf{B}/\partial t = \nabla \times \mathbf{E}$ on flux surfaces.

\subsection{Faraday's Law}
We begin by stating Faraday's law in integral form
\begin{equation}
  \oint \mathbf{E} \cdot d\mathbf{l} = -\frac{1}{c}\frac{\partial}{\partial t} \int \mathbf{B} \cdot d\mathbf{A}
\end{equation}
where the integral is around a closed contour of toroidal flux and $d\mathbf{A}$ is orthogonal to the surface defined by the flux contour.
In a MHD equilibrium $\mathbf{B}$ lies within a flux surface, making the RHS of this equation equal to zero.
This implies that the loop integral on a surface $V_l = 2 \pi R E_T$ is constant, defining the toroidal loop voltage $V_l=V_l(\rho)$.
The integral form of the equation for the poloidal electric field is
\begin{equation}
  \oint E_\rho dl_\rho = \frac{1}{c} \frac{\partial \Phi}{\partial t} = 0
\end{equation}
where the electric field loop integrates poloidally to zero because the toroidal flux $\Phi = \int \mathbf{B} \cdot d\mathbf{A}$ is constant in time on a flux surface.
The radial derivative of the loop voltage profile is given by
\begin{equation}
  \frac{\partial V_l}{\partial \rho} = \frac{2\pi}{c}\frac{\partial}{\partial t}\bigg(\frac{\partial \psi}{\partial \rho}\bigg)
\end{equation}
where $\psi = (1/2\pi)\int \mathbf{B}_p \cdot d\mathbf{A}$ is the poloidal flux from Section~\ref{sec:equilibrium} and $\partial \psi/\partial \rho$ is a flux-surface constant.
Note that since $\mathbf{B}_p$ is is tangent to the flux surfaces and $\nabla \cdot \mathbf{B} = 0$, then $\partial \psi / \partial \rho$ is a flux surface constant.

Next, we define the flux-surface averaged $\mathbf{E} \cdot \mathbf{B}$ by splitting into toroidal and poloidal components:
\begin{eqnarray}
  \langle \mathbf{E} \cdot \mathbf{B} \rangle &=& \langle E_T \cdot B_T \rangle + \langle E_p \cdot B_p \rangle \\ \nonumber
  &=& \bigg \langle \frac{V_l}{2 \pi R} \frac{R B_T}{R} \bigg \rangle + \bigg \langle E_p \frac{|\nabla \rho|}{R} \frac{\partial \psi}{\partial \rho} \bigg \rangle \\ \nonumber
  &=& \frac{(R B_T) V_l}{2 \pi} \langle 1/R^2\rangle + 2 \pi\bigg(\frac{\partial \psi}{\partial \rho}\bigg) \bigg(\frac{\partial V}{\partial \rho}\bigg)^{-1} \oint dl_p E_p
\end{eqnarray}
and the second term on the RHS is zero as $\oint dl_p E_p = 0$.
Therefore, we can define the loop voltage as
\begin{equation}
    V_l = \frac{\langle \mathbf{E} \cdot \mathbf{B} \rangle 2 \pi}{(R B_T) \langle 1/R^2 \rangle}
\end{equation}

Thus the expression for Faraday's law in terms of $\langle \mathbf{E} \cdot \mathbf{B} \rangle$ is
\begin{equation}
    \label{Faraday}
    \frac{1}{c} \frac{\partial}{\partial t}\bigg(\frac{\partial \psi}{\partial \rho}\bigg) = \frac{\partial}{\partial \rho} \bigg[\frac{\langle \mathbf{E} \cdot \mathbf{B} \rangle}{(R B_T) \langle 1/R^2 \rangle} \bigg]
\end{equation}
Next we will convert the RHS of the above with Ampere's law and Ohm's law.

\subsection{Ampere's Law}
In this section, Ampere's law is used to derive the relationship between the parallel current density and the magnetic fields, arriving at an expression for $\langle \mathbf{j} \cdot \mathbf{B}\rangle$ to be inserted into Ohm's law in terms of flux surface quantities readily available in TRANSP.
In integral form Ampere's law is
\begin{equation}
    \oint \mathbf{B} \cdot d\mathbf{l} = \frac{4\pi}{c} \int \mathbf{j} \cdot d\mathbf{A}
\end{equation}

\subsubsection{Toroidal component of Ampere's law}
The toroidal component of Ampere's law in integral form relates the integrated poloidal field to the toroidal current
\begin{equation}
    \oint B_p \cdot dl_p = \frac{4 \pi}{c} I_T
\end{equation}
and using $B_p = (1/R) \partial \psi / \partial \rho |\nabla \rho|$ gives
\begin{equation}
    \frac{\partial \psi}{\partial \rho} \oint \frac{|\nabla \rho|}{R} dl_p = \frac{4 \pi}{c} I_T
\end{equation}
This geometry factor on the LHS is already numerically integrated using
\begin{equation}
    \bigg(\frac{\partial V}{\partial \rho}\bigg)^{-1} 2 \pi \oint dl_p \cdot \frac{|\nabla \rho|}{R} = \langle|\nabla \rho|^2 / R^2\rangle
\end{equation}
giving
\begin{equation}
    I_T = \frac{c}{8 \pi^2} \bigg(\frac{\partial V}{\partial \rho}\bigg) \bigg(\frac{\partial \psi}{\partial \rho}\bigg) \langle |\nabla \rho|^2 / R^2 \rangle
\end{equation}
Now relating the current density to the gradients in the poloidal field we apply $\partial / \partial \rho$ to both sides using
\begin{equation}
    \frac{\partial}{\partial \rho} I_T = \frac{\partial}{\partial \rho} \int_0^\rho d\rho \oint dl_p \frac{J_T}{|\nabla \rho|} = \oint dl_p \frac{R}{|\nabla \rho|}\frac{J_T}{R}
\end{equation}
or
\begin{equation}
    \frac{\partial}{\partial \rho} I_T = \frac{1}{2 \pi} \bigg(\frac{\partial V}{\partial \rho}\bigg) \langle J_T / R \rangle
\end{equation}
Combining this with the expression for $I_T$ gives
\begin{equation}
    \langle J_T/R \rangle = \frac{c}{4 \pi} \bigg( \frac{\partial V}{\partial \rho} \bigg)^{-1} \frac{\partial}{\partial \rho} \bigg[ \bigg( \frac{\partial V}{\partial \rho} \bigg) \bigg( \frac{\partial \psi}{\partial \rho} \bigg) \langle | \nabla \rho|^2 / R^2 \rangle \bigg]
\end{equation}

\subsubsection{Poloidal component of Ampere's law}
The Grad-Shafranov equation $\nabla p = \mathbf{J} \times \mathbf{B}$ implies the existence of a poloidal current in the plasma that will effect the interior toroidal field.
With $\mathbf{J}$ tangential to flux surfaces and $\nabla \cdot \mathbf{J} = 0$ implies
\begin{equation}
    \frac{\partial I_p}{\partial \rho} = \frac{2 \pi R}{|\nabla p|} J_p
\end{equation}
is constant on a flux surface.
Then Ampere's Law yields
\begin{equation}
    \label{Ampere_tor}
    2 \pi \frac{\partial (R B_T)}{\partial \rho} = -\frac{4 \pi}{c} \frac{\partial I_p}{\partial \rho}
\end{equation}
When the poloidal field equation is solved and $\langle \mathbf{J} \cdot \mathbf{B} \rangle$ is known, then $\partial I_p /\partial \rho$ can be evaluated using the Grad-Shafranov equation, and $g = (R B_T) / (R B_T^{ext})$ can be checked.
Until then, the equivalent relation
\begin{equation}
    \label{Ampere_pol}
    |J_p| = - \frac{c}{4 \pi} \frac{|\nabla p|}{R} \frac{\partial}{\partial \rho}(R B_T)
\end{equation}
is useful.

\subsubsection{Parallel component of Ampere's law}
Using Ohm's law in the next section will relate $\langle \mathbf{E} \cdot \mathbf{B} \rangle$ to $\langle \mathbf{J} \cdot \mathbf{B} \rangle$, so it is desirable to express Ampere's Law accordingly.
Since
\begin{eqnarray}
    \langle \mathbf{J} \cdot \mathbf{B} \rangle &=& \langle J_T B_T \rangle + \langle J_p B_p \rangle \nonumber \\
    &=& (R B_T) \langle J_T/R \rangle + \frac{\partial \psi}{\partial \rho} \langle J_p |\nabla \rho|/R \rangle
\end{eqnarray}
we combine \ref{Ampere_tor} and \ref{Ampere_pol} to arrive at
\begin{eqnarray}
    \langle \mathbf{J} \cdot \mathbf{B}\rangle &=& \frac{c}{4\pi} (R B_T) \bigg( \frac{\partial V}{\partial \rho} \bigg)^{-1} \frac{\partial}{\partial \rho} \bigg[ \bigg( \frac{\partial V}{\partial \rho} \bigg) \bigg( \frac{\partial \psi}{\partial \rho} \bigg) \langle | \nabla \rho |^2 / R^2 \rangle \bigg] \nonumber \\
    &-& \frac{c}{4 \pi} \frac{\partial \psi}{\partial \rho} \frac{\partial}{\partial \rho}(R B_T) \langle | \nabla \rho |^2 / R^2 \rangle
\end{eqnarray}
which simplifies to 
\begin{eqnarray}
    \label{Ampere}
    \langle \mathbf{J} \cdot \mathbf{B}\rangle &=& \frac{c}{4\pi} (R B_T)^2 \bigg( \frac{\partial V}{\partial \rho} \bigg)^{-1} \times \frac{\partial}{\partial \rho} 
    \bigg[ \bigg( \frac{\partial V}{\partial \rho} \bigg) \bigg( \frac{\partial \psi}{\partial \rho} \bigg) \frac{\langle | \nabla \rho |^2 / R^2 \rangle}{(R B_T)} \bigg]
\end{eqnarray}

\subsection{Ohm's Law}
Ohm's Law is expressed as
\begin{equation}
    \label{Ohm}
    \langle \mathbf{E} \cdot \mathbf{B} \rangle = \eta_\parallel \bigg[ \langle \mathbf{J} \cdot \mathbf{B} \rangle - \langle \mathbf{J} \cdot \mathbf{B} \rangle^{ext} \bigg]
\end{equation}
where $\eta_\parallel$ is the parallel resistivity, which needs to be provided for an arbitrary geometry generalization of Spitzer and/or neoclassical resistivity.
The term $\langle \mathbf{J} \cdot \mathbf{B} \rangle^{ext}$ represents non-ohmic currents such as bootstrap, beam-driven and RF driven currents calculated independently and provided as input to the poloidal field calculation.
Through Ohm's Law (\ref{Ohm}), Faraday's Law (\ref{Faraday}) and Ampere's Law (\ref{Ampere}) we arrive at the poloidal field diffusion equation used in TRANSP as
\begin{eqnarray}
  \frac{1}{c} \frac{\partial}{\partial t} \bigg(\frac{\partial \psi}{\partial \rho} \bigg) = &&\nonumber \\
  && \frac{\partial}{\partial \rho} \bigg\{\frac{c \eta_\parallel}{4 \pi} \frac{(R B_T)}{\langle R^{-2} \rangle (\partial V/\partial \rho)} \frac{\partial}{\partial \rho} \bigg[\frac{\partial V}{\partial \rho} \frac{\partial \psi}{\partial \rho} \frac{ \langle |\nabla \rho|^2 /R^2 \rangle}{(R B_T)}\bigg]\bigg\} \nonumber \\
  &&- \frac{\partial}{\partial \rho} \frac{\eta_\parallel \langle \mathbf{J} \cdot \mathbf{B} \rangle^{ext}}{(R B_T) \langle R^{-2} \rangle}
\end{eqnarray}

In the TRANSP code this equation is solved on a fixed $\xi=\rho/\rho_{sep} = \sqrt{\Phi/\Phi_{sep}}$ grid, and therefore the following transformations are used
\begin{eqnarray}
  \rho &\rightarrow& \rho_{sep} \xi \\
  \frac{\partial}{\partial \rho} &\rightarrow& \frac{1}{\rho_{sep}}\frac{\partial}{\partial \xi} \\
  \frac{\partial}{\partial t}\bigg|_\rho &\rightarrow& \frac{\partial}{\partial t}\bigg|_\xi - \xi \dot{l}\frac{\partial}{\partial \xi}\bigg|_t
\end{eqnarray}
where the last transformation is performed noting $\dot{l} \equiv (1/\rho_{sep})d\rho_{sep}/dt = (1/2\Phi_{sep})d\Phi_{sep}/dt$.
It is also convenient to use $\iota = 1/q = 2\pi \partial \psi / \partial \Phi$ as the dependent variable.
From the definitions of $\xi$ and $\dot{l}$, 
\begin{equation}
    \frac{\partial \psi}{\partial \rho} = \frac{1}{\rho_{sep}}\frac{\partial \psi}{\partial \xi} = \frac{1}{\rho_{sep}} \frac{\iota \xi \Phi_{sep}}{\pi}
\end{equation}
and also from the definition of $\rho$, $\Phi_{sep} = \pi \rho_{sep}^2 B_0$.
Substituting for $\partial \psi/\partial \rho$ and applying $\rho \rightarrow \xi$ transformations, the equation becomes
\begin{eqnarray}
  \frac{\partial}{\partial t}\bigg|_{\xi} (\iota \xi) + (\iota \xi)\dot{l} - \xi \dot{l}\frac{\partial}{\partial \xi}(\iota \xi) \nonumber =&&\\
  && \frac{\partial}{\partial \xi} \bigg\{\frac{c \eta_\parallel}{4 \pi} \frac{(R B_T)}{\langle R^{-2} \rangle (\partial V/\partial \xi)} \frac{\partial}{\partial \xi} \bigg[\frac{\partial V}{\partial \xi}(\iota \xi) \frac{ \langle |\nabla \xi|^2 /R^2 \rangle}{(R B_T)}\bigg]\bigg\} \nonumber \\
  &&- \frac{\pi c}{\Phi_{sep}} \frac{\partial}{\partial \xi} \bigg[\frac{\eta_\parallel \langle \mathbf{J} \cdot \mathbf{B} \rangle^{ext}}{(R B_T) \langle R^{-2} \rangle}\bigg]
\end{eqnarray}
Note that all occurrences of $B_0$ and $\rho_{sep}$ have canceled out.
This equation can also be solved for $(\iota \xi)$, which in cylindrical coordinates is proportional to $B_p$.

\subsection{TRANSP options for magnetic diffusion equation}
The initial conditions for the poloidal field diffusion equation in TRANSP are crucial for accurately modeling the evolution of the current profile and magnetic field within a tokamak. These initial conditions include the specification of the initial $q$ profile and the current density profile. There are two primary options for setting these initial conditions.

The first option involves setting the initial $q$ profile, which can be derived from the interpretation of experimental measurements and magnetic equilibrium reconstruction. Within this option, there are two possible selections: either matching the total {\mc{toroidal}} plasma current {\mc{$I_p$}} to the experimental total plasma current during the poloidal field diffusion or not matching it. If the total plasma current is matched, some additional adjustments might be necessary to ensure consistency.

The second option involves setting the initial plasma current profile. In this case, the plasma density is re-normalized to match the total plasma current. This option also offers two choices: fixing the parallel current density $j_\parallel$ and evolving the parallel component of the electric field $E_\parallel$, or fixing the parallel component of the electric field and evolving the parallel current density.

By setting these initial conditions accurately, TRANSP can effectively model the time evolution of the magnetic field and current profiles, leading to better predictions and optimizations of tokamak plasma performance.

All these options require knowledge of the resistivity profiles, which can be specified using various models. It is possible to switch back and forth among these options during a run. The poloidal field diffusion equation can also be used to estimate the resistivity profile in these cases, though non-physical negative values of resistivity can sometimes result, depending on the quality of the $q$ profile data and its time derivatives.

{\mc{Magnetic diffusion in TRANSP is carried out as a separate part of the time step from the equilibrium solution (or optionally integrated with it; see option NLISODIF in Table~\ref{tab:mdiffusion} below). Changes to the plasma current profile due to flux diffusion feed into the equilibrium calculation on the following time step.}}

Key input variables related to the poloidal field diffusion equation are summarized in Table~\ref{tab:mdiffusion}.

\begin{table}[h]
    \centering
    \begin{tabular}{|l|p{10cm}|}
        \hline
        \textbf{Option} & \textbf{Description} \\
        \hline
        \texttt{NLMDIF} & Predictive Poloidal Field Diffusion: Evolves the $q$ profile over time using initial conditions from \texttt{EFIELD}, a resistivity model, and boundary conditions. \\
        \hline
        \texttt{\mc{NLISODIF}} & {\mc{Isolver Flux Diffusion: When running with Isolver, evolves the $q$ profile self-consistently with the equilibrium within Isolver in place of using TRANSP's poloidal field diffusion equation.}} \\
        \hline
        \texttt{NLQMHD} & MHD Equilibrium Analysis: Sets the $q$ profile based on the output of {\mc{the selected}} MHD equilibrium {\mc{solver (at present, only Isolver supports this option)}}. Equivalent to setting both \texttt{NLMDIF=.TRUE.} and \texttt{NLISODIF=.TRUE.}. \\
        \hline
        \texttt{NLMBPB} & Evolving q Profile Using Input Data: Evolves the $q$ profile using input data such as the ratio of poloidal to toroidal magnetic field (\texttt{Bp/Bt}) over radial position and time. \\
        \hline
        \texttt{NLQDATA} & Setting $q$ Profile from Ufile: Directly sets the $q$ profile from a QPR Ufile, which contains the desired q profile data. \\
        \hline
        \texttt{RESIS\_FACTOR} & Resistivity Scaling: Scales the resistivity profile using the specified factor. \\
        \hline
        \texttt{NLSPIZ} & Spitzer Resistivity: Uses Spitzer resistivity if set to \texttt{.TRUE.}, otherwise uses neoclassical resistivity models. \\
        \hline
        \texttt{NLPCUR} & Total Plasma Current: Matches the plasma current if \texttt{.TRUE.}, allowing for total plasma current evolution prediction if \texttt{.FALSE.}. \\
        \hline
        \texttt{NLVSUR} & Surface Voltage: Matches surface voltage data and adjusts Zeff for resistivity over time if \texttt{NLPCUR} is \texttt{.TRUE.}. \\
        \hline
        \texttt{NLINDUCT} & Inductive Coupling: Predicts plasma current through inductive coupling from poloidal field coils when using Isolver and measured coil currents (\texttt{FB\_NAMES} defined). \\
        \hline
    \end{tabular}
    \caption{Summary of magnetic diffusion equation options in TRANSP.}
    \label{tab:mdiffusion}
\end{table}

\section{Models for heating and current drive\label{sec:sources}}
TRANSP employs a uniform approach for selecting heating and current drive modules. The input parameters \texttt{EC\_MODEL}, \texttt{LH\_MODEL}, \texttt{IC\_MODEL}, and \texttt{NB\_MODEL} define the modules for Electron Cyclotron (EC), Lower Hybrid (LH), Ion Cyclotron (IC), and Neutral Beam (NB) heating and current drive computations. Currently, there is only one option available for NB heating, and the \texttt{NB\_MODEL} parameter is reserved for future use when additional modules for NB heating and current drive are implemented.

\subsection{Neutral beam injection NUBEAM model\label{sec:nubeam}}
The NUBEAM model in the TRANSP code ~\cite{goldston81,pankin04} is used for simulating the behavior of neutral beam injection (NBI) in tokamak plasmas. It determines how the neutral beams penetrate the plasma and where they deposit their energy and momentum, including the calculation of beam attenuation as the neutral atoms ionize or charge-exchange and become part of the plasma. NUBEAM tracks the resulting fast ion distribution in phase space, considering the effects of collisions with background plasma particles and electromagnetic fields. It also computes the contribution of the neutral beams to plasma heating and current drive, including both the direct energy transfer from the beams to the plasma and the indirect effects due to the fast ion population. Additionally, NUBEAM assesses the potential for beam-driven instabilities and their impact on plasma performance. The model can be used to interpret diagnostic measurements related to neutral beam injection, such as neutron emission and charge exchange recombination spectroscopy. Overall, NUBEAM provides detailed modeling of beam injection and fusion products, which is crucial for accurate predictions of plasma behavior and performance in tokamaks.

The NUBEAM model self-consistently handles classical guiding center drift orbiting and incorporates collisional and atomic physics effects during the slowing down of the fast species population, which is represented by an ensemble of Monte Carlo model particles. It includes options for sawtooth-induced phenomena, radio-frequency heating, or anomalous radial diffusion in a time-evolving plasma with nested magnetic flux surfaces represented by a numerical MHD equilibrium.

Time-stepping in the model is explicit. During each time step, the Monte Carlo deposition of new particles and the slowing down of the Monte Carlo ion ensemble occur against a fixed plasma for a prescribed interval. After this interval, the state of the Monte Carlo model is recorded, and control returns to the plasma model, allowing it to evolve in time as described in Sec.~\ref{sec:tgrids}. This cycle of Monte Carlo deposition, slowing down, and plasma evolution continues until the simulation is complete.

The model uses a technique called {\it{goosing}}\mc{~\cite{pankin04}} to deal with the difference in time scales between fast guiding center drift orbit bounces and slower physical processes like collisions, atomic physics, or anomalous diffusion. \mc{Instead of explicitly computing every orbit bounce, NUBEAM introduces a \textit{goosing factor}, a timestep multiplier that artificially accelerates these processes, significantly reducing computational cost while maintaining accuracy. Control parameters, such as \texttt{FPPCON}, \texttt{CXPCON}, and \texttt{GOOCON}, regulate the frequency of collision and atomic physics evaluations per orbit. If an orbit's travel time exceeds twice the estimated bounce time, the goosing factor is immediately updated to ensure accuracy. This method reduces computational load by two to three orders of magnitude, and a backup mechanism ensures proper handling of off-midplane trapped orbits, which may not cross the midplane in certain tokamak equilibria}. 

In this section, following the description of the NUBEAM grid, \mc{atomic physics options, and fusion products modeling, the focus will shift to the advancements in the NUBEAM model made over the past two decades since the last comprehensive review was published~\cite{pankin04}. These advancements include the implementation of neutron rate feedback control algorithm for fast ion diffusivity}, a 3D halo neutral model, enhancements in radio-frequency (RF) capabilities, the development of the \textit{kick model} for energetic particle-driven instabilities such as \mc{Alfv\'en} modes (AEs)~\cite{mp_2014}, and the incorporation of the Resonance Line Broadened Quasi-linear (RBQ) model for fast ion distribution relaxation due to \mc{Alfv\'enic} eigenmodes~\cite{GorelenkovPoP19}. Additionally, recent optimization efforts for GPU architecture utilizing OpenACC libraries are described.

\subsubsection{Monte Carlo grid}
\begin{wrapfigure}{l}{0.55\textwidth}\vspace{-9mm}
    \includegraphics[width=0.5\textwidth]{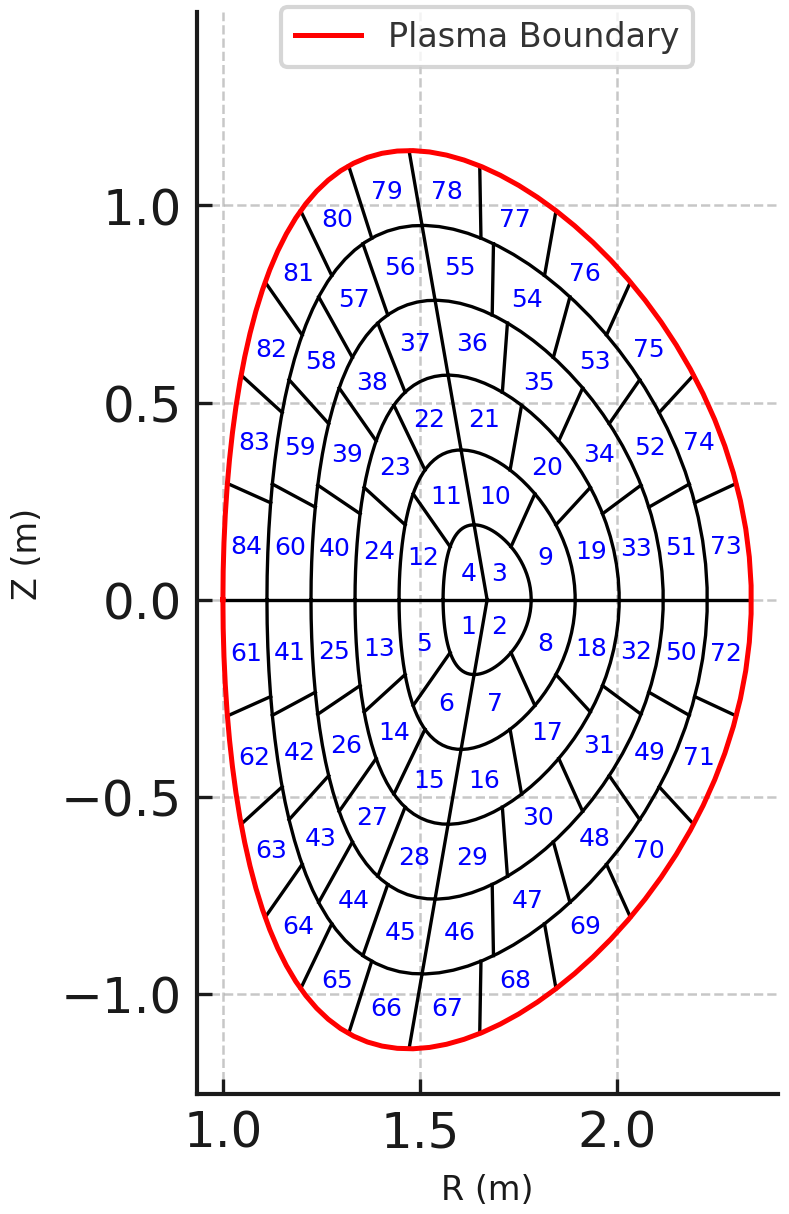} 
    \vspace{-9mm}\singlespacing
    \caption{\mc{Schematic view of Monte Carlo grid in the NUBEAM model with the zone numbers. The plasma boundary is shown in red.}} \label{fig:mcgrid}
\end{wrapfigure}

The NUBEAM model employs a Monte Carlo (MC) grid, which is an irregular 2D spatial grid developed to capture the 2D binned data in MC calculations. \mc{ It is constructed of ``zone rows'' that are aligned with the flux coordinates and equally spaced in $\xi$ as described in Section~\ref{sec:sgrid}; each zone row is subdivided into a different number of zones equally spaced in the equilibrium poloidal angle coordinate $\theta$, with fewer subdivisions for zone rows near the axis, and more for zone rows near the edge. This results in a set of zones all with roughly equal volume and cross-sectional area, ensuring consistency in MC summation statistics. A schematic view of the MC grid with zone numbers is shown in Fig.~\ref{fig:mcgrid}}. This grid is used for fast ion distribution functions and for certain spatially 2D profiles output by NUBEAM, such as beam halo thermal neutral sources, beam-target and beam-beam fusion rates.

Poloidal zones are stored contiguously, with the poloidal zone index increasing with the $\theta$ coordinate, which is oriented counter-clockwise in the plasma cross-section drawn to the right of the machine axis of symmetry. However, there are options to reverse the storage order to be consistent with a clockwise-oriented poloidal angle coordinate, as preferred by some codes. For both up-down symmetric and up-down asymmetric data, the starting point for $\theta$ zone indexing can be specified as 0, -$\pi$, or default, which is 0 for up-down symmetric geometry and -$pi$ for up-down asymmetric geometry. These choices are controlled by arguments in the \mc{MC-grid data interface routines that generate the 2D MC grid and map physics quantities to this grid}.

\subsubsection{Atomic physics options}
\begin{figure}[hb]
    \centering
   \includegraphics[width=0.54\textwidth]{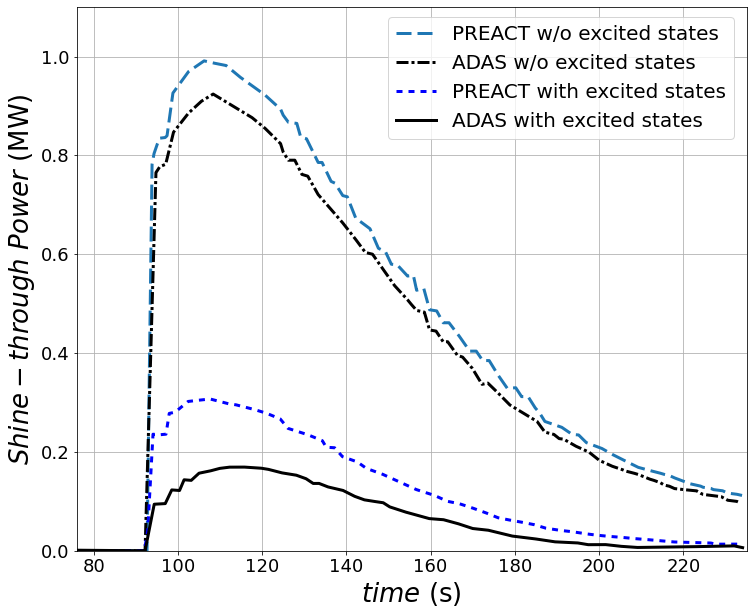} 
	\caption{Differences in the shine-through power simulated by the NUBEAM model for an ITER discharge using ADAS~\cite{adas} and PREACT~\cite{janev87} atomic physics models. This figure also demonstrates the importance of including excitation state effects in NUBEAM when the target plasma density is high, as these effects significantly impact the deposition profile~\cite{janev89}. \mc{In these TRANSP simulations, with Negative Neutral Beam injection having an energy of 1 MeV and a beam power of 33 MW, it is assumed that the flat electron density profile increases linearly over time. The modeling is discussed in references~\cite{budnyNF2008,budnyNF2009,budnyNF2012,budnyNF2016}.}\label{fig:dep}}
\end{figure}

To improve deposition profiles and provide a reliable halo neutral source for diagnostic simulations, the atomic physics in NUBEAM was enhanced. This enhancement involved fully implementing ADAS atomic physics data~\cite{adas} for ground and excited states of hydrogen-like and helium beams, as well as target ions with $Z$ = 1 to 10. The main physics issue prompting this enhancement beyond the legacy PREACT atomic physics data~\cite{janev87} in TRANSP was the density dependence of beam stopping data provided by ADAS, which accounts for collisional excitation effects on beam neutral atoms. \mc{Two approaches are available in NUBEAM for estimating the correction factor. Users can either apply the Janev-Boley-Post fit~\cite{janev89} to account for the enhancement of excited-state cross sections or calculate the factor within NUBEAM as the ratio of the total beam stopping cross section to that obtained from atomic physics processes in the ground-state model. This enhancement factor is then applied to the ionization cross sections. These approximations enable NUBEAM to incorporate appropriately enhanced cross sections, rather than relying solely on effective beam stopping cross sections.} The difference in shine-through profiles computed with PREACT and ADAS atomic data is shown in Fig.~\ref{fig:dep}.

\mc{The most consistent atomic physics model in TRANSP/NUBEAM is the ADAS310~\cite{adas} program, implemented as a library. ADAS310 computes the excited-state population structure, as well as effective ionization and recombination coefficients for hydrogen-like atoms or hydrogenic ions in an impure plasma. This method ensures that each plasma species is treated independently of the others.}

\mc{Combinations of input variables are given in Table~\ref{tab:atom}. The importance of using proper atomic physics data is shown in Fig.~\ref{fig:dep} for the  shine-through profiles computed with PREACT and ADAS atomic data.}

 \begin{table}[]
 \centering
 \begin{tabular}{|c|c|c|}
 \hline
 \textbf{NSIGEXC} & \textbf{ LEV\_NBIDEP = 1 } & \textbf{LEV\_NBIDEP = 2} \\
 \texttt{ }   & \textbf{PREACT} & \textbf{ADAS}  \\ \hline
 \textbf{0}&\multicolumn{2}{c|}{ ground state model} \\ \hline
 \textbf{1}    & \multicolumn{2}{c|} {NUBEAM calculates enhancement factor} \\ \hline
 \textbf{2}     &\multicolumn{2}{c|} { Janev-Boley-Post parameterization for enhancement factor~\cite{janev89}}\\ \hline
 \textbf{3}   & - & excitation state model \\ \hline
 \end{tabular}
 \caption{\mc{Input parameter settings for atomic physics data in TRANSP / NUBEAM.}}
 \label{tab:atom}
 \end{table}

\subsection{\mc{Fusion product modeling in NUBEAM}} 
\mc{The TRANSP NUBEAM code may simultaneously model multiple fast ion species per run.  Fast ion  species can be either neutral beam species, fusion product species, RF - minorities which are treated as as Maxwellian species in NUBEAM). Table~\ref{tab:fus} presents a list of fusion products available for modeling in TRANSP/NUBEAM.}
\begin{table}[]
\centering
\begin{tabular}{|l|l|}
\hline
\textbf{Fusion Product} & \textbf{Fusion Reaction} \\ \hline
$\mathbf{^{4}_{2}He} ({\it{alphas}})$
& $^{2}_{1}\text{D} + ^{3}_{1}\text{T} \rightarrow  ^{4}_{2}\text{He} + \text{n }+ \text{17.6 MeV}$ \\
& $^{2}_{1}\text{D} + ^{3}_{2}\text{He} \rightarrow  ^{4}_{2}\text{He} + \text{p} + \text{18.35 MeV}$ \\
& $^{3}_{1}\text{T} + ^{3}_{1}\text{T} \rightarrow  ^{4}_{2}\text{He} + \text{2n} + \text{11.3 MeV}$ \\ \hline

$\mathbf{^{3}_{2}He}$ 
& $^{2}_{1}\text{D} + ^{2}_{1}\text{D} \rightarrow  ^{3}_{2}\text{He} + \text{n} + \text{3.27 MeV}$ \\ \hline

$\mathbf{^{3}_{1}T} ({\it{tritons}}) $ 
& $^{2}_{1}\text{D} + ^{2}_{1}\text{D} \rightarrow  ^{3}_{1}\text{T} + \text{p}+ \text{4.03 MeV}$ \\ \hline

$\mathbf{p^+}\text{(3.02 MeV)}$
& $^{2}_{1}\text{D} + ^{2}_{1}\text{D} \rightarrow  ^{3}_{1}\text{T}  + \text{p}  + \text{4.03MeV}$ \\ \hline

$\mathbf{p^+}\text{(14.7 MeV)}$
& $^{2}_{1}\text{D} + ^{3}_{2}\text{He} \rightarrow  ^{4}_{2}\text{He} + \text{p}  + \text{18.35MeV}$ \\ \hline
\end{tabular}
\caption{\mc{Available fusion products for modeling in NUBEAM, including conventional and RF-driven fusion reactions.
\label{tab:fus}}}
\end{table}

\mc{NUBEAM calculates neutron emission from all possible combinations of fast ion species-fast ion species and fast ion species-thermal ions in background plasma reactions. It not only estimates the total number of neutrons or fusion reactions on a reaction-by-reaction basis but also provides a breakdown of rates based on the reagent types (fast ions vs. thermal ions). The NUBEAM output profiles for neutron emission are flux surface-averaged and presented as 2D profiles showing poloidal variation on the MC grid. Neutron emission from reactions with RF minorities is computed using the FPP code, while TRANSP calculates the neutron rate for the background plasma. The accuracy of NUBEAM's neutron emission calculations has been validated against experimental measurements in several tokamaks, as described in Refs.~\cite{Hellesen:2010,Sperduti:2021,Stancar:NF2021,Stancar:NF2023}.}

\subsection{\mc{Neutron rate feedback control for fast ion diffusivity}}
\mc{Neutron rate feedback control in TRANSP, developed by Dr. M.D. Boyer, dynamically adjusts anomalous fast ion diffusivity to match measured neutron flux. This algorithm provides a self-consistent way to refine fast ion transport modeling by ensuring agreement between simulated and experimental neutron rates. By modifying fast ion diffusivity based on real-time neutron diagnostics, this approach improves predictive accuracy in TRANSP simulations, particularly in cases where discrepancies in neutron emission indicate deviations in fast ion confinement or anomalous transport effects. The feedback mechanism operates through a proportional-integral (PI) controller, applying corrections while maintaining stability through saturation constraints and anti-windup mechanisms. The feedback control algorithm is enabled by setting \texttt{NLFB\_NEUTCNTRL} to TRUE in the TRANSP input.} 

\mc{The feedback control algorithm operates by computing the error between the simulated and measured neutron rates. This error is then used in a PI control loop, where the proportional term adjusts the diffusivity in response to immediate changes, and the integral term accumulates corrections over time to eliminate steady-state errors. To prevent excessive corrections, the control action is bounded by predefined limits. An anti-windup mechanism is also implemented to avoid control saturation when the adjustment exceeds these bounds. The final computed diffusivity scale factor is then applied to the fast ion transport model, ensuring consistency between measured and simulated neutron flux. This control loop ensures that the neutron rate used in power balance calculations remains consistent with experimental measurements while maintaining stability in fast ion transport modeling. }

\mc{The anomalous fast ion diffusivity in the model is computed as:}
\mc{\begin{equation}D_a = U(t) D_a^{prof}(x,t)\end{equation}}
\mc{where $D_a^{prof}(x,t)$  represents a spatial profile that may change over time based on NUBEAM input \texttt{DIFB}, and $U(t)$ is a time-varying scale factor for the profile. This scale factor is adjusted during the run to align the measured and calculated neutron rates. The profile shape $D_a^{prof}(x,t)$ is recalculated each time the input for fast ion diffusivity is read as:} 
\mc{\begin{equation}D_a^{prof} = s_a \hat{D}(x,t) / \hat{D}|_{x=0}\end{equation}}
\mc{where $\hat{D}$ is input profile that is set by the TRANSP input \texttt{DIFB}, and $s_a$ is a scale factor for fast ion diffusivity that is set by the TRANSP input \texttt{FB\_NEUTCNTRL\_DIF}. If  $\hat{D}|_{x=0} = 0$, then $D_a^{prof}$  is defined as a flat profile with all values equal to $s_a$.}

\mc{To prevent the diffusivity from becoming negative or excessively large, limits are imposed on the magnitude of $U(t)$. Let these limits be denoted as $U_{max}$ and $U_{min}$. The saturated output $U_{sat}$, is given by: }

\mc{\begin{equation}
U_{sat}(t) =
\begin{cases} 
U_{max}, & \text{if } U_{unsat} > U_{max} \\ 
U_{unsat}, & \text{if } U_{min} \le U_{unsat} \le U_{max} \\ 
U_{min}, & \text{if } U_{unsat} <  U_{min} 
\end{cases}
\end{equation}}
\mc{The limits $U_{max}$ and $U_{min}$ are set using the TRANSP inputs \texttt{FB\_NEUTCNTRL\_UMIN} and 
\texttt{FB\_NEUTCNTRL\allowbreak\_UMAX} correspondingly. The unbounded control variable before saturation is applied, $U_{unsat}(t)$, consists of a pre-programmed component, a feedback component, and an anti-windup component:}

\mc{\begin{equation}U_{unsat}(t) = \frac{\hat{D}|_{x=0}}{s_a} + k_pu_c(t) + k_i \int_{0}^{t}u_c - k_{aw} \int_{0}^{t}u_{aw}\end{equation}}

\mc{Here, $u_c = (n_c - n_m) / s_n$, where  $s_n$ is a scale factor \texttt{FB\_NEUTCNTRL\_NEUT}, $n_m$ is the measured neutron rate \texttt{XNEUTD} , $n_c$ is calculated neutron rate \texttt{TOTNTNS}, $u_{aw} = U_{unsat}(t) - U_{sat}(t)$, and $k_p$, $k_i$ and $k_{aw}$ are set by the TRANSP inputs \texttt{FB\_NEUTCNTRL\_KP}, \texttt{FB\_NEUTCNTRL\_KI}, and \texttt{FB\_NEUTCNTRL\_KAW} correspondingly. The proportional gain $k_p$ controls the immediate response of the system, ensuring fast correction to neutron discrepancies. The integral gain $k_i$ accumulates past errors to correct for long-term biases, while the anti-windup gain $k_{aw}$ prevents excessive integration when the control output saturates.}

\mc{All the parameters described above are specified as input variables in TRANSP.}

\subsubsection{3D halo neutral model}
The original beam-in-a-box model in NUBEAM~\cite{pankin04} simulates neutral beam injection and the formation of halo neutrals within a defined 3D Cartesian domain aligned with the neutral beam footprint. This ``box'' captures primary beam neutrals and \mc{first generation} halo neutrals formed through charge exchange interactions . The box's dimensions and subdivisions are adjustable, allowing detailed tracking of neutral particles as they ionize or exit the simulation domain, ensuring efficient and accurate modeling of beam deposition and halo evolution. \mc{In this model neutrals were distributed over the entire plasma volume}.

The new 3D halo beam model~\cite{medley16} implemented in NUBEAM offers several key improvements over the original beam-in-a-box model originally available in TRANSP.  \mc{While the original model distributed halo neutrals over the entire plasma volume,} \mc{this} 3D halo model localizes \mc{all} neutrals around the neutral beam footprint, providing a more accurate spatial distribution, \mc{ and tracks} multiple generations of halo neutrals through successive charge exchange and ionization events until they either ionize or exit the simulation domain, resulting in a more realistic representation of halo neutral behavior and density profiles \mc{ locally}.

The 3D halo model uses a detailed Monte Carlo simulation to follow the trajectories of halo neutrals, considering spatial and velocity distributions, which offers a higher level of detail and spatial resolution compared to the original model. Furthermore, the 3D halo neutral model in TRANSP has been benchmarked against the Fast-Ion D-Alpha simulation (FIDAsim) code~\cite{heidbrink11}, demonstrating excellent agreement in spatial profiles and magnitude of beam and halo neutral densities, as well as \mc{neutral particle analyzers (NPA)}  energy spectra. This validation ensures the reliability and accuracy of the 3D model compared to the older model. By accurately modeling the spatial distribution of halo neutrals, the 3D halo model significantly improves the reliability of diagnostic simulations, such as those for \mc{NPAs}, leading to better interpretation and prediction of experimental data. These advancements make the 3D halo model a more sophisticated and accurate tool for simulating neutral beam injection and its effects on tokamak plasmas compared to the original beam-in-a-box model. \mc{A complete list of parameters for 3D neutral halo simulation in NUBEAM is presented in Table~\ref{tab:3D}. More details on parameter settings can be found in Sec. 2 of~\cite{medley16}. The 3D halo model calculates NPA fast particle flux and pitch angle (at which fast ions can reach the NPA diagnostic), beam and halo neutral densities, attenuation factor for neutrals, energy spectrum, and emissivity for the NPA sightline.}

\begin{table}[]
 \centering
 \begin{tabular}{|l|l|c|}
 \hline
 \textbf{Parameter} & \textbf{Description} & \textbf{Units} \\ \hline
 \multicolumn{3}{|c|}{\textbf{Input Parameters}} \\ \hline
 \texttt{LEVMOD\_HALO = 2}&{ trigger 3D halo model in NUBEAM}& {} \\ \hline
 \texttt{NBSBOX}    & {number of beam sources (boxes) to calculate neutral density locally} & {} \\ \hline
 \texttt{NBEBOX}     &{energy fraction of beam source to fill box}& {} \\\hline
 \texttt{ XBOXHW}   & {box half-width}&{cm}\\  \hline
 \texttt{ YBOXHW}   & {box half-height}&{cm}\\  \hline
 \texttt{ YBOXHW}   & {box half-height}&{cm}\\  \hline
 \texttt{ XLBOX1}   & {box starting point (distance from beam grid)}&{cm}\\  \hline
 \texttt{ XLBOX2}   & {box stopping point (distance from beam grid) }&{cm}\\  \hline
 \texttt{ NYBOX}   & {number of zones in y-direction in 3D box}& {} \\  \hline
 \texttt{ NLBOX}   & {number of zones in 3D box length}& {} \\  \hline
 \texttt{ NDEPBOX}   & {number of neutral tracks to follow, Monte Carol statistics control}& {} \\  \hline
 \texttt{ NSPLT}   & {maximum splitting along each neutral track}& {} \\  \hline
 \texttt{ NSPLT\_KIN}   & {maximum allowed kinetic splitting at each CX launch point}& {} \\  \hline
 \texttt{ NSPLT\_GEO}   & {maximum allowed geometrical splitting along each neutral track}& {} \\  \hline
 \texttt{ NSPLT\_GEN}   & {descendant halo neutral generations to be followed}& {} \\ \hline
 \multicolumn{3}{|c|}{\textbf{Output Variables}} \\ \hline
 \texttt{ DCXFL$j$(t)}   & {fast ion flux for detector $j$} & {s\textsuperscript{-1}eV\textsuperscript{-1}cm\textsuperscript{-2}} \\ \hline
\texttt{ DCXDT$j$(xs,t)}   & {thermal ion distribution for detector $j$} & {eV\textsuperscript{-1}cm\textsuperscript{-3}} \\ \hline
\texttt{ DCXDF$j$(xs,t)}   & {fast ion distribution for detector $j$} & {eV\textsuperscript{-1}cm\textsuperscript{-3}} \\ \hline
\texttt{ DCXDRT$j$(xs,t)}   & {thermal ion CX rate  for detector $j$} & {s\textsuperscript{-1}} \\ \hline
\texttt{ DCXATT$j$(xs,t)}   &{fast neutrals attenuation for detector $j$} &  \\ \hline
\texttt{ DCXDFL$j$(xs,t)}   &{fast neutrals emissivity for detector $j$} & {s\textsuperscript{-1}eV\textsuperscript{-1}cm\textsuperscript{-3}} \\ \hline
\texttt{ DCXIMF$j$(xs,t)}   &{fast neutrals inv mean free path for detector $j$} & {s\textsuperscript{-1}} \\ \hline
\texttt{ DCXPA$j$(xs,t)}   &{fast neutrals pitch angle for detector $j$} &  \\ \hline
\texttt{ DCXBN0$j$(xs,t)}   &{beam neutrals for detector $j$} & {cm\textsuperscript{-3}}  \\ \hline
\texttt{ DCXFN0$j$(xs,t)}   &{fast neutrals for detector $j$} & {cm\textsuperscript{-3}}  \\ \hline
 \texttt{ DCXVN0$j$(xs,t)}   &{volume thermal neutrals for detector $j$} & {cm\textsuperscript{-3}}  \\ \hline
 \texttt{ DCXCN0$j$(xs,t)}   &{cold thermal neutrals for detector $j$} & {cm\textsuperscript{-3}}  \\ \hline
 \texttt{ DCXVG$g$\_$j$(xs,t)}   &{volume generation $g$ thermal neutrals for detector $j$} & {cm\textsuperscript{-3}}  \\ \hline
 \end{tabular}
 \caption{\mc{Input parameters to set and trigger the 3D halo model in NUBEAM, and TRANSP outputs to observe the 3D halo effect. Here \texttt{xs} denotes the distance along NPA sightlines.}}
 \label{tab:3D}
 \end{table}

\subsubsection{Fast ion RF interaction model in NUBEAM}
A model that describes  the interaction between RF waves and fast ions \mc{(that includes beam ion and fusion products)} in tokamak plasmas has been implemented in NUBEAM~\cite{bertelli17,park12}. The quasi-linear diffusion coefficient for cyclotron absorption~\cite{kennel66} is a key element in the kinetic ion equation, which describes how the ion distribution function evolves under RF wave influence. The interaction between RF waves and particles is described using Fourier amplitudes of the wave field and Bessel functions. The model considers different components of the electric field, assuming a discrete wave spectrum. The kinetic equation for ion velocity distribution includes quasi-linear heating effects, incorporating terms for parallel and perpendicular diffusion with their respective coefficients. The model accounts for the effective interaction time of particles with the wave-particle resonance, which depend on the particle's turning points relative to the resonance layer and involve the particle's magnetic moment and the spatial gradient of the magnetic field. The corrected components of fast ion velocities that account for cyclotron heating effects can be expressed~\cite{xu91} as
\begin{align}
v_{\perp}^{\prime 2} &= v_{\perp}^2+4\left(1-\frac{k_{\|} v_{\|}}{\omega}\right)^2 \alpha_{r f} \Delta t+4 \sqrt{3}\left(R_1-0.5\right)\left(1-\frac{k_{\|} v_{\|}}{\omega}\right) \frac{\omega}{k_{\|}}\left|\frac{k_{\|} v_{\perp}}{\omega}\right| \sqrt{2 \alpha_{r f} \Delta t} \\ \notag
v_{\|}^{\prime} &= v_{\|}+2 \frac{k_{\|}}{\omega}\left(1-\frac{k_{\|} v_{\|}}{\omega}\right) \alpha_{r f} \Delta t+2 \sqrt{3}\left(R_1-0.5\right)\left|\frac{k_{\|} v_{\perp}}{\omega}\right| \sqrt{2 \alpha_{r f} \Delta t}
\end{align}
where $$\alpha_{r f}=\frac{\pi q^2}{4 m^2}\left|E_{+}\right|^2 \sum_{n=-\infty}^{n=+\infty} \delta\left(\omega-k_{\|} v_{\|}-n \Omega\right)\left|J_{n-1}\left(\frac{k_{\perp} v_{\perp}}{\Omega}\right)\right|^2,$$ 
$\left|E_{+}\right|^2=\frac{1}{2}\left|E_x+i E_y\right|^2$, $R_1$ is a random number ranging from 0 to 1, $\Delta t$ is the advance time in the integration of the particle equations of motion, $\omega$ is the wave frequency, $k_\parallel$ and $k_\perp$ are the wave vector along and perpendicular to the magnetic field, $\Omega$  is the gyro-frequency for each species at the instantaneous particle position, and $n$ is the resonance number.

The model for interaction between RF waves and fast ions introduces effects that are important for improved understanding of plasma heating and current drive. The model can be applied for the optimization of tokamak plasma heating and current drive, for the analysis of fast ion behavior, and for improving the design of efficient RF systems with predictive simulations.
\mc{TRANSP namelist parameters to control the RF operator in NUBEAM are presented in Table~\ref{tab:RF}. Note that the cyclotron resonance heating (ICRH) model TORIC-5 should be turned on in TRANSP to create wavefield information for NUBEAM. The fast ion distribution function will be the most relevant output to examine for ICRH effects on fast ions.}

\begin{table}[]
 \centering
 \begin{tabular}{|l|l|c|}
 \hline
 \textbf{Parameter} & \textbf{Description} & \textbf{Units} \\ \hline
\multicolumn{3}{|c|}{\textbf{Input Parameters}} \\ \hline
\texttt{NLFI\_MCRF=.TRUE.}&{to initiate (R,Z) integrator and RF model in NUBEAM} &\\ \hline
 \texttt{NR\_ORBRZ}    & {grid points for R - coordinate}& \\ \hline
  \texttt{NZ\_ORBRZ}    & {grid points for Z - coordinate} &\\ \hline
 \texttt{ORBRZ\_ACC}     &{(R,Z) orbit integrator error control}&\\ \hline
 \multicolumn{3}{|c|}{\textbf{Output Variables}} \\ \hline
 \texttt{MCRFPTOT(t)} & total RF power to beam, fusion ions & W\\ \hline
 \texttt{MCRFP(x,t)} & profile for  RF power to beam, fusion ions & W\\ \hline
\end{tabular}
 \caption{\mc{Input parameters to activate the RF model in NUBEAM, and related TRANSP output variables}}
 \label{tab:RF}
 \end{table}

\subsubsection{\mc{\textit{Kick model}} for fast ion transport by Alfv\'en instabilities}
To account for resonant fast ion transport induced by \mc{Alfv\'en} and other MHD instabilities, NUBEAM has been updated to include a physics-based reduced model known as the \textit{kick model} \cite{mp_2014,mp_ppcf_2017}.

The \textit{kick model} introduces a \textit{transport probability matrix},  $p(\Delta E, \Delta P_\phi | E,P_\phi,\mu)$, for each instability or set of instabilities  included in the simulation. This matrix is defined in terms of the constants of motion variables: energy $E$, canonical toroidal angular momentum $P_\phi$, and magnetic moment $\mu$~\cite{rw_book}.  For each region in phase space defined by $(E,P_\phi,\mu) $, $p(\Delta E, \Delta P_\phi)$ represents the probability of changes (or \textit{kicks}) in $E$ and $P_\phi$ experienced by energetic ions due to their interaction with the instability.

The transport probability matrices are computed numerically using the particle-following code ORBIT~\cite{ORBIT}, utilizing mode structures from the NOVA code~\cite{cheng_novak,gyfu_1992,gorelen_1999}. This method allows the model to accurately represent the effects of resonant wave-particle interactions, which are important for understanding the dynamics of fast ions in the presence of MHD instabilities.

The \textit{kick model} improves the ability to distinguish between fast ions based on their phase space coordinates, such as energy and pitch angle, representing a major advancement over the previous diffusive and convective transport models used in TRANSP. The capability to model transport as discrete steps (or kicks) in phase space rather than continuous diffusion provides a more precise and detailed picture of fast ion behavior under the influence of Alfv\'enic modes and other instabilities.

One of the key features of the \textit{kick model} is its physics-based approach, which makes it more accurate and reliable compared to earlier models that relied on simplified assumptions. By defining a transport probability matrix that describes the likelihood of energy and momentum changes for fast ions, the model allows for detailed tracking of ion trajectories and interactions with instabilities. The numerical computation of this matrix using ORBIT and NOVA ensures that the transport coefficients are based on realistic and detailed simulations.
\mc{As detailed in the Appendix of~\cite{mp_ppcf_2017}, transport matrix calculations are performed outside of TRANSP, using a representative mode amplitude close to that measured experimentally. The amplitude can then be prescribed as a function of time in a TRANSP run via a Ufile, for example to account for time-varying modes.}

The model's ability to discriminate between \mc{the effects of instabilities on} fast ions based on their phase space coordinates \mc{and species } is crucial for accurately predicting the effects of instabilities on ion confinement and transport~\mc{\cite{kick_multispec_2022} and} ~\mc{\cite{tep_ppcf_2023}}. Implementing the \textit{kick model} within the NUBEAM module of TRANSP provides a robust framework for both interpretive and predictive analyses of tokamak discharges, including the effects of MHD instabilities on fast ion confinement and neutral beam current drive.

Since its implementation, the \textit{kick model} has been extensively used to quantitatively assess the impact of Alfv\'en eigenmodes (AEs) on fast ion behavior~\cite{mp_ppcf_2017}. It has proven particularly valuable in predicting the degradation of fast ion confinement due to MHD instabilities, thereby enabling better optimization of plasma performance and stability in tokamak experiments. The model's detailed insights into the development and validation processes have demonstrated its effectiveness in capturing the complex dynamics of fast ion transport in the presence of MHD instabilities. 

\subsubsection{Resonance line Broadened Quasi-linear (RBQ) model }
The objective of the Resonance Broadened Quasi-linear (RBQ) model~\cite{GorelenkovPoP19} is to accurately simulate the relaxation of the hot ion distribution function in the presence of multiple Alfv\'enic instabilities. The RBQ model aims to provide an efficient framework for understanding and predicting the behavior of energetic particles (EP) in tokamak plasmas, particularly in scenarios where Alfv\'enic modes significantly influence the fast ion transport.

In the RBQ model implemented in NUBEAM, EP diffusion and/or convection are specified for each grid cell. The ions diffuse and/or move convectively according to given values of $D_{\bar{P}_{\varphi}\bar{P}_{\varphi}}$ and $D_{\bar{\mathcal{E}}\bar{\mathcal{E}}}$. At a specific point $(\bar{P}_{\varphi}, \bar{\mathcal{E}})$ within a grid cell, the particle diffuses with these coefficients while simultaneously experiencing convective motion characterized by $\boldsymbol{C}$, the convective EP motion coefficient $\boldsymbol{C}=C_{\bar{P}_{\varphi}}+C_{\bar{\mathcal{E}}}$. Diffusion coefficient matrices $D_{\bar{P}_{\varphi}\bar{P}_{\varphi}}$ and $D_{\bar{\mathcal{E}}\bar{\mathcal{E}}}$  as well as the convective motion coefficients $\boldsymbol{C}$ are calculated by the RBQ code within the quasi-linear framework~\mc{\cite{GorelenkovNF24}}.

Given a time step $\Delta t$, the resonant ion undergoes diffusive changes in the $\bar{P}_{\varphi}$ \mc{and  $\bar{\mathcal{E}}$} direction\mc{s}, expressed as $\sigma_{\bar{P}_{\varphi}} = \pm\sqrt{D_{\bar{P}_{\varphi}\bar{P}_{\varphi}}\Delta t}$, \mc{$\sigma_{\bar{\mathcal{E}}} = \pm\sqrt{D_{\bar{\mathcal{E}}\bar{\mathcal{E}}}\Delta t}$,  correspondingly,} with the sign\mc{s determined randomly}.  Additionally, convective motion induces changes in $\bar{P}_{\varphi}$ by $C_{\bar{P}_{\varphi}}\Delta t$ and in $\mathcal{E}$ by $C_{\bar{\mathcal{E}}}\Delta t$. The time step $\Delta t$ can be optimized for better simulation accuracy. The random steps in NUBEAM constant-of-motion variables are inherent to its Monte-Carlo technique. The model assumes that change in \mc{$\Delta\mu$ is} negligible ($\Delta\mu = 0$), while changes in $\Delta\bar{\mathcal{E}}$ and $\Delta\bar{P}_{\varphi}$ are significant ($\Delta\bar{P}_{\varphi} \neq 0$ \mc{and $\Delta\bar{\mathcal{E}} \neq 0$}).

The RBQ model \mc{extends} earlier quasi-linear (QL) models by incorporating realistic Alfv\'en eigenmode structures and pitch\mc{-}angle scattering \mc{effects in the constant-of-motion space formulation}~\mc{\cite{duarte_prl,GorelenkovNF24}. A key advancement is the inclusion of a window function for each fast ion resonating with the mode, refining the treatment of wave-particle interactions.}  Unlike the local perturbative-pendulum approximation, the RBQ model employs an iterative procedure to \mc{account for} eigenstructure variations within resonances, \mc{often leading to a different} saturation level than predicted by uniform mode structures. This \mc{enhanced approach} has been applied to a DIII-D plasma with an elevated $q$-profile, where beam ion \mc{transport exhibits stiff characteristics~\cite{GorelenkovPoP19}. I}nitial results \mc{demonstrate the model's }robustness, numerical efficiency, and predictive capability\mc{, making it a powerful tool for studying} fast ion relaxation in \mc{both present-day} and future burning plasma experiments.

The RBQ model involves a system of quasi-linear diffusion equations, including boundary conditions, governing the ion distribution function's evolution under the influence of multiple modes. The resonant frequency approach forms the basis of the RBQ methodology, accounting for the resonance broadening induced by multiple modes. The model captures the nonlinear resonant ion dynamics, providing accurate predictions of resonance broadening and saturation levels.

Simulation results demonstrate the RBQ model's effectiveness in predicting fast ion behavior and Alfv\'enic mode saturation~\cite{GorelenkovNF24}. Compared to classical models and other diffusion models, the RBQ model captures complex EP dynamics more accurately, providing detailed distribution functions for subsequent plasma analysis. This makes the RBQ model a significant advancement in simulating energetic particle dynamics in the presence of multiple \mc{Alfv\'en} instabilities, offering a robust and efficient tool for fusion plasma research and optimization.

\subsubsection{NUBEAM optimization for GPU architecture}
Driven by the need for rapid analysis and experimental planning in control rooms, this work draws inspiration from the present-day requirements of quick turnaround between shots. The NUBEAM model has undergone modifications to leverage the power of GPU architecture, thereby accelerating the simulation of fast ion population deposition, orbiting, and slowing down in tokamak plasmas~\cite{gorelenkova23}. The newly developed model has been rigorously tested on two supercomputers, Traverse at PPPL and Perlmutter at NERSC, which utilize NVIDIA V100 SXM2 GPUs and NVIDIA A100 GPUs, respectively.

\mc{The optimization of NUBEAM for GPU architectures has been a multi-step process involving significant code restructuring to improve parallel processing. Initially, OpenMP was introduced alongside MPI with the expectation that utilizing shared memory would enhance performance without requiring hardware modifications. This restructuring involved a complete reorganization of time and particle loops to enable parallel execution, along with analyzing and defining large number of global variables as thread-private to take advantage of threading. However, it was found that OpenMP introduced load-balancing challenges, particularly in cases where only a small number of Monte Carlo particles were involved in certain processes, such as secondary fast-ion species. In contrast, MPI provided better scalability for such cases by distributing work across processes more effectively.}

\mc{The transition to GPU computing required a separate effort to restructure the code for efficient offloading to GPUs, rather than simply reverting to MPI. This redesign was necessary to fully leverage GPU parallelism and improve overall computational performance.}

Building on the modifications made for OpenMP, OpenACC was integrated with MPI to take advantage of the power of GPU architectures. This involved revising the logic for Monte Carlo (MC) markers from orbiting due to losses, thermalization, \mc{neutralization} and recycling, implementing a switch from GPU to CPU calculations, when the number of MC markers falls below a threshold, replacing the Random Number Generator (RNG), and removing legacy Fortran code not supported by OpenACC, with CUDA libraries to enhance performance.

These efforts culminated in a significant reduction in runtime for simulations. For instance, the simulation of the DIII-D Super-H discharge for 200 ms using NUBEAM with 160K MC particles demonstrated a substantial speedup. The results show a notable performance improvement of approximately 10 times on Traverse, utilizing four GPUs and 32 MPI processes for a NUBEAM simulation with 1\mc{6}0K Monte Carlo particles. Moreover, Perlmutter exhibited even better performance, achieving a speedup of approximately 20 times.

\subsection{RF models for heating and current drive\label{sec:rf}}
The RF modules in TRANSP are developed externally to PPPL and are documented in multiple references. This section provides citations for these modules and a brief description of the available options for each module.

\subsubsection{EC heating and current drive models}
The options for EC heating and current drive include the TORAY-GA~\cite{cohen87,lin97}, TORBEAM~\cite{poli18}, and GENRAY~\cite{smirnov03} models. In addition, an option to include an IMAS compatible module is implemented by setting \texttt{EC\_MODEL}=`IMAS' in the TRANSP input as discussed in Sec.~\ref{sec:imas}. In this case, the name of the EC H\&CD model is \mc{retrieved} from the IMAS database. At this time, the only supported EC H\&CD model through this option is the IMAS-compatible TORBEAM model. When setting up TRANSP runs, specific inputs for the TORAY-GA, TORBEAM, and GENRAY models should be provided  \mc{through the TRANSP namelist and the TORAY-GA and GENRAY namelists}, unless users prefer to rely on the default settings. References to the EC H\&CD models are specified by the TRANSP input parameters \texttt{TORAY\_TEMPLATE} and \texttt{GENRAY\_TEMPLATE}. Additionally, general parameters for all EC H\&CD models are set directly through the TRANSP input file. These input parameters and related outputs for EC H\&CD models are summarized in Table~\ref{tab:ec}.

\begin{table}[] 
\centering
\begin{tabular}{|c|l|c|}
\hline
\textbf{Parameter} & \textbf{Description} & \textbf{Units} \\ \hline
\multicolumn{3}{|c|}{\textbf{Input Parameters}} \\ \hline
\texttt{NANTECH}   & Number of gyrotrons (maximum 10, default 1) & - \\ \hline
\texttt{DTECH}     & Frequency of calls to the EC model & - \\ \hline
\texttt{TECHON}    & Times when ECRH is turned on & s \\ \hline
\texttt{TECHOFF}   & Times when ECRH is turned off & s \\ \hline
\texttt{POWECH}    & Power to the plasma & W \\ \hline
\texttt{FREQECH}   & Frequency of the injected wave & Hz \\ \hline
\texttt{XECECH}    & Radial coordinate of the launching point & cm \\ \hline
\texttt{ZECECH}    & Vertical coordinate of the launching point & cm \\ \hline
\texttt{PHECECH}   & Toroidal angle location of source & degrees \\ \hline
\texttt{THETECH}   & Polar launching angle of the central ray & degrees \\ \hline
\texttt{PHAIECH}   & Azimuthal launch angle of the central ray & degrees \\ \hline
\texttt{RFMODECH}  & Fraction of power launched in O-mode & - \\ \hline
\multicolumn{3}{|c|}{\textbf{Output Variables}} \\ \hline
\texttt{FE\_OUTTIM} & Times for additional outputs from the EC model & s \\ \hline
\texttt{PEECH(x,t)} & Profile of total power deposition to electrons & W/cm\textsuperscript{3} \\ \hline
\texttt{PEECHj(x,t)} & Profile of power deposition to electrons from gyrotron $j$ & W/cm\textsuperscript{3} \\ \hline
\texttt{ECCUR(x,t)} & Profile of total Electron Cyclotron current drive & A/cm\textsuperscript{2} \\ \hline
\texttt{ECCURj(x,t)} & Profile of Electron Cyclotron current drive from gyrotron $j$ & A/cm\textsuperscript{2} \\ \hline
\texttt{PECIN} & Total injected Electron Cyclotron power & W \\ \hline
\texttt{PECINj} & Injected Electron Cyclotron power from gyrotron $j$ & W \\ \hline
\texttt{THETECH$j$} & Polar launching angle of the central ray for gyrotron $j$ & degrees \\ \hline
\texttt{PHAIECH$j$} & Azimuthal launch angle of the central ray for gyrotron $j$ & degrees \\ \hline
\end{tabular}
\caption{Input and output parameters for EC models in TRANSP.}
\label{tab:ec}
\end{table}

TORBEAM running through the IMAS interface relies on the parameters set in the \texttt{ec\_launchers} IDS. 

\subsubsection{\mc{I}C heating and current drive models}
The TORIC model~\cite{brambilla99} is the only available option for the IC heating and current drive. Two versions of TORIC are currently available in TRANSP. TORIC-5~\cite{wright04} is \mc{available} through a legacy interface and TORIC-6 is available through the IMAS interface, when \texttt{IC\_MODEL} is set to `IMAS' in the TRANSP input.  Both versions compute current drive directly by the wavefield for antennas with a non-symmetric $n_\phi$ spectrum when \texttt{MSYM\_NPHI($j$)} is set to zero for antenna $j$. For symmetric spectra (\texttt{MSYM\_NPHI}($j$)=1, the driven current is zero. To use TORIC's current drive in TRANSP, \texttt{NMOD\_ICRF\_CUR} needs to be set to one and the user can optionally scale the current with the anomalous multiplier \texttt{XANOM\_CURIC}. For a non-symmetric launch spectrum, parameters \texttt{NUM\_NPHI}, \texttt{NNPHI}, and \texttt{WNPHI} can be used to control the definition of the antenna wave $n_\phi$ launch spectrum for each antenna. The parameter \texttt{NUM\_NPHI} specifies the number of $n_\phi$ values, \texttt{NNPHI} provides the actual $n_\phi$ values, and \texttt{WNPHI} sets the corresponding weights for these values. Other input variables and selected output variables related to the ICRF model is given in Table~\ref{tab:ic}.

\begin{table}[h!]
\centering
\begin{tabular}{|c|l|c|}
\hline
\textbf{Variable} & \textbf{Description} & \textbf{Units} \\ \hline
\multicolumn{3}{|c|}{\textbf{Inputs}} \\ \hline
\texttt{NMDTORIC}  & Number of poloidal modes in TORIC & - \\ \hline
\texttt{RFARTR}    & Distance from antenna to Faraday shield & cm \\ \hline
\texttt{RFARTR\_A} & Antenna-specific distance to Faraday shield (uses RFARTR if set to -1.0) & cm \\ \hline
\texttt{ANTLCTR}   & Effective antenna propagation constant & - \\ \hline
\texttt{NFLRTR}    & Controls ion FLR contributions: Use 1 to include & - \\ \hline
\texttt{NFLRETR}   & Controls electron FLR contributions: Use 1 to include & - \\ \hline
\texttt{FLRFACTR}  & Adjustment factor for ion FLR terms in TORIC & - \\ \hline
\texttt{NBPOLTR}   & Inclusion of poloidal magnetic field \mc{in ICRF calculation}& - \\ \hline
\texttt{NQTORTR}   & Inclusion of toroidal broadening in plasma dispersion function & - \\ \hline
\texttt{NCOLLTR}   & Inclusion of collisional contributions in plasma dispersion function & - \\ \hline
\texttt{ENHCOLTR}  & Enhancement factor for electron collisions when NCOLL is included & - \\ \hline
\texttt{ALFVNTR}   & Parameters for ad hoc collisional broadening of \mc{Alfv\'en} and ion-ion resonances & - \\ \hline
\multicolumn{3}{|c|}{\textbf{Outputs}} \\ \hline
\texttt{ICCUR}         & ICRF driven current & A/cm\textsuperscript{2} \\ \hline
\texttt{PIICH}         & ICRF ion heating & W/cm\textsuperscript{3} \\ \hline
\texttt{PEICH}         & ICRF electron heating & W/cm\textsuperscript{3} \\ \hline
\texttt{NMINI}         & Total ICRF minority density & N/cm\textsuperscript{3} \\ \hline
\texttt{PBWAV\_$X$}      & RF power to $X$ beam: wave-code deposition & W/cm\textsuperscript{3} \\ \hline
\texttt{PBQSL\_$X$}      & RF power to $X$ beam: FPP quasi-linear operator & W/cm\textsuperscript{3} \\ \hline
\texttt{PBQLN\_$X$}      & RF power to $X$ beam: FPP quasi-linear operator renormalized & W/cm\textsuperscript{3} \\ \hline
\texttt{BRFRAT\_$X$}     & RF to $X$ beam: Ratio of power wave deposition to FPP quasi-linear operator & - \\ \hline
\texttt{BRKPRHO\_$X$}    & RF to $X$ beam: ratio  of average wave deposition to FPP contribution & - \\ \hline
\end{tabular}
\caption{Input and output variables for the TORIC ICRF model. In this table, $X$ represents different beams including various isotopes of Hydrogen (H, D, and T), Helium (He$^3$ and He$^4$) and fusion products. FPP stands for Fokker-Planck-Poisson formalism.}
\label{tab:ic}
\end{table}

\subsubsection{LH heating and current drive models}

The GENRAY code~\cite{smirnov03} with CQL3D~\cite{harvey92} as a Fokker-Planck solver; and the LSC-SQ module that is a modified version of the LSC code~\cite{ignat94,ignat00} are the options for Lower Hybrid heating and current drive. The inputs to these models are set with separate input files specified with the \texttt{GENRAY\_TEMPLATE} and \texttt{LSC\_TEMPLATE} parameters in the TRANSP input respectively. The outputs from the LH H\&CD models in TRANSP include several key metrics. The total power to ions and electrons is measured, along with the power absorbed and not absorbed in the plasma. Power input on each antenna and power absorbed for each antenna are recorded, as well as the driven current, electron heating, and ion heating for each antenna. Specific outputs include the total LH power to ions \texttt{PLHI}, electrons \texttt{PLHE}, absorbed in plasma \texttt{PLHABS}, not absorbed in plasma \texttt{PLHREF}, and the power and current metrics for each individual antenna.

\mc{
\section{Radiation Models\label{sec:radiation}}
TRANSP includes models for calculating radiative power losses in tokamak plasmas, 
incorporating bremsstrahlung, impurity line radiation, and cyclotron radiation. 
These losses can be included in the electron power balance and used as standalone 
diagnostic outputs. The selection and configuration of radiation modeling are 
controlled through several namelist parameters.

If measurements of radiated power are available from the shot of interest,
they can be provided to TRANSP via a bolometry UFILE by setting \texttt{PREBOL}
and \texttt{EXTBOL} appropriately in the TRANSP namelist.
The file can either provide 1D ``wide angle'' data in watts/time for an
assumed flat profile, or a 2D radiation profile vs time.
Another namelist variable, \texttt{PRFAC}, can optionally be used to set the
fraction of the thermal neutral eflux power to subtract out of the power
radiated profile before using that profile in the electron energy balance.

In the absence of supplied bolometer data, TRANSP is capable of calculating
power radiated due to bremsstrahlung, impurity line radiation, and cyclotron
radiation, and incorporating these into the power balance.
Bremsstrahlung radiation results from electron-ion collisions and is calculated 
for both bulk plasma and impurity species if enabled. Impurity line radiation originates 
from bound-bound transitions of impurity species, with its magnitude dependent on the impurity 
composition and density. Cyclotron radiation is emitted due to electron gyro-motion and 
depends on the magnetic field strength and plasma density, with the 
net power loss affected by the wall reflectivity parameter.

The master namelist variable \texttt{NPRAD} determines how radiation is calculated.
Setting \texttt{NPRAD} to -1 turns off the radiation loss term entirely, while
a value of zero suppresses the theory-based calculation in favor of the
bolometry UFILE data.
$\texttt{NPRAD}=1$, the default setting, uses bolometry data for radiation when
the electron temperature $T_e$ is an input quantity, but switches to the theory
calculation when $T_e$ is predicted.
A value of 2 forces the theoretical calculation for all cases.
Other input options and selected output variables related to radiated power
are given in Table~\ref{tab:rad}. 

\begin{table}[h!]
\centering
\begin{tabular}{|c|p{10.5 cm}|c|}
\hline
\textbf{Variable} & \textbf{Description} & \textbf{Units} \\ \hline
\multicolumn{3}{|c|}{\textbf{Inputs}} \\ \hline
\texttt{NLRAD\_BR} & \texttt{.TRUE.} to calculate bremsstrahlung by impurity species
(total is always calculated) & - \\ \hline
\texttt{NLRAD\_LI} & \texttt{.TRUE.} to calculate impurity line radiation & - \\ \hline
\texttt{NLRAD\_CY} & \texttt{.TRUE.} to calculate cyclotron radiation & - \\ \hline
\texttt{VREF\_CY} & Wall reflectivity in cyclotron radiation formula (default 0.9) & - \\ \hline
\multicolumn{3}{|c|}{\textbf{Output Multiplots}} \\ \hline
\texttt{PBOLO} & Radiated power used broken down by type & $W/cm^3$ \\ \hline
\texttt{CPBOLO} & Comparison of supplied and calculated radiated power &
$W/cm^3$ \\ \hline
\texttt{PBOLOS} & Calculated radiated power by each impurity species & $W/cm^3$ \\ \hline
\texttt{PBS\_<Id>} & Calculated line and bremsstrahlung radiation for species
with symbol \texttt{<Id>} & $W/cm^3$ \\ \hline
\texttt{PBSLI\_<Id>} & Calculated line radiation for all charge states for
species with symbol \texttt{<Id>} & $W/cm^3$ \\ \hline
\texttt{PBSBR\_<Id>} & Calculated bremsstrahlung radiation for all charge
states for species with symbol \texttt{<Id>} & $W/cm^3$ \\ \hline
\texttt{PBS\_<Id>\_<C>} & Calculated line and bremsstrahlung radiation for
species with symbol \texttt{<Id>} and charge state C & $W/cm^3$ \\ \hline
\end{tabular}
\caption{Input and output variables for TRANSP radiated power calculations.
	\label{tab:rad}}
\end{table}

The radiated power density in $W/cm^3$ due to bremsstrahlung is calculated by
TRANSP according to
\begin{equation}
P_{Br} = 1.8408 \times 10^{-32} Z_{eff} n_e^2 \sqrt{T_e}
\label{eq:brem}
\end{equation}
where $Z_{eff}$ is the effective atomic number of the plasma, $n_e$ is the
electron number density in $cm^{-3}$, and $T_e$ is the electron temperature in eV.

Line radiation power in TRANSP is determined for each charge species present
based on the local electron temperature and density using lookup tables in the
ADAS atomic physics library~\cite{adas}.
The total line radiation power density is then computed as a sum over these
species, weighted by the species' number density profiles.

The cyclotron radiation power density in $W/cm^3$ is computed according to the
formula
\begin{equation}
P_{cyc} = 1.3\times 10^{-21} (T_e B_0)^{5/2}
\sqrt{\frac{n_e (1 - \texttt{VREF\_CY})(1 + \frac{570 a}{R_0 \sqrt{Te}})}{a}}
\label{eq:cycrad}
\end{equation}
where $n_e$ is the electron number density in $cm^{-3}$, $T_e$ is the electron
temperature in eV, $B_0$ is the vacuum magnetic field on axis in $T$, $R_0$ and $a$ are the major and minor radii of the plasma respectively in $m$, and
\texttt{VREF\_CY} is the wall reflectivity fraction in Table~\ref{tab:rad}.

}

\section{Events\label{sec:events}}
\mc{TRANSP includes models specifically developed for interpretive simulations, designed to handle large-scale macroscopic events such as sawtooth crashes and pellet injections while avoiding reliance on experimental data at the times of these events. These models were originally implemented in TRANSP's interpretive mode to accurately reconstruct plasma behavior from experimental measurements. When predictive capabilities were introduced in TRANSP, new models that support predictive simulations were introduces as well. For example, the models for triggering of sawteeth have been adapted to work in predictive mode, enabling simulations of sawtooth-induced profile evolution. Similarly, models for pellet injections support predictive simulations, allowing users to assess the impact of fueling scenarios on plasma performance.}

\subsection{Models for sawteeth\label{sec:sawteeth}}
There are several models and parameter choices available in TRANSP to accurately simulate and analyze sawtooth crash events. 

\begin{table}[]
 \centering
 \begin{tabular}{|c|l|}
 \hline
 \textbf{MODEL\_SAWTRIGGER} & \textbf{the sawtooth trigger model } \\\hline
  \textbf{-1}& {sawteeth prescribed by \texttt{UFILE}:  \texttt{PRESAW} and  \texttt{EXTSAW}} \\ \hline
  \textbf{0}& {Prescribed sawtooth period starting at  \texttt{T\_SAWTOOTH\_ON(j)}} \\ 
  \ & {and continuing with  \texttt{SAWTOOTH\_PERIOD}} \\ \hline
 \textbf{1}    &Park-Monticello model ~\cite{park1989}\\ \hline
 \textbf{2}   & Porcelli model~\cite{porcelli96}\\ \hline
 \end{tabular}
 \caption{\mc{Input control parameter for sawtooth trigger models in TRANSP namelist.}}
 \label{tab:saw}
 \end{table}

The sawtooth model can be activated by setting the \texttt{NLSAW}\mc{\texttt{\_TRIGGER}} variable to \texttt{.TRUE.}. Sawtooth event times can be defined using  \mc{different models. The namelist control  parameter for these models is shown in Table~\ref{tab:saw}. With  \texttt{MODEL\_SAWTRIGGER=-1}, the \texttt{UFILE} input file specified in TRANSP namelist variables \texttt{PRESAW} and \texttt{EXTSAW} will be used for the sawtooth trigger times.  This \texttt{UFILE} can be} created by a pattern recognition \mc{code ($e.g.$, ``sawtoo'')} based on analysis of a trace of x-ray or \mc{Electron Cyclotron Emission} diagnostic (ECE) data. Discontinuities in these data must be preserved during data interpretation and analysis in TRANSP.  \mc{The ``sawtoo'' code can be built using the ``ufiles'' library, with its repository available on Princeton University's GitHub website.} 

\mc{For \texttt{MODEL\_SAWTRIGGER=0}, t}he parameter \texttt{DTSAWD} is used for setting data safety intervals around sawteeth, while the parameters \texttt{T\_SAWTOOTH\_ON} and \texttt{SAWTOOTH\_PERIOD} are used for controlling the timing of sawtooth events. Switches \texttt{NLSAWE}, \texttt{NLSAWI}, \texttt{NLSAWB}, \texttt{NLSAWF}, and \texttt{NLSAWIC} control the activation of the sawtooth model for electrons, ions, beam fast ions, fusion product fast ions, and ICRF minority fast ions respectively.

The Kadomtsev~\cite{kadomtsev75} and Porcelli~\cite{porcelli96}  models are theory-based models \mc{available} in TRANSP to analyze sawtooth events. The Kadomtsev model ensures the safety factor $q\ge 1$ throughout, while the Porcelli model can result in $q< 1$ near the axis. \mc{Setting control parameter \texttt{NLSAW=.TRUE.} in the TRANSP namelist will turn on the Kadomtsev sawtooth model}. The variable \texttt{NMIX\_KDSAW} controls which model is used, with two options for the Kadomtsev model and two for the Porcelli model. Specifically, possible values of \texttt{NMIX\_KDSAW} are defined as follows:
\begin{itemize}
\item \texttt{NMIX\_KDSAW}: 1 -- standard Kadomtsev model;
\item \texttt{NMIX\_KDSAW}: 2 -- Kadomtsev model with full particle mixing instead of Helical flux matching method. Predicted densities are fully flattened;
\item \texttt{NMIX\_KDSAW}: 3 -- Porcelli model with two mixing regions, one for the ``island'' around $q=1$, and a second one for the axial region inside the island annulus;
\item \texttt{NMIX\_KDSAW}: 4 -- Porcelli model with a single mixing region for the predicted plasma species, covering both the $q=1$ island and the axial region.
\end{itemize}
The fraction of ion and electron energy affected by sawtooth events can be adjusted using variables \texttt{XSWID1}, \texttt{XSWID2}, and \texttt{XSWFRAC\_DUB\_TE}. \mc{The \texttt{XSWID1} input variable controls the prediction of the response of ion temperature after the sawtooth.  For the default value, \texttt{XSWID1=0.0}, the sawtooth models compute the change in the ion energy density profile.  Then the new ion temperature profile is found by dividing the actual post sawtooth ion density determined by measured data.  Setting \texttt{XSWID1} to 0.0 conserves total ion energy. However, the new ion temperature profile can have an odd shape if the measured density change is different from what the Kadomtsev/Porcelli model would have predicted, or if the size of the sawtooth modeled by TRANSP is different from what actually happened in the experiment. As an alternative, the input variable \texttt{XSWID1} can be set to 1.0 to use the Kadomtsev/Porcelli prediction  for change of ion temperature and the data for change of ion density. Setting \texttt{XSWID1=1.0} does not strictly conserve ion energy.  Values of \texttt{XSWID1} between 0.0 and 1.0 give a sliding control between these two model options.}

\mc{The input parameter \texttt{XSWID2} controls the prediction of the response of the plasma current profile to the sawtooth event.  The default, \texttt{XSWID2=0.0}, gives the full Kadomtsev/Porcelli predicted change in the current profile, i.e. (for full Kadomtsev) $q=1$ on axis and $q>1$ off axis, after the sawtooth. Setting \texttt{XSWID2=1.0} causes there to be no current mixing, $i.e.$, the current and $q$ profile are modeled as unchanged by the sawtooth event. Setting \texttt{XSWID2} between 0 and 1 gives a sliding control to select a model that ``averages'' between these two model options.}

\mc{The input parameter \texttt{XSWFRAC\_DUB\_TE} controls what fraction of the change in poloidal field energy density due to the sawtooth is applied to the 
thermal electrons.  This is of importance only if the electron temperature is being predicted. If \texttt{XSWFRAC\_DUB\_TE=1.0}, and if the \textit{ad hoc} current mixing somehow produces a field that contains more energy than before the sawtooth, this will reduce the electron temperature. In some cases with a very low density this may cause the temperature to go negative, crashing the code. Note that the change in para/diamagnetic toroidal field energy should also be taken into account, but this is not available at the sawtooth time, because the MHD equilibrium is not immediately recalculated.}

The theory-based models can be used both for interpretive and predictive TRANSP modeling. 

In the Kadomtsev model,  the crash is triggered by the ideal MHD instability that leads to the formation of a magnetic island at the $q=1$ surface and its subsequent growth due to resistive diffusion. The Porcelli model is more advanced compared to the Kadomtsev model. It suggests that sawtooth crashes are triggered by the growth of a magnetic island at the $q=1$ surface due to resistive instabilities and takes into account the fast particles and other kinetic effects. The models have been previously compared~\cite{bateman06}.

The sawteeth can interact with fast ion species that are tracked with the RF or NBI modules. In particular, TRANSP includes the ICRF minority ion sawtooth model.  The model works in conjunction
with the ICRF wave and Fokker Planck codes~\cite{smithe89}.  The model replaces the minority ion distribution function with its volume averaged value from the magnetic axis to the mixing radius; outside this radius the distribution function is unchanged.  This is done at each sawtooth event when there is a non-thermal minority ion distribution; it does not
work before the ICRF turns on, or after the minority ions have thermalized.  Energy moments of the modified distribution function show this flattening while the minority particle density
is not affected by this model because TRANSP legislates the minority density as determined by input parameters.

\subsection{Models for pellets\label{sec:pellets}}
TRANSP accounts for the timing of pellet injections by setting hazard time intervals in the TRANSP input. These intervals exclude data points around the injection times to prevent inaccuracies in the analysis caused by the transient effects of pellet ablation. This approach ensures that the interpretive analysis remains robust and reliable, even in the presence of rapid changes induced by pellet injections. The data are extrapolated around pellet injection times, which are specified with two arrays \texttt{TPELDA(1)} and \texttt{TPELDA(2)} for last valid time preceding the pellet injection and first valid time following the pellet injections. The \texttt{NPEL} and \texttt{TPEL} parameters in the TRANSP input define the number and timing of pellet injections, respectively, ensuring accurate temporal resolution of the events.

In TRANSP, there are several ablation models~\cite{parks77,houlberg79, parks05,kuteev84,milora83,houlberg88,macaulay94} that can be specified using the \texttt{KPELLET} parameter. The pellet ablation model defines how the pellet material transitions into plasma. This information is important both for interpretive and for predictive simulations with TRANSP. Each ablation model is tailored to different scenarios and pellet compositions. Models 0--5 are used for hydrogenic species and models 6--11 are used for pellets with impurities. Some additional details about the pellet model in TRANSP are give in Table~\ref{tab:pellets}.

\begin{table}[h!]
\centering
\begin{tabular}{|c|l|p{8cm}|}
\hline
\textbf{KPELLET} & \textbf{Pellet Ablation Model Identifier} & \textbf{Description} \\ \hline
0 & NGS model & Neutral Gas Shielding model with electron distribution and elliptical shielding effect. \\ \hline
1 & Milora's NGS model & Milora's NGS model, assuming single electron energy and spherical shielding. \\ \hline
2 & NGPS & Neutral Gas Pressure Shielding model with electron distribution and $1~mm$ neutral layer. \\ \hline
3 & Macaulay's hydrogenic model & Macaulay's model specifically designed for hydrogenic pellets. \\ \hline
4 & Kuteev's hydrogenic model & Kuteev's hydrogenic pellet ablation model. \\ \hline
5 & Parks hydrogenic model & Parks' model for hydrogenic pellets. \\ \hline
6 & Parks hydrogenic model arbitrary $Q$ & A variant of Parks' hydrogenic model allowing arbitrary heat flux (Q). \\ \hline
10 & Parks impurity model & Parks' ablation model for impurity pellets, adapted from the hydrogenic version. \\ \hline
11 & Kuteev impurity model & Kuteev's model for impurity pellets, focusing on non-hydrogenic materials. \\ \hline
12 & Advanced impurity model& New model for impurity pellets by P. Parks. \\ \hline
\end{tabular}
\caption{Pellet ablation models used in TRANSP identified by \texttt{KPELLET} parameter.}
\label{tab:pellets}
\end{table}

In TRANSP, the pellet injector's configuration involves several parameters that determine its starting position and angles. The radial starting point is set by \texttt{PLRSTA}, while \texttt{PLYSTA} defines the vertical starting point. The toroidal tilt angle, which influences the injector's rotation around the torus, is specified by \texttt{PLPHIA}. Additionally, the poloidal tilt angle, affecting the injector's tilt relative to the poloidal plane, is determined by \texttt{PLTHEA}. These parameters enable precise control over the injection geometry, impacting the pellet's trajectory and deposition profile within the plasma.

The \texttt{APEL} parameter \mc{is an array specifying} the atomic weight of the material \mc{for each pellet}, accommodating different isotopes of Hydrogenic and impurity species. This flexibility allows researchers to model various experimental conditions and their impacts on plasma behavior. Additionally, the ablation of the pellet material into the plasma is a critical factor in determining the subsequent plasma dynamics, influencing factors such as density and temperature profiles. Some ablation models support multiple isotopes with a maximum of two isotopes in the mixture. The atomic weight of the second pellet material is specified by \mc{array parameter} \texttt{APEL2}, and its fraction out of the total pellet material is determined by \texttt{FPEL2}. By default, \texttt{FPEL2} is set to 0, indicating a pure pellet with atomic weight equal to \texttt{APEL}.

\mc{The pellet models in TRANSP, controlled by the \texttt{KPELLET} parameters, are not yet fully verified and are currently disabled. However, users interested in contributing to model validation may request access on an individual basis for verification and validation studies. Until full validation is completed, this option will only be enabled for users who specifically request these capabilities. A comprehensive description of the model validation process, including inputs and outputs, will be provided in a separate paper detailing the verification and validation efforts.}

\section{Predictive capabilities\label{sec:predictive}}
TRANSP has an option to advance plasma profiles for electron and ion temperatures, electron density, ion thermal density, total impurity density and toroidal rotation velocity by solving the following set of equations:
\begin{itemize}
\item Ion energy conservation equation
\begin{equation}
\frac{\partial}{\partial t}\left[\frac{3}{2} V^{\prime} n_i k T_i\right]
+ \frac{\partial}{\partial \rho}\left[V^{\prime} \left\langle \left| \nabla \rho \right|^2 \right\rangle n_i k \left( \chi_i \frac{\partial T_i}{\partial \rho} - T_i v_i \right)\right]
- \dot{\xi} \frac{\partial}{\partial \rho}\left[\rho V^{\prime} \frac{3}{2} n_i k T_i \right]
= S_i V^{\prime}\label{eq:ti}
\end{equation}
\item Electron energy conservation equation
\begin{equation}
\frac{\partial}{\partial t}\left[\frac{3}{2} V^{\prime} n_e k T_e\right]
+ \frac{\partial}{\partial \rho}\left[V^{\prime} \left\langle \left| \nabla \rho \right|^2 \right\rangle n_e k \left( \chi_e \frac{\partial T_e}{\partial \rho} - T_e v_e \right)\right]
- \dot{\xi} \frac{\partial}{\partial \rho}\left[\rho V^{\prime} \frac{3}{2} n_e k T_e \right]
= S_e V^{\prime}\label{eq:te}\end{equation}
\item Angular momentum conservation equation\\
\begin{align}
& \frac{\partial}{\partial t}\left[V^{\prime} \sum n_i m_i\left\langle R^2\right\rangle \omega\right]+\frac{\partial}{\partial \rho}\left[V^{\prime} \sum n_i m_i\left\langle R^2|\nabla \rho|^2\right\rangle\left(\chi_{\varphi} \frac{\partial \omega}{\partial \rho}-\omega \frac{v_{\varphi}}{\rho}\right)\right] \\
& -\dot{\xi} \frac{\partial}{\partial \rho}\left[V^{\prime} \rho \sum n_i m_i\left\langle R^2\right\rangle \omega\right]=S_\omega V^{\prime}\label{eq:vtor} 
\end{align}
\item Electron density conservation equation
\begin{equation}
\frac{\partial}{\partial t}\left[V^{\prime} n_e\right]+\frac{\partial}{\partial \rho}\left[V^{\prime}\left\langle|\nabla \rho|^2\right\rangle\left(n_e v_e-D_e \nabla n_e\right)\right]-\dot\xi \frac{\partial}{\partial \rho}\left[\rho V^{\prime} n_e\right]=S_e V^{\prime}\label{eq:ne}
\end{equation}
\item Total thermal plasma density conservation equation
\begin{equation}
\frac{\partial}{\partial t}\left[V^{\prime} \sum_i n_i\right]+\frac{\partial}{\partial \rho}\left[V^{\prime}\left\langle|\nabla \rho|^2\right\rangle\left(\sum_i n_i V_i-\sum_i D_i \nabla n_i\right)\right]-\dot\xi \frac{\partial}{\partial \rho}\left[\rho V^{\prime} \sum_i n_i\right]=\sum_i s_i V^{\prime}\label{eq:nmain}
\end{equation}
\item Total impurity plasma density conservation equation
\begin{equation}
\frac{\partial}{\partial t}\left[V^{\prime} \sum_{j, q} n_j^q\right]
+ \frac{\partial}{\partial \rho}\left[V^{\prime} \left\langle \left| \nabla \rho \right|^2 \right\rangle \left(\sum_{j, q} n_j^q v_j^q - \sum_{j, q} D_j^q \nabla n_j^q \right)\right]
- \dot{\xi} \frac{\partial}{\partial \rho}\left[\rho V^{\prime} \sum_{j, q} n_j^q\right]
= \sum_{j, q} s_j^q V^{\prime}\label{eq:nimp}
\end{equation}
\end{itemize}
where $S_e$, $S_i$, $S_j^q$, $S_{te}$, $S_{ti}$, and $S_\omega$ are the source terms, which include radiation loss, neutral gas source, edge source, NBI, ICRF, ECH, LHW sources, as well as fusion reaction source terms.
$D_e$, $D_i$, $D_j^q$, $\chi_e$, $\chi_i$, and $\chi_{\phi}$ are the particle diffusivities for electron, ions and impurities, thermal conductivity, and momentum diffusivity correspondingly,  $v_e$, $v_i$, $v_j^q$ , $V_{te}$, $V_{ti}$, and $v_{\phi}$ are the convective velocities, $\rho=\sqrt{\Phi / {\pi B_0}}$, $\dot\xi=\frac{1}{2 \Phi_{\lim }}$, $\frac{d \Phi_{\lim }}{d t} \quad$ and $\Phi_{\lim }$ is the enclosed toroidal flux. 

For solving these transport equations, TRANSP utilizes PT\_SOLVER, a modular, parallel, multi-regional solver. PT\_SOLVER employs the finite difference method to discretize the transport equations and uses the Newton iteration method\mc{~\cite{jardin2008}}, to solve tri-diagonal finite difference equations. Currently, PT\_SOLVER is verified for advancing the electron and ion temperatures, electron density, and toroidal velocity profiles~\mc{\cite{budnyIAEA12,abbate24}}. Verification of PT\_SOLVER for impurity and total thermal density is ongoing. Charge neutrality, $n_e = \sum_i n_i + \sum_{j} Z_j n_j$, and $Z_{\text{eff}} = \sum_{j} Z_j n_j^2$ are used to constrain the density prediction. For instance, if electron density is predicted, the total thermal density and impurity density are derived \mc{using the value of $Z_{\text{eff}}$}.

PT\_SOLVER in TRANSP is activated by setting \texttt{LPREDICTIVE\_MODE=3} in the TRANSP input. Prediction of different plasma profiles is enabled by resetting the input parameters \texttt{LPREDICT\_TE} for the electron temperature, \texttt{LPREDICT\_TI} for the ion temperature, \texttt{LPREDICT\_NE} for the electron density, \texttt{LPREDICT\_NMAIN} for the total thermal ion density, \texttt{LPREDICT\_NIMP} for the total impurity density, and \texttt{LPREDICT\_PPHI} for the toroidal velocity from zero to one. A selection of theory-based models for the anomalous and neoclassical transport is implemented in PT\_SOLVER to compute the particle and toroidal momentum diffusivities, thermal conductivities, and pitch velocities. PT\_SOLVER allows up to three regions with different combinations of anomalous and neoclassical models: the ``axial'', ``confinement", and ``edge" regions. This separation is motivated by the need for different transport assumptions near the magnetic axis and in the near H-mode pedestal region compared to the main plasma bulk. At least one region, usually the confinement region, must be defined. PT\_SOLVER interfaces with the rest of TRANSP through PLASMA\_STATE, a data interface developed at PPPL for sharing data between codes and modules. PT\_SOLVER is a portable module that can run standalone and has a separate namelist from the rest of TRANSP. 

The PT\_SOLVER regions are defined through a combination of several input parameters. The first parameter is \texttt{XIMIN\_CONF}. It defines the boundary of the axial region. If this parameter is set to a nonzero positive number, the axial region is defined from $\xi=0$ to $\xi=\texttt{XIMIN\_CONF}$. The parameters \texttt{XIBOUND}, \texttt{XNBOUND} and \texttt{XPHIBOUND} define the confinement boundaries for temperature, density and toroidal velocity predictions. The edge region covers the remaining region from $\xi=\texttt{X*BOUND}$ to $\xi=1$.  

The selection of anomalous transport models in PT\_SOLVER includes GLF23~\cite{waltz97}, TGLF~\cite{staebler07}, MMM (Multi Mode model)~\cite{rafiq13}, RLW (Rebut-Lallia-Watkins model)~\cite{rebut89}, COPPI (Coppi-Tang model)~\cite{jardin93}, CDBM (Current Diffusive Ballooning Mode Model)~\cite{fukuyama95,takei07}, USER, and Paleo (Paleoclassical)~\cite{callen05} models. The TGLF and MMM models, which are still under development, are regularly updated in PT\_SOLVER, with version 9.1.2 being the current implementation as of May 2024. The neoclassical models available in PT\_SOLVER are NEOCH (Kapisn model), NEOGK (NEO model)~\cite{belli08,belli12}, and NCLASS~\cite{houlber97}, though the NEOGK model is not fully verified. The neoclassical Kapisn model is based on the modified Chang-Hinton model~\cite{chang86}. All models, except the USER model, are well documented elsewhere. The USER model is a simple analytical model designed for plasma regions or regimes lacking a robust theory-based model, and is briefly described in Sec.~\ref{sec:usermodel}. Users can update the combination of models in PT\_SOLVER through the PT\_SOLVER input file. 

The GLF23, TGLF, MMM and NEOGK models require separate input files \mc{that are referenced from the TRANSP input file}. All other models are initialized with parameters set in the TRANSP input file. \mc{The transport models that require separate input files have default input files in the TRANSP distribution. If the user does not specify model inputs, TRANSP applies the default inputs defined in these files. These defaults have been developed in collaboration with the transport model developers and are designed to be suitable for typical simulations across various tokamaks. Careful examination of the model inputs and configuration of simulation-specific files is strongly recommended to ensure accuracy. Since these transport models are developed externally and function as independent modules within TRANSP, their input requirements may evolve outside the control of TRANSP developers. Consequently, a detailed description of these inputs is not provided in this paper. When updates to these models are integrated into TRANSP, the corresponding template files and documentation are revised accordingly and made available on the TRANSP webpage.}

\mc{In the example PT\_SOLVER input file below, a combination of different models for anomalous and neoclassical transport is used:}
\begin{center}
\fontsize{6}{7}\selectfont
\renewcommand{\baselinestretch}{0.8} 
\begin{minipage}{0.48\textwidth}
\begin{verbatim}
2
&pt_updates
 pt_times=-1.0, 0.5
/

&pt_controls
 pt_update_time = -1.

 ! Sawtooth model
 pt_sawtooth%active = .false.
 ! 1 - PTSAW-1; 2 - PTSAW-2; 3 - PTSAW-3; 4 - USER
 pt_sawtooth%version = 4  
 pt_sawtooth%xsaw_bound = 0.1
 ! Te, Ti, Pphi, Ni, Nz, Ne
 pt_sawtooth%xanom  = 1.0, 1.0, 1.0, 1.0, 1.0, 1.0  

 ! Anomalous transport model
 ! Possible selections : GLF23, TGLF, MMM, RLW, 
 !                       COPPI, CDBM, USER, Paleo
 pt_axial%glf23%active = .true.
 pt_confinement%glf23%active = .true.
 pt_edge%glf23%active = .true.
 
 ! Te, Ti, Pphi, Ni, Nz, Ne
 pt_axial%glf23%xanom  = 1.0, 1.0, 1.0, 1.0, 1.0, 1.0  
 pt_confinement%glf23%xanom  = 1.0, 1.0, 1.0, 1.0, 1.0, 1.0  
 pt_edge%glf23%xanom  = 1.0, 1.0, 1.0, 1.0, 1.0, 1.0  
 
 ! 1 - NCEXB; 2 - DMEXB; 3 - TGYROEXB; 
 ! 4 - TRANSPEXB
 pt_axial%exb%version  = 3 
 pt_confinement%exb%version  = 3 
 pt_edge%exb%version  = 3 
\end{verbatim}
\end{minipage}
\hfill
\begin{minipage}{0.48\textwidth}
\fontsize{6}{7}\selectfont
\renewcommand{\baselinestretch}{0.8} 
\begin{verbatim}
 ! Neoclassical transport model
 ! Possible selections : NEOCH, NEOGK, NCLASS, 
 !                      GTCNEO
 pt_axial%neoch%active = .true.
 pt_confinement%neoch%active = .true.
 pt_edge%neoch%active = .true.

! Te, Ti, Pphi, Ni, Nz, Ne
 pt_axial%neoch%xanom  = 1.0, 1.0, 1.0, 1.0, 1.0, 1.0  
 pt_confinement%neoch%xanom  = 1.0, 1.0, 1.0, 1.0, 1.0, 1.0  
 pt_edge%neoch%xanom  = 1.0, 1.0, 1.0, 1.0, 1.0, 1.0  

! Residuals
 pt_residual%RES_TE = 1D-3
 pt_residual%RES_TI = 1D-3
 pt_residual%RES_NE = 1D-3
 pt_residual%RES_NIMP = 1D-3
 pt_residual%RES_NMAIN = 1D-3
 pt_residual%RES_Pphi = 1D-3

 ! Number of iterations
 NEWTON_ITERATES = 100
 THETA_IMPLICIT = -1D0
/

&pt_controls
 pt_update_time = 0.5

 ! Anomalous transport model
 ! Possible selections : GLF23, TGLF, MMM, RLW, 
 !                       COPPI, CDBM, USER, Paleo
 pt_confinement%tglf%active = .true.
/
\end{verbatim}
\end{minipage}
\normalsize
\end{center}

In this example, the number of updates is specified with the first number in the input file and update times are set in the first namelist block. The first update time is always -1 and the first PT\_SOLVER namelist is activated as soon as PT\_SOLVER is activated in the TRANSP namelist. The \texttt{pt\_axial}, \texttt{pt\_confinement}, and \texttt{pt\_edge} data structures in the input defines the models anomalous and neoclassical model and their parameters for the axial, confinement and edge regions in PT\_SOLVER. The \texttt{xanom} defines scaling factors for different channels of transport. When used for anomalous and neoclassical transport, both the diffusivities and convective velocities for a particular channel of transport are scaled by the \texttt{xanom} factor.
PT\_SOLVER provides four options for setting the $E\times B$ flow shear model for the anomalous transport.
The first two options, \texttt{NCEXB} and \texttt{DMEXB}, are based on the NCLASS and neoclassical expressions of flows. Two other options, \texttt{TGYROEXB} and \texttt{TRANSPEXB}, are based on the Waltz-Miller~\cite{waltz99,waltz97} and Hahm-Burrell~\cite{hahm95} formulations for the $E\times B$ flow shear rates.
\begin{align}
\omega_{E \times B}^{WM} =& \frac{r}{q} \frac{d}{d r}\left(\frac{E_r}{B_\theta R}\right)\notag\\
\omega_{E \times B}^{HB}=&\frac{c R B_\theta}{B} \frac{d}{d r}\left(\frac{E_r}{B_\theta R}\right)
\end{align}
where $E_r$ is the radial electric field, $B_\theta$ is the poloidal component of the magnetic field $B$, and $R$ is the major radius. Different anomalous transport models in PT\_SOLVER require specific formulations for ExB flow shear. The GLF23 and TGLF models expect the Waltz-Miller formulation, while the MMM model requires the Hahm-Burrell formulation.

It is possible to use a combination of several anomalous transport models within a single region in PT\_SOLVER. This approach is useful when theory-based models have deficiencies, and additional transport is required, which can be provided using the USER model described in Sec.~\ref{sec:usermodel}. For instance, some transport models have known shortcomings near the magnetic axis. The USER model can address this additional transport need. Alternatively, a sawtooth model can enhance anomalous transport with a large \texttt{xanom} factor when conditions for sawteeth or interchange modes are met. PT\_SOLVER offers four sawtooth model options in its input file:

\begin{enumerate}
\item Sawtooth model enhancing transport where $q<1$;
\item Interchange model;
\item Combination of sawtooth and interchange models;
\item User-defined sawtooth model.
\end{enumerate}

The second and third options are only available with the TGLF model for anomalous transport. \mc{The sawtooth model in PT\_SOLVER was introduced for the TGLF model, which has a known transport shortfall near the magnetic axis in the region $\xi < 0.2$. This shortfall is believed to be caused by sawtooth and interchange modes, both of which are accounted for in the PT\_SOLVER sawtooth model. The model enhances the effective diffusivity, flattening temperature and density profiles, meaning its effect on current and $q$-profiles is indirect. 
While the PT\_SOLVER sawtooth model can be applied to other transport models, this is not recommended, as more accurate and detailed models, such as the Kadomtsev or Porcelli models described in Sec.~\ref{sec:sawteeth}, are available through the TRANSP input file. However, these models do not include triggering conditions for interchange modes.}

\mc{Multiple transport models can be combined to utilize the best available components from each, providing a more accurate description of transport processes in tokamak plasmas. For example, if a user wants to predict plasma profiles in the confinement region using the TGLF model while also including the micro-tearing mode (MTM) model~\cite{rafiq16} from MMM, this can be achieved by setting}
\begin{verbatim}
pt_confinement%tglf%active = .true.
pt_confinement%mmm%active = .true.
\end{verbatim}  
\mc{in the PT\_SOLVER input file. To prevent double counting contributions from ITG/TEM and ETG modes, these components can be disabled in the MMM namelist file, ensuring that only MTM is included.}


\mc{The PT\_SOLVER sample input file, presented earlier in this section, illustrates the use of different transport models across various plasma regions and the ability to modify transport models over time. In this example, the neoclassical Chang-Hinton model and anomalous GLF23 models are used from the initial time when the PT\_SOLVER is enabled. At $t=0.5$~s, specified by the input parameters \texttt{pt\_times} and \texttt{pt\_update\_time}, PT\_SOLVER updates the anomalous transport model in the confinement region to TGLF. The remaining PT\_SOLVER input} variables include the number of Newton iterations (\texttt{NEWTON\_ITERATES}), the implicitness parameter (\texttt{THETA\_IMPLICIT}), and the maximum residuals for six transport channels specified using the input parameters \texttt{pt\_residual\%$\Upsilon$}, where $\Upsilon$ should be replaced with \texttt{RES\_TE} \texttt{RES\_TI}, \texttt{RES\_NE}, \texttt{RES\_NMAIN}, \texttt{RES\_NIMP}, and \texttt{RES\_PPHI} for different transport channels. The implicitness parameter is used to compute the diffusivities and convective velocities for the current iteration as the following:
\mc{$$
y^j   = \theta_{implicit} \hat{y}^j + (1-\theta_{implicit}) y^{j-1}
$$}
where $\theta_{implicit}$ is the implicitness parameter, \mc{$\hat{y}^j$ are the diffusivities and convective velocities computed in transport models at the current iteration, and} $y^{j-1}$ and $y^j$ are the diffusivities and convective velocities from the previous and current iterations \mc{in PT\_SOLVER}, respectively. The default value set with $\texttt{THETA\_IMPLICIT}=-1$ enables an implicitness parameter that depends on the iteration number and usually works reasonably well.

\subsection{USER model for anomalous transport\label{sec:usermodel}}

The USER model sets profiles for anomalous diffusivity which change linearly between the boundaries of the axial, confinement, and boundary regions. This model can be enabled by setting \texttt{pt\_[area]\%user\%active=.True.} in the PT\_SOLVER namelist. For instance, setting \texttt{pt\_confinement\%user\%active=.True.} enables the USER model in the confinement region.

The diffusivity values at the boundaries, measured in $cm^2\ s$, are set using the variables \texttt{USER\_CHI0} and \texttt{USER\_CHI1} for six channels of anomalous transport. These channels are:
\begin{enumerate}
\item Electron energy channel
\item Ion energy channel
\item Toroidal angular momentum channel
\item Main plasma ion particle channel
\item Impurity ion particle channel
\item Electron particle channel
\end{enumerate}

\begin{figure}
    \includegraphics[width=0.50\textwidth]{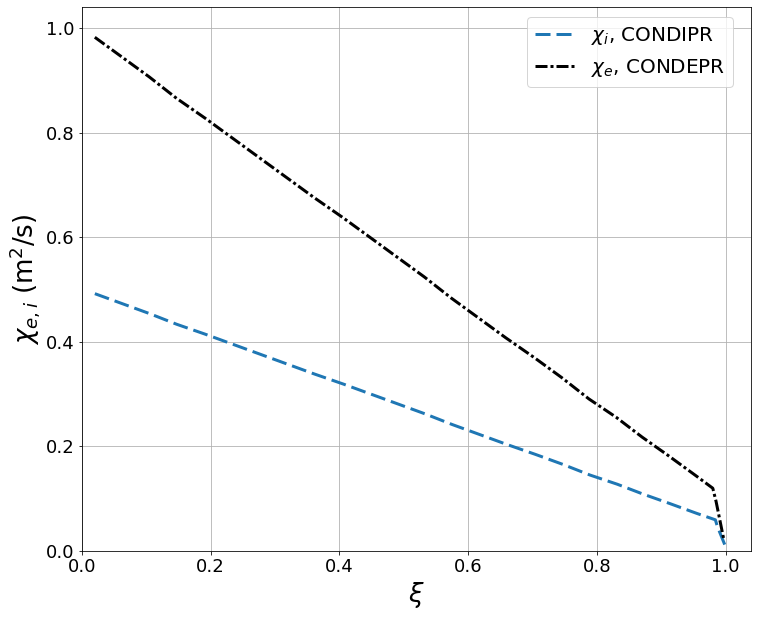}
    \justify
    \caption{Ion and electron thermal diffusivities defined with the USER model.} \label{fig:usermodel}
\end{figure}

All channels are disabled by default. To enable a channel, use the \texttt{NCHAN} variable. For example, the settings:

\begin{verbatim}
NCHAN = 1,1,4*0
USER_CHI0 = 1.e4, 5e3, 4*0.
USER_CHI1 = 1.e3, 5e2, 4*0.
\end{verbatim}

will set the diffusivity profiles for the electron and ion energy channels only, and these profiles will have the shapes shown in Fig.~\ref{fig:usermodel}.

\subsection{Alternative option to predict the rotation profile}

TRANSP offers an additional option for predicting plasma rotation, developed for interpretive runs where toroidal rotation measurements are unavailable but an estimate of rotation velocities is desired. \mc{This option can be used independently of PT\_SOLVER, meaning it does not require the predictive solver to be enabled. Additionally, it can be used alongside PT\_SOLVER if desired. The option to predict toroidal rotation profiles in interpretive mode was primarily introduced for cases where experimental toroidal rotation measurements are unavailable. To prevent conflicts, care must be taken not to enable toroidal rotation prediction in both PT\_SOLVER and this option simultaneously. }The toroidal momentum diffusivity can be specified in one of three ways, controlled by the \texttt{NVPHMOD} input parameter:

\begin{enumerate}
    \item \textbf{Separate input file for toroidal momentum diffusivity}: When \texttt{NVPHMOD}=1, the diffusivity is specified in an input file.
    \item \textbf{Fraction of interpretive ion thermal diffusivity}: When \texttt{NVPHMOD}=2, the diffusivity is set to a fraction of the interpretive ion thermal diffusivity, with the Prandtl number provided by the \texttt{XKFVPH} parameter.
    \item \textbf{Constant value for toroidal momentum diffusivity}: When \texttt{NVPHMOD}=3, a constant diffusivity is specified using the \texttt{XKPINP} input parameter.
\end{enumerate}

The plasma rotation profile is predicted from $\xi=0$ to $\xi=\texttt{XPHIBOUND}$. For $\xi > \texttt{XPHIBOUND}$, TRANSP uses experimental data if available. 

A value of zero for \texttt{NVPHMOD} indicates that the provided rotation data will be analyzed, but no predictive calculations will be performed. In this mode, TRANSP will utilize the existing rotation data as-is without attempting to generate new rotation profiles or make any adjustments based on theoretical models. This option is useful when the goal is to interpret and understand the given data without introducing additional predictive profiles.

\subsection{Defining the boundary conditions for predictions}

The transport equations (\ref{eq:ti})-(\ref{eq:nimp}) require boundary conditions that must take into account the regions of validity of theory-based models. While some anomalous and neoclassical transport models can be used from the magnetic axis to the separatrix in Ohmic and L-mode discharges, they are generally unsuitable for the steep gradients encountered in an H-mode pedestal. This must be considered when setting the boundary conditions for predictive simulations using the TRANSP parameters \texttt{X$\Upsilon$BOUND} (where $\Upsilon$ represents \texttt{I}, \texttt{N}, and \texttt{PPHI} for temperature, density, and toroidal rotation). Options include using experimental profiles or selecting theory-based pedestal models. These options are described below.

The controls for determining the $T_e$ boundary condition during $T_e$ profile prediction are governed by the parameters \texttt{MODEEDG} and \texttt{TEEDGE}. When \texttt{MODEEDG} is set to 2, \texttt{TEEDGE} specifies the electron temperature at the boundary specified with the parameter \texttt{XIBOUND}. For \texttt{MODEEDG} set to 3, the input data for $T_e$ specifies the edge electron temperature data, which can be set either using external data files, parametrized profiles, or IMAS. If \texttt{MODEEDG} is set to 5, one of the pedestal models is used to compute the boundary condition. Similarly, for ion temperature $T_i$, \texttt{MODIEDG} determines the boundary condition: \texttt{MODIEDG}=1 and \texttt{MOD0ED}=1 use \texttt{TIEDGE}; \texttt{MODIEDG}=2 uses \texttt{TIEDGE}; \texttt{MODIEDG}=3 (default) uses $T_e$ input data at the edge; and \texttt{MODIEDG}=4 uses $T_i$ input data at the edge specified with external files, IMAS, or parametrized profile. The option \texttt{MOD0ED} is implemented to set the ion temperature at the boundary from the recycling temperature. When \texttt{MOD0ED} is set to 1, the edge recycling source energies use time-dependent recycling temperature data if provided; otherwise, they use the \texttt{TIEDGE} input to set the recycling neutral temperatures. If time-dependent recycling temperature data is provided, \texttt{MOD0ED} is enforced to be 1. For warm source energies, \texttt{MOD0ED}=2 uses \texttt{TI0FRC} (a fraction of central ion temperature); \texttt{MOD0ED}=3 uses electron temperature data at the edge; and \texttt{MOD0ED}=4 uses ion temperature data at the edge.

The boundary conditions for the density $n_e$ profile prediction are controlled by \texttt{MODNEDG}. When \texttt{MODNEDG} is set to 2, the edge density is specified by \texttt{EDGENE}. With \texttt{MODNEDG} set to 3, the input data for $n_e$ is taken from an external input file, parametrized profiles or IMAS. When \texttt{MODNEDG} is set to 5, the pedestal model is used to compute the boundary condition. Similarly, the boundary conditions for the toroidal rotation $\Omega_\phi$ profile prediction are controlled by \texttt{MODOMEDG}. When \texttt{MODOMEDG} is set to 2, the edge angular velocity is specified by \texttt{OMEDGE}. With \texttt{MODOMEDG} set to 3, the input data are set from the external file, parametrized dependence or IMAS. There is currently no pedestal model for angular velocity.

When the pedestal models are enabled by setting the parameters \texttt{MODEEDG},  \texttt{MODIEDG}, and \texttt{MODNEDG} to 5, the options for the pedestal models are activated. It should be noted that the pedestal models can be used for one or multiple plasma quantities. The first and most trivial pedestal option sets the pedestal height and width to pre-set values. To impose user-defined pedestal widths and heights from the namelist in TRANSP, set \texttt{NMODEL\_PED\_WIDTH = 0} and \texttt{NMODEL\_PED\_HEIGHT = 0}, respectively. For pedestal widths, use \texttt{TEPEDW} for electron temperature, \texttt{TIPEDW} for ion temperature, and \texttt{XNEPEDW} for electron density, with positive values in cm and negative values in $\xi$. For pedestal heights, use \texttt{TEPED} and \texttt{TIPED} for electron and ion temperatures (default 100 eV), and \texttt{XNEPED} for electron density (default $10^{12}$ cm$^{-3}$).

To use a predictive model for the pedestal pressure and width, set \texttt{NMODEL\_PED\_HEIGHT = 1} and \texttt{NMODEL\_PED\_WIDTH = 1}. Since some models predict both pressure and width, it is recommended to set both flags to 1 as results may be overwritten. When the pedestal height is predicted, the calculated density and temperature can be rescaled using \texttt{SCALE\_TEPED} for electron temperature, \texttt{SCALE\_TIPED} for ion temperature, and \texttt{SCALE\_NEPED} for electron density. All pedestal models calculate the pedestal pressure, with the pedestal temperature derived from the given pedestal density using the expression $p_\text{ped}/2k_B n_{ped}$, where $k_B$ is the Boltzmann constant, $p_{ped}$ and $n_\text{ped}$ are the pressure and density at the top of the pedestal. The resulting pedestal temperature depends on how the pedestal density is calculated.

The options to calculate pedestal density are determined by the \texttt{LPED(2)} parameter. For \texttt{LPED(2)=1}, the pedestal density $n_{\text{ped}}$ is set to 0.71 times the line-averaged density.  For \texttt{LPED(2)=3}, the pedestal density is calculated as $C  n_{\text{sep}}$, where \( C \) is set by the input variable \texttt{CPED(4)} and $n_\text{sep}$ is the density at the separatrix. For \texttt{LPED(2)=4}, the pedestal density and width are set at the location of the maximum profile gradient. Lastly, when \texttt{LPED(2)=2}, the pedestal density and width are obtained from a fit over the density profile using a specific parametrization:
\begin{equation}
\begin{split}
n_e(\xi) = n_{e,0} & \left\{ \left(1-r_2\right) \left( c_1 \left[ H(1-\xi) \left( 1-\xi^\alpha \right)^\beta \right] + \right. \right. \\
& \left. \left.  c_2 \left[ \tanh \left(2 \frac{1-\xi_{\text{mid}}}{1-\xi_{\text{ped}}}\right) - \tanh \left(2 \frac{\xi-\xi_{\text{ped}}}{1-\xi_{\text{ped}}}\right) \right] \right) + r_2 \right\}
\label{eq:tanh}
\end{split}
\end{equation}
where
\begin{align*}
c_2 &= \left( \frac{r_1-r_2}{1-r_2} \right) \frac{1}{2 \tanh (1)}\\
c_1 &= 1 - c_2 [1 - \tanh (1)]
\end{align*}
Here, $r_1$ is the ratio of the density at the pedestal to the central value ($ n_{e, \text{ped}} = r_1 n_{e, 0} $), and $r_2$ is the ratio of the density at the separatrix to the central value ($n_{e, \text{sep}} = r_2 n_{e, 0}$). The fit returns the width of the pedestal in normalized poloidal flux coordinates and the density at the pedestal location, with the mid-pedestal location defined as half a pedestal width to the right of the pedestal and the top defined as half a pedestal width to the left. \mc{When \texttt{LPED(2)=2}} is selected, TRANSP will fit the density profile regardless of whether the density is prescribed or predicted. 

There are  three sets of options to calculate the pedestal \mc{pressure or temperature} for the TRANSP pedestal model. \mc{When the pedestal pressure is specified within the model, the temperature is determined using the pedestal density values based on the selected parameters. Conversely, when the pedestal temperature is specified, the pressure is computed using this temperature and the density from the pedestal density model.} 

The first option \mc{to calculate the pedestal pressure or temperature} is based on five models developed by T. Onjun~\cite{onjun02}. The models include:
\begin{itemize}
\item \texttt{LPED(3)=1}: Pedestal model using pedestal width based on magnetic and flow shear stabilization.
\item \texttt{LPED(3)=2}: Pedestal model using pedestal width based on flow shear stabilization.
\item \texttt{LPED(3)=3}: Pedestal temperature model using pedestal width based on normalized pressure.
\item \texttt{LPED(3)=11}: Pedestal temperature model based on thermal conduction.
\item \texttt{LPED(3)=12}: Modified pedestal temperature model based on thermal conduction.
\end{itemize}

The second set of options is based on MHD stability calculations. These options are based on the work of P. Maget~\cite{maget13}. For \texttt{LPED(3)=21} and \texttt{LPED(3)=22}, the pedestal width must be specified in cm using \texttt{CPED(3)}, or derived from a fit over the pedestal density profile if \texttt{CPED(3)=0.0} and \texttt{LPED(2)=2} are set; these are designed specifically for ITER \mc{because these models have been developed using equilibrium and stability analyses for ITER's operational scenarios, including pedestal stability constraints governed by bootstrap current, peeling-ballooning stability, and pedestal width effects on pressure limits}. Options \texttt{LPED(3)=213} and \texttt{LPED(3)=223} combine \texttt{LPED(3)=3} for the pedestal width with \texttt{LPED(3)=21} for the height, using the pedestal model for the width regardless of \texttt{LPED(2)} settings, also intended exclusively for ITER applications.

Finally, there are options for the H-mode pedestal pressure based on the EPED model~\cite{snyder11}. For \texttt{LPED(3)=23}, \mc{the model is based on the EPED1 model with the boundary conditions predicted by the SOLPS4 code for ITER~\cite{polevoi15}. Unlike the standard EPED1 model, which assumes a degradation of pedestal pressure with decreasing pedestal density, the modified EPED1 model accounts for ITER-specific edge physics effects, including the inefficiency of neutral fueling at the pedestal and the saturation of separatrix density due to divertor plasma detachment and strong plasma recombination.} For \texttt{LPED(3)=100}, the pedestal model uses multi-dimensional interpolation over parameters input to the EPED calculations. This lookup table, derived specifically for ITER, consists of over 6000 EPED calculations. Option \texttt{LPED(3)=111} uses a neural network trained on the EPED calculations for ITER, DIII-D, C-Mod, JET, and KSTAR~\cite{meneghini17}. There are several versions of EPED neural network (EPED-NN) models in TRANSP. The original version of the EPED-NN model can be selected by setting \texttt{EPED1NN\_MODEL} to 1. The so-called \mc{``multiroot''} version of the EPED-NN that supports both the H-mode and Super H-mode discharges is selected by setting \texttt{EPED1NN\_MODEL} to 2, and the updated version of the multiroot EPED-NN can be triggered with \texttt{EPED1NN\_MODEL}=-1. To use the tanh-fit for the pedestal pressure profile, as described for density profile fitting in Eq.~(\ref{eq:tanh}), set \texttt{LPED(9)} to 1. This option is available only when using the EPED model for the pedestal pressure.

\section{TRANSP tools and utilities\label{sec:data}}
Over the years a large collection of tools and utilities for TRANSP input preparation, output analysis and visualization has been created. Here we focus on several tools for preparation of TRANSP runs and data analysis. 

\subsection{Preparation for TRANSP runs}
TRANSP pre-processing tools are essential for preparing and validating input data before running simulations with the TRANSP code. These tools streamline the workflow, ensuring that the input data is accurate, consistent, and correctly formatted for the simulation. 

Key functionalities of these tools include data preparation and validation, where utilities convert experimental data into the input formats required by TRANSP, such as Ufile or IMAS, and check the integrity and consistency of the input data. These utilities are specific to experimental devices and depend on how and where the experimental data are stored. Over the years, a large number of tools have been developed to access and convert experimental data to TRANSP input formats for different tokamaks. The most recent addition includes tools developed within the OMFIT project~\cite{omfit}, which map experimental data from various tokamaks to quantities needed to initiate a TRANSP run. For equilibrium, it is common to use magnetic equilibria reconstructed with the EFIT code~\cite{lao85}, and the \texttt{scrunch2} tool from the TRANSP distribution is used to convert EFIT files into the moment representation used in TRANSP (see Sec.~\ref{sec:equilibrium}). After the experimental data are prepared and saved either in Ufiles or IMAS, a set of TRANSP pre-processing tools is used to ensure data consistency and to interpolate the data for times relevant to the simulation. In particular, the \texttt{pretr} tool collects general information about the run, such as the tokamak name, run year, and shot number, and verifies if the tokamak is supported by TRANSP. This tool also verifies some basic logical consistency of the input and defines if the simulation will be serial or parallel. After completing the \texttt{pretr} verification, the next tool, \texttt{trdat}, processes data from Ufiles or IMAS and collects inputs for all models used in the TRANSP simulation, such as GENRAY, CQL3D, TGLF, and NEO. The processed experimental data are stored in a single NETCDF file that is used to pass all this information to TRANSP. Users often prefer to use a set of higher-level scripts that incorporate all these steps.

To run TRANSP, basic scripts are utilized, starting with \texttt{tr\_start}, which must be executed from the directory containing the TRANSP input file. This script creates the MDSplus tree for input data and writes a request file, using environment variables for your email address, preferred editor, and information on the MDSplus server. Following this, \texttt{tr\_send} submits the run to the PPPL queue or via \texttt{globus-job-submit} for non-PPPL users. To clean up, \texttt{tr\_cleanup} queues a request to remove all traces of a run, aborting it if still active. The \texttt{tr\_look} script writes interim MDSplus output into a specific directory on \texttt{transpgrid.pppl.gov}, with options to archive active or aborted runs or generate CDF files only for non-PPPL users. Finally, \texttt{tr\_halt} suspends a run immediately or at a specified time. These scripts are essential for managing and executing TRANSP simulations effectively.

\subsection{Data analysis and visualization\label{sec:rplot}}

\begin{wrapfigure}{l}{0.55\textwidth}
    \includegraphics[width=0.54\textwidth]{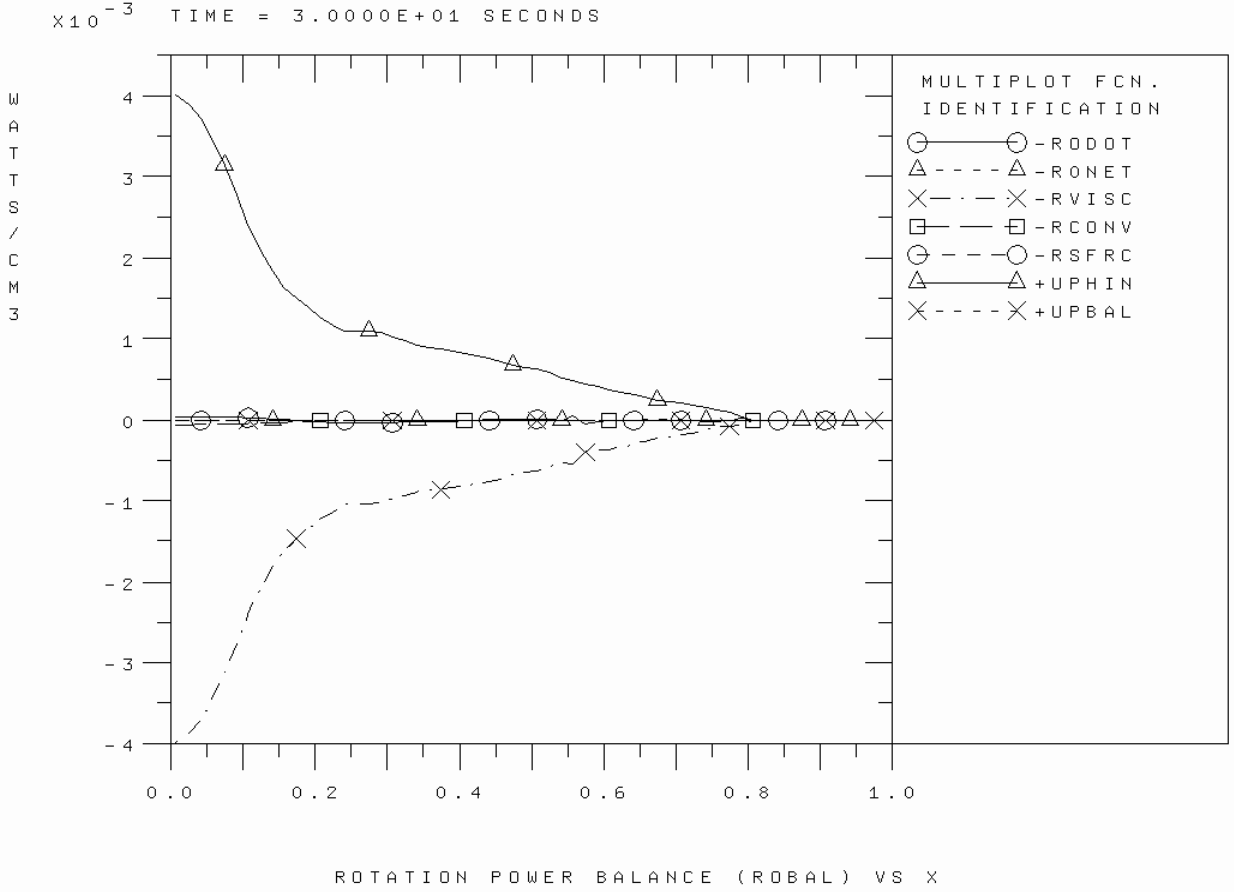} 
    \vspace{-7mm}\singlespacing
    \caption{Contributions to energy balance from rotation that includes the total rotational energy input \texttt{UPHIN}, gain of the rotational energy \texttt{RODOT}, viscous rotational energy loss \texttt{RVISC}, charge-exchange rotational energy loss \texttt{R0NET}, convective rotational energy loss \texttt{RCONV}, and rotational source friction \texttt{RSRRC} for the STAR tokamak~\cite{brown23}. The rotational energy balance is shown as \texttt{UPBAL}.} \label{fig:rbalance}
\end{wrapfigure}

The \texttt{rplot} utility offers a comprehensive suite of functions designed to facilitate the analysis and visualization of TRANSP results. This tool enables users to effectively manipulate data and generate plots that are essential for understanding plasma behavior in simulations. A powerful feature of \texttt{rplot} is the use of multigraphs, which represent predefined selections of plots combined to provide a comprehensive understanding of specific features or effects. Examples of multigraphs include those used for the analysis of particle and power balances, enabling the simultaneous visualization and comparison of various related parameters. A multigraph for contributions from rotation to the energy balance \mc{in the simulations for STAR tokamak~\cite{brown23}} is shown in Fig.~\ref{fig:rbalance}. \mc{It illustrates the contributions to the energy balance from sources and sinks associated with plasma rotation. Specifically, these results show that while the total energy input \texttt{UPHIN} is not negligible, it is largely offset by the viscous rotation energy loss \texttt{RVISC}, resulting in a small net contribution to the overall energy balance from rotation-related terms}.

Additionally, \texttt{rplot} allows users to define custom functions, manipulate data from TRANSP output, perform area, volume, and flux integrals, create new multigraphs, compare data from multiple TRANSP runs, plot MHD equilibria, extract data, and perform other advanced analyses. These capabilities make \texttt{rplot} an invaluable tool for researchers working with plasma simulations. Despite its advanced data manipulation capabilities, the visualization capabilities in \texttt{rplot} are rather basic. It includes support for 1D and 2D plots, selection of scaling options for axes, and grid representation. However, the \texttt{rplot} tool does not support color figures and does not allow changing fonts or customizing legends on the plots. More advanced visualization tools come from the TRANSP user community. Examples of these tools include TRANSP visualization capabilities in OMFIT~\cite{omfit,grierson18} or other tools developed in Python.

\subsection{Extracting equilibria from TRANSP results }
TRANSP offers options to evolve equilibria as described in Sec.~\ref{sec:equilibrium}. This information can be a valuable input for various codes such as MHD stability codes or gyrokinetic codes. In addition to a recently developed option to save equilibrium information in the \texttt{equilibrium} IDS in IMAS (see Sec.~\ref{sec:imas}), the \texttt{trxpl} tool provides an option to convert equilibrium information from TRANSP output files to a number of different formats, including EFIT geqdsk files~\cite{lao85} and NETCDF files. The \texttt{trxpl} tool can retrieve this information either from TRANSP output files or from MDSplus databases containing TRANSP results. It can access \texttt{rplot} functionality to manipulate the data.

The equilibrium data can be saved in various formats, including Fourier moments, EFIT, ESI (Equilibrium Spline Interface)~\cite{li17}, and PSIPQGRZ used by JSOLVER~\cite{delucia80}. The tool also provides options to generate profiles for the guiding center ORBIT~\cite{white84} and gyrokinetic GS2~\cite{gs2} codes and extract information relevant to kinetic equilibrium reconstruction. \mc{The \texttt{trxpl} tool includes an option to save selected equilibrium-related quantities in Ufile format (see Table~\ref{tab:trxpl_vars})}.
\begin{table}[h]
    \centering
    \begin{tabular}{|l|l|l|}
        \hline
        \textbf{Name} & \textbf{Units} & \textbf{Description} \\
        \hline
        \texttt{PHITOR}  & $Wb$       & Toroidal magnetic flux \\
        \texttt{PSI}     & $Wb/rad$   & Poloidal magnetic flux \\
        \texttt{G}       & $T\cdot m$ & Poloidal current function \\
        \texttt{PMHD}    & $Pa$       & MHD pressure \\
        \texttt{Q}       & -        & Safety factor \\
        \texttt{R}       & $m$        & Major radius \\
        \texttt{Z}       & $m$        & Vertical position \\
        \texttt{BPHI}    & $T$        & $B_\phi$ component of magnetic field \\
        \texttt{BR}      & $T$        & $B_r$ component of magnetic field \\
        \texttt{BZ}      & $T$        & $B_r$ component of magnetic field \\
        \texttt{BMOD}    & $T$        & Total magnetic field magnitude \\
        \hline
    \end{tabular}
    \caption{\mc{List of equilibrium variables available in \texttt{trxpl} available for output in the Ufile format.}}
    \label{tab:trxpl_vars}
\end{table}

Additionally, it can save a subset of information needed for the IMAS \texttt{equilibrium} IDS. The data can be saved in ASCII, NETCDF, or binary formats.

\subsection{Exporting the fast ion distribution function from TRANSP output}

The NUBEAM model in TRANSP (see Sec.~\ref{sec:nubeam}) always calculates the fast ion distribution function. TRANSP calculates many quantities derived from the fast-ion distribution function, but does not save it by default. The fast ion distribution function quantities to be saved in output need to be specified in the TRANSP \mc{input \texttt{SELAVG} with a list of names to be averaged prior to output. By default, the output is not averaged.} For example, setting:
\begin{verbatim}
SELAVG='FBM BMVOL BDENS2 EBA2PL EBA2PP' 
\end{verbatim}
in the TRANSP input will save the distribution function \texttt{FBM}, the volume of differential toroidal tube \texttt{BMVOL}, the total number of fast ions \texttt{BDENS2}, and the average parallel \texttt{EBA2PL} and perpendicular \texttt{EBA2PP} energies of the fast ions. These quantities are obtained by averaging the MC statistics over time specified with the TRANSP input variable \texttt{AVGTIM} \mc{at certain time slices specified by the \texttt{OUTTIM} TRANSP namelist control parameter}. The user has the option to save it as a particle version or the guiding center version using \texttt{NLFBMFLR} in the TRANSP input. Setting \texttt{NLFBMFLR} to \texttt{.TRUE.} will use the distribution function at the particle finite Larmor radius location, and setting it to \texttt{.FALSE.} will use the distribution function at the guiding center location.  The distribution function is saved as external ASCII files at the times specified with the parameter \texttt{OUTTIM}.

Several tools have been developed to access, manipulate, and visualize the fast ion distribution function and related quantities. The \texttt{get\_fbm} tool is commonly used to access and convert the distribution function to the NETCDF format. The distribution function $f$ is saved as a function of the coordinates $R$, $Z$, energy $E$, and pitch angle $p = v_\parallel / v$. This data can be used for beam analysis and coupled with other codes. For example, the number density of fast ions as a function of $R$ and $Z$ can be computed as:
\begin{equation}\label{eq:nnbi}
n_{NBI} = {\int_0^{\infty} \int_{-1}^{1} f(E, p, R, Z) \, dE \, dp}
\end{equation}

The distribution function can also be saved in the IMAS format.

\begin{figure}
\centering
\sbox0{%
  \begin{minipage}[b]{.35\textwidth}
  \includegraphics[width=\textwidth]{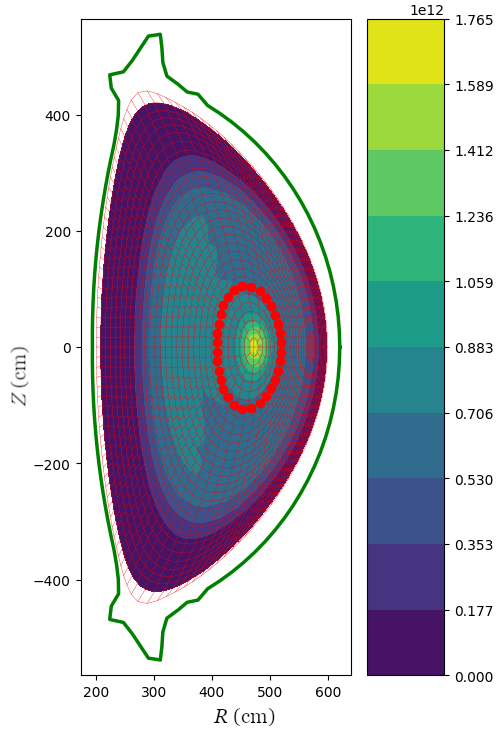}
  \vfill
  (a)
  \end{minipage}}
\usebox{0}\hfill
\begin{minipage}[b][\ht0][s]{.6\textwidth}
\includegraphics[width=\textwidth]{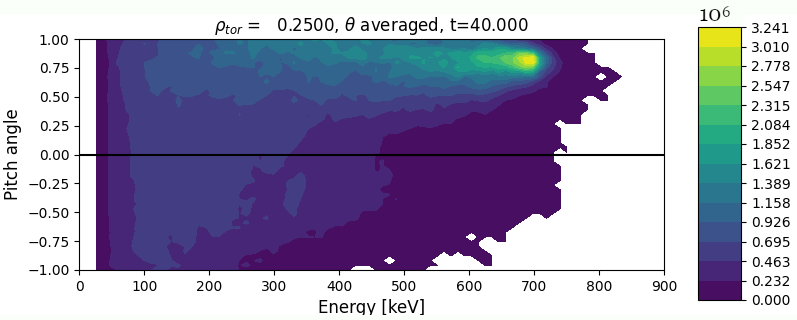}\\
(b)\\
\caption{Distributions of fast NB Deuterium ions as a function of $R$ and $Z$  (a) and as a function of pitch angle $p$ and energy $E$ for the flux surface $\xi=0.25$ (b) in the STAR tokamak~\cite{brown23}.\label{fig:fbm}}
\end{minipage}
\end{figure}

\mc{Another useful tool for accessing and visualizing the fast ion distribution function from the NUBEAM model in TRANSP is FBM. Developed by Dr. G. Tardini at the Max-Planck-Institut für Plasmaphysik in Garching, it is publicly available on GitHub~\cite{FBM}. The distribution of fast NB Deuterium ions as a function of $R$ and $Z$, as defined in Eq.~(\ref{eq:nnbi}) is shown in Fig.~\ref{fig:fbm}a. Additionally, the distribution of the Deuterium ions as a function of $p$ and $E$ for the flux surface marked with a bold red curve in Fig.~\ref{fig:fbm}a is shown in Fig.~\ref{fig:fbm}b.}

\section{Interface to IMAS\label{sec:imas}}
The ITER Integrated Modeling (IM) Program, initiated in 2014, established strategic programmatic requirements and coordination with ITER members~\cite{IMASprogramme,pinches18}. Its primary roles are to support Plasma Operation and Plasma Research while engaging the community through voluntary contributions, the Integrated Modeling Expert Group (IMEG), and the International Tokamak Physics Activity (ITPA) topical groups. For Plasma Operation, the IM Program has developed and systematically executed a set of physics modeling tools used for pulse validation before operation, pulse design and validation between shots, and comprehensive plasma reconstruction post-shot. These efforts have been crucial for ensuring ITER's operational efficiency and safety. For Plasma Research, a broader set of high-fidelity physics modeling tools is employed both pre- and post-operation, focusing on single time slice analysis of microscopic behavior. The Integrated Modeling \& Analysis Suite (IMAS) provides a comprehensive environment of codes, libraries, frameworks, and databases to support both operation and research.

As a significant integrated modeling tool for tokamak plasmas, the TRANSP code is being used for the IM Program and has been integrated into the IMAS framework. This enables seamless access to experimental databases and facilitates integration with other modules, codes, and frameworks, thereby enhancing TRANSP's capabilities of modeling, verification and validation. Its roles encompass interpretive transport analysis and predictions, supporting the assessment of ITER scenarios and Heating and Current Drive (H\&CD) requirements for Fusion Pilot Plant design. These capabilities in TRANSP can be used for other tokamaks as well, and for other fusion devices with sufficient extension of the IMAS framework. Incorporating IMAS in TRANSP was done with this goal. The input and output of IMAS-enabled libraries are stored in Interface Data Structures (IDSs) of IMAS, which enables on-the-fly code coupling, avoiding the conventional approach of code modification to accommodate native input/output data.

Integrating the IMAS in TRANSP required substantial enhancements and compatibility improvements. TRANSP supports two levels of IMAS integration. The first level involves tools for pre- and post-processing TRANSP data in IMAS format, such as the \texttt{imas2transp} tools that access the IMAS database and prepare data in conventional TRANSP input formats. The second level includes the direct integration of IMAS libraries within TRANSP, allowing direct access to input data from the IMAS databases and saving TRANSP output directly to these databases. The enhancements are done such that the existing workflow is unaffected. Users can select the interface via namelist entries as described below. The TRANSP output in IMAS format is mainly saved in the IDSs used for coupling with external modules. More comprehensive conversion of TRANSP output to IMAS format is done using a standalone \texttt{transp2imas} translator tool written in Python. The conventional TRANSP output has reduced storage compared to internal run-time data, which is written directly to IDSs. It is worth noting that, since the former is used as input for \texttt{transp2imas} its converted IDSs share the same limitation. The translator is being used for testing and developments and also for re-running old runs using their output data.

The TRANSP-IMAS interface also has several templates for coupling with external modules, including those for sources such as ECH, ICRF, and NBI. These interfaces have been tested with the TORBEAM module for EC heating and current drive~\cite{poli18}, the TORIC module for IC heating and current drive~\cite{brambilla99,brambilla09}, and the NUBEAM module for NB heating and current drive~\cite{pankin04}. The TORBEAM, TORIC, and NUBEAM modules support both IMAS and legacy interfaces. Some modules are implemented using both interfaces to verify the templates for heating modules. For instance, the TORBEAM module can be used by setting \texttt{EC\_MODEL='TORBEAM'} or \texttt{EC\_MODEL='IMAS'}, invoking the same version of the TORBEAM module. Similarly, the TORIC module can be used in TRANSP by setting \texttt{IC\_MODEL='TORIC'} or \texttt{IC\_MODEL='IMAS'}. However, selecting \texttt{IC\_MODEL='TORIC'} invokes TORIC v5, while selecting \texttt{IC\_MODEL='IMAS'} invokes TORIC v6, the implementation of which is still being verified.

The IMAS interfaces in TRANSP have also been tested by coupling the TRANSP and gyro-fluid reduced resistive MHD FAR3D code~\cite{spong21}, which accounts for the effects of energetic particles. The results of a legacy file-based interface have been verified against those of the IMAS-based interface~\cite{breslau22}.

The IMAS interface in TRANSP remains optional due to the IMAS license agreement, which prevents distributing the IMAS-integrated TRANSP code to users who have not signed the license with the ITER Organization. Without the IMAS interface, certain modules, like TORIC-6, are unavailable. When compiled with the IMAS interface, the TRANSP input files specify IMAS parameters: \texttt{IMAS\_TOK\_ID}, \texttt{IMAS\_REF\_SHOT}, and \texttt{IMAS\_REF\_RUNID} set the tokamak name, reference shot number, and run ID, respectively. \texttt{IMAS\_OUT\_RUNID} sets the output run ID. The inputs \texttt{IMAS\_TOK\_ID} and \texttt{IMAS\_REF\_SHOT} are optional. They need to be set only if the reference shot is different from the current shot. The parameter \texttt{IMAS\_BACKEND} defines the backend. Currently supported backends are  `MDSPLUS', `HDF5', and `ASCII'. Additionally, there is also a provision to read a pulse schedule from the appropriate IDS by setting \texttt{IMAS\_PULSE\_SET} to 1.

\section{Best practices in TRANSP software engineering\label{sec:practices}}
Adherence to best software engineering practices is essential to maintain robust and reliable research software projects, particularly for long-lived ones such as TRANSP. Software engineering practices also evolve over time, requiring the TRANSP development team to periodically refresh their approach. Although the TRANSP development team has long been committed to providing a consistent and trustworthy code, without constant effort, legacy projects will accumulate a significant amount of technical debt. By the late 2010s, it was recognized that the TRANSP software engineering approaches had not kept up with best practices. This led to a code that was difficult to develop and had major portability issues. In response, a significant effort was undertaken to improve the software engineering practices used by the development team. This included the removal of deprecated and ineffective physics modules from the code base; spinning out sections of the code base useful to other projects into their own code repositories, and removal of third-party codes that could be linked to as standalone libraries. Not only did this lead to a leaner and decluttered code base, it allowed for the formalization of a TRANSP dependency environment that could easily be repeated on other high-performance computing centers. The TRANSP home-grown build system was also deprecated in favor of a simpler GNU Make-based build system that allowed for parallel build and reduced the compilation time from over six hours to about one. These efforts led to a much more robust and portable code base that is much more efficient to develop and maintain.

Another significant effort with regards to software engineering was the adoption of modern Continuous Integration and Continuous Deployment (CI/CD) techniques. To enable this, TRANSP was moved to a GitHub repository which is mirrored to a local PPPL GitLab installation used to manage the CI/CD. Pushing a new branch to the GitHub repository automatically triggers a build on the PPPL cluster with both the Intel and the GNU compilers. An extensive suite of deterministic regression tests are run if the builds are successful. Development branches are required to pass the automated CI/CD before being merged into the main branch. The modernization of CI/CD with TRANSP has been focused on enabling the team to refactor and write large sections of the code with minimal breakage and has significantly improved the reliability of the code.
As new capabilities are added to the code, additional test cases are added to the CI/CD test suite.
Additional automated steps are run on production branches for the continuous deployment of artifacts, such as the installation of TRANSP and related codes and libraries on local computing clusters and the creation of software containers.

\section{Summary\label{sec:summary}}
TRANSP has evolved significantly since its inception in the late 1970s, becoming an indispensable tool for tokamak plasma analysis. It supports a wide range of tokamak configurations and performs thousands of simulations annually, aiding both current and future fusion energy experiments. The code's ability to handle time-dependent evolution of plasma profiles, along with its integration with IMAS, underscores its advanced capabilities in both interpretive and predictive modeling. The continuous improvements and integration of new models for heating, current drive, and handling of large-scale events enhance its utility. TRANSP's role in the broader fusion research community is vital, providing insights and supporting the optimization of tokamak operations. The adoption of modern software engineering practices further ensures its robustness and adaptability for future advancements.

The predictive and interpretive capabilities in the TRANSP code are discussed. The modules for sources, equilibrium, poloidal field diffusion equations are shared between the predictive and interpretive workflows. The power, particle, and momentum balance analyses used for interpretive analysis are explained in detail in this paper. It also discusses the derivation and implementation of the poloidal field diffusion equation within TRANSP. TRANSP offers three main options for advancing the representation of plasma MHD equilibrium over time. The first option involves reading equilibrium profiles from an external code, performing interpolation as necessary without recomputing the equilibrium internally. The second option is using the fixed-boundary inverse equilibrium solver TEQ, which solves for the equilibrium at each geometry timestep based on pressure and q profiles, adjusting the edge q to match the total plasma current. The third and most versatile option is the ISolver free boundary solver, which maintains a detailed model of the geometry and material characteristics of the poloidal field coil set. ISolver can perform a least-squares fit for the PF coil currents to match a prescribed plasma boundary or solve a circuit equation incorporating coil current data, feedback circuits, and induced vessel currents. ISolver can also advance the $q$ profile self-consistently, modeling the inductive coupling of the coils and vessel to the plasma.

TRANSP features a variety of modules for simulating heating and current drive in tokamak plasmas. The NUBEAM model handles neutral beam injection, detailing beam deposition and associated effects. For RF heating and current drive, TRANSP integrates externally developed models like TORAY-GA, TORBEAM, and GENRAY, with TORBEAM being IMAS-compatible. The TORIC model, available in two versions, computes current drive from the wavefield and supports non-symmetric antenna spectra. Lower Hybrid heating and current drive are managed using the GENRAY code and CQL3D Fokker-Planck solver, supplemented by the LSC-SQ module. These models provide detailed metrics on power distribution and driven current, allowing TRANSP to effectively simulate the complex interactions and effects of various heating and current drive sources in tokamak plasmas. \mc{TRANSP incorporates radiated power density models for bremsstrahlung, line radiation, and cyclotron radiation to ensure accurate calculation of the electron energy balance.} Large-scale events such as sawtooth crashes and pellet injections are addressed, showing how TRANSP handles these occurrences.

TRANSP incorporates advanced predictive capabilities developed since the 1990s, utilizing theory-based transport models for both anomalous and neoclassical transport. These models enable time-dependent simulations to forecast tokamak behavior under various operational scenarios, thereby assisting in optimizing plasma performance. The predictive framework shares many components with the interpretive framework, facilitating the addition of predictive features. Predictive TRANSP employs modular solvers such as PT\_SOLVER to advance plasma profiles for electron and ion temperatures, electron density, ion thermal density, total impurity density, and toroidal rotation velocity. The predictive transport equations include ion and electron energy conservation, angular momentum conservation, and density conservation for electrons and ions. These equations are solved using finite difference methods and the Newton iteration method. The predictive mode supports various theory-based models, including GLF23, TGLF, MMM, RLW, COPPI, and CDBM, along with neoclassical models like NEOCH, NEOGK, and NCLASS. This integration allows for accurate simulation of plasma behavior, providing critical insights for the design and operation of future fusion devices. 

The paper also touches on recent efforts to integrate TRANSP with the ITER Integrated Modeling and Analysis Suite (IMAS), enhancing data access and integration. The integration of TRANSP with the ITER Integrated Modeling and Analysis Suite (IMAS) represents a significant advancement, enhancing its capabilities for tokamak plasma analysis. IMAS provides a comprehensive environment of codes, libraries, frameworks, and databases to support both operation and research in fusion energy. TRANSP's integration into the IMAS framework allows for seamless access to experimental databases and facilitates integration with other modules and codes. This integration enhances TRANSP's modeling, verification, and validation capabilities, supporting the assessment of ITER scenarios and heating and current drive (H\&CD) requirements for fusion pilot plant design. The TRANSP output in IMAS format is primarily saved in the Interface Data Structures (IDSs) used for coupling with external modules. A standalone tool, \texttt{transp2imas}, written in Python, is used for more comprehensive conversion of TRANSP output to IMAS format. This conversion ensures that the output can be utilized for further analysis and integration with other simulation tools. TRANSP also includes several templates for coupling with external modules, including those for sources such as ECH, ICRF, and NBI. These interfaces have been tested with modules like TORBEAM, TORIC, and NUBEAM, supporting both IMAS and legacy interfaces. Overall, the IMAS integration significantly enhances TRANSP's utility for the global tokamak research community, providing advanced tools for data analysis and simulation, and supporting the optimization of tokamak operations.

The TRANSP development team has adopted best software engineering practices and modern Continuous Integration (CI) techniques to maintain a robust and reliable code. By the late 2010s, TRANSP required significant modernization due to development difficulties and portability issues. Efforts included removing deprecated physics modules, modularizing useful sections, and eliminating redundant third-party codes. The home-grown build system was replaced with a GNU Make-based system, improving compilation times and portability. TRANSP was migrated to a GitHub repository mirrored on PPPL's GitLab, automating builds and regression tests for stable code integration. This modernization has enhanced code reliability, facilitated extensive refactoring, and enabled the continuous addition of new capabilities, including automated deployment of software containers and installations on local clusters.

There are several topics not covered in this paper, including models for neutral transport and recent efforts on core-edge integration. These subjects will be addressed in future publications.

\section*{Acknowledgments}

The authors would like to express their deepest gratitude to Dr. Francesca Poli from the ITER Organization for her inspiration and encouragement in writing this paper. We sincerely thank Dr. Jeff Lestz from General Atomics for carefully reading this manuscript and for his constructive comments and valuable suggestions, which have helped improve the clarity and quality of this work. The research described in this paper was conducted at Princeton Plasma Physics Laboratory, a national laboratory operated by Princeton University for the United States Department of Energy under Prime Contract No. DE-AC02-09CH11466. The United States Government retains a non-exclusive, paid-up, irrevocable, worldwide license to publish or reproduce the published form of this manuscript, or allow others to do so, for United States Government purposes. 
The development and testing of the GPU optimized NUBEAM library was completed using resources of the National Energy Research Scientific Computing Center, which is supported by the Office of Science of the U.S. Department of Energy under Contract No. DE-AC02-05CH11231.

The data that supports the findings of this study are openly available at  \url{https://doi.org/10.34770/eb8a-4x71}.

\bibliography{manuscript}
\end{document}